\documentclass{aa}  

\usepackage{graphicx}
\usepackage[varg]{txfonts}
\usepackage{natbib}
\usepackage{booktabs}
\usepackage{longtable} 
\usepackage[maxfloats=256]{morefloats}
\usepackage{hyperref}
\begin{document} 

   \title{The MEGARA view of outflows in LINERs}

   \author{L. Hermosa Mu{\~n}oz \inst{1,2}
        \and S. Cazzoli\inst{1}
        \and I. M{\'a}rquez\inst{1}
        \and J. Masegosa\inst{1}
        \and M. Chamorro-Cazorla\inst{3,4} 
        \and A.~Gil de Paz\inst{3,4}
        \and {\'A}.~Castillo-Morales\inst{3,4} 
        \and J. Gallego\inst{3,4}
        \and E. Carrasco\inst{5}
        \and J. Iglesias-P{\'a}ramo\inst{1}
        \and M.~L.~Garc{\'i}a-Vargas\inst{6}
        \and P. G{\'o}mez-{\'A}lvarez\inst{6}
        \and S. Pascual\inst{3,4}
        \and A. P{\'e}rez-Calpena\inst{6}
        \and N. Cardiel\inst{3,4}
        }

   \institute{
        Instituto de Astrof\'isica de Andaluc\'ia - CSIC, Glorieta de la Astronom\'ia s/n, 18008 Granada, Spain \\
              \email{lhermosa@cab.inta-csic.es}
        \and
        Centro de Astrobiolog\'ia (CAB, CSIC-INTA), ESAC Campus, 28692 Villanueva de la Ca{\~n}ada, Madrid, Spain
        \and
        Departamento de F{\'i}sica de la Tierra y Astrof{\'i}sica, Universidad Complutense de Madrid, E-28040 Madrid, Spain
        \and
        Instituto de F{\'i}sica de Part{\'i}culas y del Cosmos IPARCOS, Facultad de Ciencias F{\'i}sicas, Universidad Complutense de Madrid, E-28040 Madrid, Spain
        \and
        Instituto Nacional de Astrof{\'i}sica, {\'O}ptica y Electr{\'o}nica, Luis Enrique Erro No.1, C.P. 72840, Tonantzintla, Puebla, Mexico
        \and 
        FRACTAL S.L.N.E.. C/ Tulip{\'a}n 2, p13, 1A, 28231, Las Rozas de Madrid, Spain
             }

   \date{Received M DD, YYYY; accepted M DD, YYYY}

 
  \abstract{
  Feedback processes, in particular those driven by outflows, are believed to play a major role in galaxy evolution. Outflows are believed to be ubiquitous in all Active Galactic Nuclei (AGNs), although their presence in low luminosity AGNs, in particular, for Low-Ionisation Nuclear Emission line Regions (LINERs), has only started to be explored. Their properties (geometry, mass and energetics) are still far from being properly characterised.}
  {The main goal is to use integral field spectroscopic data from the MEGARA instrument, at the Gran Telescopio Canarias (GTC), to analyse a small sample of nine LINERs, candidates of hosting ionised gas outflows. We aim to study the main emission lines in the optical wavelength range to identify their properties and physical origin.}
  {We obtained data cubes in several bands at the lowest (R$\sim$6000) and highest (R$\sim$20000) spectral resolution of MEGARA. We modelled and subtracted the stellar continuum to obtain the ionised gas contribution, and then fitted the emission lines to extract their kinematics (velocity and velocity dispersion). We identified outflows as a secondary component in the emission lines and obtained their main properties.}
  {The primary component of the emission lines was typically associated to gas in the galactic disc. For some objects, there is an enhanced-$\sigma$ region typically co-spatial with the secondary component. We associated it to turbulent gas produced due to the interaction with the outflows. We find signatures of outflows in six LINERs, with mass outflow rates ranging from 0.004 to 0.4 $M_{\sun}$\,yr$^{-1}$ and energy rates from $\sim$\,10$^{38}$ to $\sim$\,10$^{40}$\,erg\,s$^{-1}$. Their mean electronic density is 600\,cm$^{-3}$, extending to distances of $\sim$\,400\,pc at an (absolute) velocity of $\sim$\,340\,km\,s$^{-1}$ (on average). They tend to be compact and unresolved, although for some sources they are extended with a bubble-like morphology.}
  {Our results confirm the existence of outflows in the best LINER candidates identified using previous long-slit spectroscopic and imaging data. These outflows do not follow the scaling relations obtained for more luminous AGNs. For some objects we discuss jets as the main drivers of the outflows.}


   \maketitle
%
\section{Introduction}
\label{Sec:Intro}
   \noindent Outflows are believed to be ubiquitous in galaxies that host an active galactic nuclei (AGN) \citep[e.g.][]{Veilleux2005,Morganti2017,Veilleux2020}. They are thought to be powered either by the activity of the supermassive black hole (SMBH), or triggered by the kinetic energy injected by the jet, although in reality they could be a consequence of both processes acting together \citep{Veilleux2005,Fabian2012,Kormendy2013,Heckman2014}, or even some contribution from nuclear star formation. Outflows have been extensively studied in luminous AGNs, as they are more powerful and thus easy to identify \citep[e.g.][]{Fluetsch2019}. However, they are also present in lower-luminosity AGNs such as Low-Ionisation Nuclear Emission-line Regions (LINERs) \citep{Masegosa2011,Dopita2015,Cazzoli2018,HM2020,Cazzoli2022,HM2022}. LINERs are the largest population of AGNs in the Local Universe \citep{Heckman1980,Ho1997,Marquez2017}, and thus ideal targets for doing spatially resolved studies. 

   \noindent \cite{HM2022} (hereafter HM22) made the largest systematic study searching for outflows in LINERs by obtaining narrow band imaging data for 70 nearby (z$<$0.025) galaxies. The targets were confirmed to be genuine AGNs through the detection of a compact X-ray source in their centres \citep{GM2009,Marquez2017,Agostino2023}. This imaging information allows to detect, or at least hint, the presence of outflows as extended ionised gas structures emerging from the nuclear region of the galaxies \citep[see e.g.][]{Pogge2000,Masegosa2011}. However, the presence of extended, filamentary ionised gas may be related to processes other than outflows, such as past interactions or mergers \citep[e.g.][]{Raimundo2021}. To estimate the reliability of the ionised gas distribution as a tracer of outflows, HM22 complemented the images with all the kinematical information available in the literature for the galaxies. This is considered a more reliable proof for outflows, since outflow signatures in the emission or absorption lines are easily identified as an additional, typically broad and blueshifted, kinematical component affecting the line profiles \citep[e.g.][]{Davies2014,Harrison2016,Mingozzi2019,Davies2020,Cazzoli2020,HM2020,Raimundo2021,Cazzoli2022}. By combining imaging and kinematical information, HM22 reported a detection rate of $\sim$50\% of outflows in LINERs.
   However, the kinematical data in that work came from various instruments, covering different wavelength ranges at different spectral and spatial resolutions, which may affect the detection and characterisation of the outflows to its full extension. 
   This can be specially problematic for the detections with long slit spectroscopic data, since the signatures of outflows will not be retrieved if the slit is not aligned with the relevant feature. Thus, uniform integral field spectroscopic (IFS) data are needed to establish unified criteria and confirm (or not) the reported detection rate.
   
   \noindent To that aim, we performed a pilot study with a small sample of nine LINERs that were the best candidates to host an outflow either by imaging data or by long-slit kinematic information \citep[see][HM22]{Masegosa2011,Cazzoli2018,HM2020}. We obtained IFS data with the MEGARA (\textit{Multi-Espectr{\'o}grafo en GTC de Alta Resoluci{\'o}n para Astronom{\'i}a}) instrument, at the Gran Telescopio Canarias (GTC), for all the targets. We have already published our first study on the characterisation of outflows for the prototypical LINER NGC\,1052, finding evidence for an outflow with both MUSE, at the Very Large Telescope (VLT), and MEGARA/GTC data (see \citealt{Cazzoli2022}, hereafter C22). In this paper, we analyse optical IFS data for the remaining eight galaxies with the aim to: (i) model the emission lines of the ionised gas to detect and confirm the possible outflow signatures; (ii) characterise the main properties (kinematics, energetic, mass, etc.) of the outflows; and (iii) compare the information from the kinematic signatures and the imaging data from HM22.

   \noindent This paper is organised as follows: in Sect.~\ref{Sect:sample_data} we describe the sample selection and data; in Sect.~\ref{Sect:Datareduction} we describe the data reduction process and the analysis procedure, with the modelling of both the stellar continuum and the emission lines per each cube, and the estimation of outflow parameters. In Sect.~\ref{Sect:Results} we present the kinematic maps for both stellar and gaseous components resulting from our analysis. We discuss the results, the outflow detection and their properties, and compare to previous works in Sect.~\ref{Sect:Discussion}. Finally, the conclusions are presented in Sect.~\ref{Sect:Future}. Throughout the paper we adopt the cosmological parameters $H_{0}$\,=\,67.7 km\,s$^{-1}$\,Mpc$^{-1}$, $\Omega_{\rm m}$\,$=$\,0.31 and $\Omega_{\rm \Lambda}$\,$=$\,0.69 \citep{Planck2018}.
   
\section{Sample and data}
\label{Sect:sample_data}

   \noindent We analysed a total of nine LINERs (see Table~\ref{Table:galaxieslisted}). All the objects were confirmed as true AGNs based on the detection of a hard X-ray unresolved central source \citep{GM2009}. The parent sample of our data set was extracted from \cite{Cazzoli2018} and \cite{HM2020}, where the long-slit spectroscopic properties of 21 and nine type 1.9 and 2 LINERs were characterised, respectively; and from HM22, where narrow band imaging data for 70 LINERs were analysed. We selected targets for which \cite{Cazzoli2018,HM2020} and HM22 reported the detection of non rotational motions with the spectroscopic information and/or morphological signatures in the ionised gas that could be ascribed to outflows, presumably associated to the AGN. 
   
\begin{figure}
      \includegraphics[width=\columnwidth]{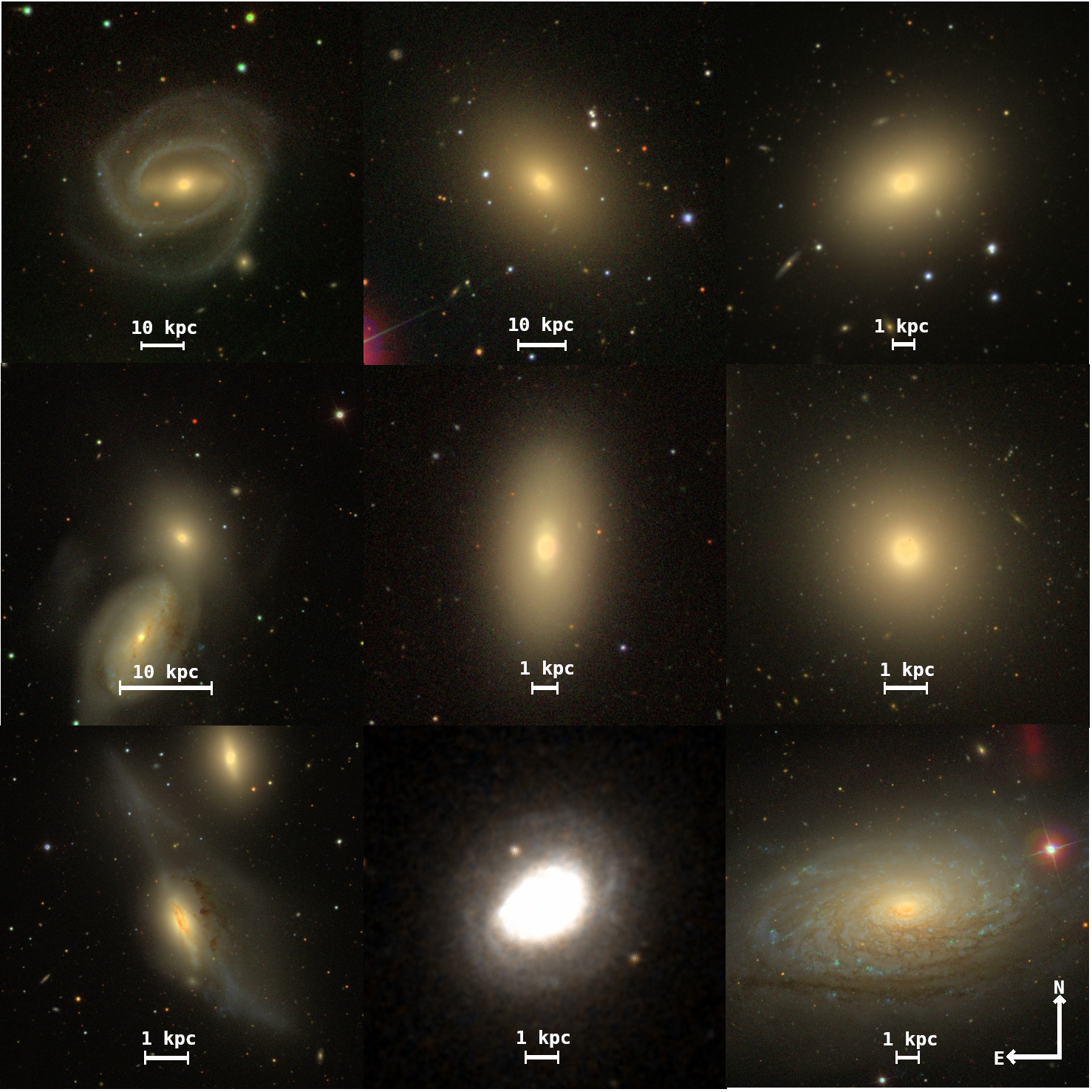}
   \caption{Optical images of the nine LINERs observed with MEGARA. From left to right, top to bottom: NGC\,0266, NGC\,0315, NGC\,1052, NGC\,3226, NGC\,3245, NGC\,4278, NGC\,4438, NGC\,4750 and NGC\,5055. All images have been obtained from the Sloan Digital Sky Survey DR9, except for NGC\,4750, that is from the Digitalized Sky Survey. The physical scale is indicated for all images; north is up and east to the left.}
   \label{Fig5:AllGalaxies}
   \end{figure}

   \noindent We gathered IFS data with the MEGARA instrument for nine targets, one of which was NGC\,1052, analysed in detail in C22. The complete information of the galaxies is summarised in Table~\ref{Table:galaxieslisted}. The data and observations are described in Sect.~\ref{SubSec:datagather} and Table~\ref{Table:obslog}. We show as a reference the optical images of the host galaxies in Fig.~\ref{Fig5:AllGalaxies}.
  
   \subsection{Data gathering}
   \label{SubSec:datagather}
   
   \noindent We observed the galaxies with the MEGARA instrument \citep{GildePaz2016,GildePaz2018,Carrasco2018}, attached to the GTC telescope (10.4 m), at Roque de los Muchachos observatory in La Palma, Spain. We used MEGARA in its Large Compact Bundle Integral Field Unit (IFU) mode, that is composed by a total of 623 fibres (567 for the scientific target, and 56 for obtaining sky-background spectra), each fibre covers a spatial scale of 0.62\arcsec\,, that gives a total projected field-of-view (FoV) on sky of 12.5\arcsec$\times$11.3\arcsec. MEGARA uses several Volume Phase Holographics (VPHs) available at different wavelengths with spectral resolutions from R$\sim$6000 to R$\sim$20000. In this work we use three different VPHs at the lowest (LR-B, LR-V and LR-R) and one at the highest resolution (HR-R).
   
   \noindent The VPHs, following the MEGARA cookbook\footnote{\url{http://www.gtc.iac.es/instruments/megara/media/MEGARA_cookbook_1I.pdf}}, have the following characteristics: (i) \textit{LR-B:} wavelength range of 4330-5200\,\AA\,with an average resolution R\,$=$\,6060 (i.e. 0.23\,\AA\,pix$^{-1}$); (ii) \textit{LR-V:} wavelength range of 5140-6170\,\AA\,with an average resolution R\,$=$\,6080 (i.e. 0.27\,\AA\,pix$^{-1}$); (iii) \textit{LR-R:} wavelength range of 6100-7300\,\AA\,with an average resolution R\,$=$\,6100 (i.e. 0.31\,\AA\,pix$^{-1}$); and (iv) \textit{HR-R:} wavelength range of 6440-6840\,\AA\,with an average resolution R\,$=$\,18700 (i.e. 0.09\,\AA\,pix$^{-1}$) \citep[for a detailed description, see][]{Chamorro-Cazorla2022}.
   At the redshift of our targets (\textit{z}$<$0.02), in the wavelength range covered from LR-B to LR-R, we can detect the emission lines [O\,III]$\lambda\lambda$4959,5007\,\AA\AA, H$\beta \lambda$4861\,\AA, [S\,II]$\lambda\lambda$6716,6731\,\AA\AA, [N\,II]$\lambda\lambda$6548,6584\,\AA\AA, H$\alpha \lambda$6563\AA\,and [O\,I]$\lambda\lambda$6300,6363\,\AA\AA, and the absorption lines of NaI$\lambda\lambda$5890,5896\,\AA\AA. 

   \noindent The data were gathered in a total of 10 nights (PIs: S. Cazzoli and A. Gil de Paz). We obtained two (three) individual exposures for VPH per galaxy, with typical exposure times of 1200\,s (900\,s) for the data obtained through the MEGADES (MEGARA Galaxy Disk Evolution Survey; see \citealt{Chamorro-Cazorla2022}) survey (open time). The average airmasses were 1.22 (LR-B), 1.25 (LR-V) and 1.3 (LR-R) (see Table~\ref{Table:obslog}).
   
   \noindent We note that for the galaxies in the MEGADES sample (see Table~\ref{Table:obslog}), they were observed with the LR-B, LR-V, LR-R and/or HR-R VPHs, as the whole wavelength range is needed to fulfil the scientific objectives of the survey \citep{Chamorro-Cazorla2022}. Nevertheless, the data obtained through our open time program of GTC (see Table~\ref{Table:obslog}) only uses LR-R and LR-V as these have typically higher sensitivity and signal-to-noise (S/N) per spaxel than that of LR-B \citep{Chamorro-Cazorla2022}. More importantly, the wavelength coverage of LR-V and LR-R is sufficient for obtaining a proper stellar modelling and a robust identification of the kinematical properties of the emission lines associated to the ionised gas, that may be related to outflows (see Sect.~\ref{Sect:Datareduction}).

\begin{table*}
  \caption{\label{Table:galaxieslisted} General properties for all the 9 LINERs analysed in this work. }
  \centering
  \begin{tabular}{clccccccrrr}
    \hline \hline
    \# & ID & RA (2000) & DEC (2000) & Morphology & z & Scale & \textit{i} & V mag & $R_{\rm e}$ & log(L$_{\rm bol}$) \\
    & & (hh mm ss) & (dd mm ss) &  &  & (pc per \arcsec) & (deg) &  & (\arcsec) & (erg\,s$^{-1}$) \\
    & (1) & (2) & (3) & (4) & (5) & (6) & (7) & (8) & (9) & (10) \\
   \hline
    \hline 
    1& NGC 0266 & 00 49 47.80 & +32 16 39.79   & SB(rs)ab     & 0.0155           & 351           & 16 & 11.8 &  --  & -- \\
    2& NGC 0315 & 00 57 48.88 & +30 21 08.81   & E+           & 0.0165           & 370           & 57 & 11.6 & 25.1 & 43.30 \\
    3& NGC 1052 & 02 41 04.80 & $-$08 15 20.75 & E4           & 0.005\phantom{1} & 108           & 70 & 11.0 & 21.9 & 42.77 \\
    4& NGC 3226 & 10 23 27.01 & +19 53 54.68   & E2 pec       & 0.0044           & \phantom{1}86 & 69 & 12.9 & 30.9 & 42.33 \\
    5& NGC 3245 & 10 27 18.39 & +28 30 26.56   & SA0$^0$(r)   & 0.0044           & \phantom{1}90 & 62 & 10.8 & 25.1 & 42.29 \\
    6& NGC 4278 & 12 20 06.82 & +29 16 50.72   & E1-2         & 0.0021           & \phantom{1}42 & 16 & 10.2 & 31.6 & 42.53 \\
    7& NGC 4438 & 12 27 45.59 & +13 00 31.78   & SA(s)0/a pec & 0.0002           & \phantom{1}29 & 73 & 10.9 &  --  & $<$42.36 \\
    8& NGC 4750 & 12 50 07.27 & +72 52 28.72   & (R)SA(rs)ab  & 0.0054           & 129           & 32 & 12.1 &  --  & 41.51 \\
    9& NGC 5055 & 13 15 49.33 & +42 01 45.40   & SA(rs)bc     & 0.0017           & \phantom{1}40 & 55 & 8.6  &  --  & 41.09 \\
    \hline
    \end{tabular} \\
    \tablefoot{Columns indicate: (2) RA and (3) DEC: coordinates; (4) Morphology, (5) redshift and (6) scale based on their distances to the Local Group barycentre, as given by NED; (7) inclination angle and (8) V magnitude from HyperLeda; (9) effective radius from \cite{Cappellari2011}, \cite{Veale2017} (NGC\,0315) and see C22 for NGC\,1052; and (10) bolometric luminosity estimated from \cite{GM2009b}, except for NGC\,4750, estimated from \cite{Younes2012}.}
\end{table*}

    \begin{table*}
        \caption{Observing log of MEGARA/GTC data.}
        \label{Table:obslog}
        \centering
        \begin{tabular}{lccccc}
        \hline \hline
        Galaxy & Observing date & VPH & Exposure time & Airmass & Seeing  \\ 
         & (yyyy-mm) &  & (s) &  & (\arcsec) \\
        (1) & (2) & (3) & (4) & (5) & (6) \\ \hline
        NGC\,0266         & 2020-08 & LR-V   & 3$\times$900  & 1.02 & 1.0 \\ 
                          &         & LR-R   & 3$\times$900  & 1.20 & 0.9 \\ 
        NGC\,0315         & 2019-09 & LR-V   & 3$\times$900  & 1.08 & 1.1 \\ 
                          & 2019-10 & LR-R   & 3$\times$900  & 1.18 & 1.1 \\ 
        NGC\,3226         & 2018-12 & LR-V   & 3$\times$1200 & 1.15 & 0.8 \\ 
                          &         & LR-R   & 3$\times$1200 & 1.68 & 0.6 \\ 
        NGC\,3245         & 2021-02 & LR-V   & 3$\times$900  & 1.60 & 1.0 \\ 
                          &         & LR-R   & 3$\times$900  & 1.15 & 0.9 \\ 
        NGC\,4278$^{(*)}$ & 2019-05 & LR-B   & 3$\times$1200 & 1.01 & 0.6 \\ 
                          &         & LR-V   & 3$\times$1200 & 1.07 & 0.5 \\ 
                          &         & HR-R   & 3$\times$1200 & 1.22 & 0.5 \\ 
        NGC\,4438         & 2019-01 & LR-V   & 3$\times$1200 & 1.06 & 1.0 \\ 
                          &         & LR-R   & 3$\times$1200 & 1.05 & 1.0 \\ 
        NGC\,4750$^{(*)}$ & 2019-05 & LR-B   & 3$\times$1200 & 1.43 & 0.7 \\ 
                          &         & LR-V   & 3$\times$1200 & 1.40 & 0.7 \\ 
                          &         & LR-R   & 3$\times$1200 & 1.55 & 0.6 \\
        NGC\,5055$^{(*)}$ & 2020-01 & LR-V   & 3$\times$1200 & 1.64 & 1.2 \\ 
                          &         & LR-R   & 3$\times$1200 & 1.31 & 1.2 \\ 
        \hline
        \end{tabular} \\
        \tablefoot{Columns indicate: (1) object; (2) date of the observation; (3) VPHs used; (4) exposure time; (5) mean airmass per VPH; and (6) mean seeing per target. $^{(*)}$ indicates the objects observed within the MEGADES survey (see Sect.~\ref{SubSec:datagather}).}
    \end{table*}

\section{Data reduction and data handling}
\label{Sect:Datareduction}

\noindent The data from MEGARA/GTC were reduced following the MEGARA Data Reduction Pipeline \citep[v0.11][]{Pascual2019,MegaraDRP2020ACM}. All the software and packages throughout the analysis are used in a \textsc{python} environment. The full process is described in \cite{GildePaz2018,Cazzoli2020} and \cite{Chamorro-Cazorla2022}; here we briefly describe the main steps. The pipeline applies the standard data reduction procedures, as bias subtraction, flat-field correction and wavelength and flux calibration (with the spectrophotometric standard stars observed each night). The pipeline considers the information from all the fibres individually, to trace them over the FoV and apply for each one a specific illumination correction. For one object (NGC\,4750) the last bundle (covering 20 fibres) of the LR-R cube was not properly traced due to a poor spectrograph focus and thus there are some missing fibres in the north part of the final cube (see Sect.~\ref{SubSec:ResultsGas} and \citealt{Chamorro-Cazorla2022} for more details). Finally, we combine the individual exposures using a median to create the final \textit{RSS} file\footnote{A Raw Stacked Spectra (\textit{RSS}) file is the final product of the data reduction; an image containing the spectra of the 623 fibres.} for each VPH and each target. For two objects (namely NGC\,0266 and NGC\,0315), the different observing blocks (i.e. exposures) for a given VPH were not observed during the same night, so the individual exposures needed to be shifted in spatial position. In these cases, to obtain the final combined cube, we created a mosaic with the task \textsc{hypercube} under the \textsc{megara-tools} package \citep[v.0.1.0;][]{megaratools2020}. After that, for all targets we transformed the final \textit{RSS} file to a data cube by applying a regularisation grid to obtain 0.4\arcsec\,square pixels \citep[see][]{Cazzoli2020}. \\

\noindent Once the cubes were fully reduced, we combined the LR-V and LR-R data cubes for all the targets. The procedure was the same we used for NGC\,1052 (see C22); the wavelength range corresponding to the LR-R VPH (see Sect.~\ref{SubSec:datagather}) lacked from sufficient stellar features to obtain a good modelling of the stellar component of the galaxies. Hence we decided to combine LR-R and LR-V data cubes to maximise the wavelength coverage and increase the quality of the stellar continuum fitting, specially in the LR-R range (see Fig.~\ref{Fig5:examplePPXF}). Given the difference in resolution between both VPHs (0.27\AA/pix vs 0.31\AA/pix), we degraded the LR-V resolution to that of the LR-R.

\noindent For two targets (namely NGC\,4278 and NGC\,4750) we additionally obtained LR-B data cubes (see Table~\ref{Table:obslog}). At their redshifts, the emission lines available are located at the reddermost part of the available wavelength range. This implies that these data cubes have sufficient stellar features to obtain a proper stellar modelling without combining with the other VPHs (i.e. LR-R \& LR-V). Finally, NGC\,4278 was observed with both the lowest (LR-B and LR-V) and highest resolution mode of MEGARA (HR-R instead of LR-R). The spectral coverage of HR-R is small ($\Delta\lambda \sim$400\,\AA; see Sect.~\ref{SubSec:datagather}) and the continuum featureless, so the stellar subtraction was not performed, as for NGC\,7469 in \cite{Cazzoli2020}. \\

\noindent For two objects (namely NGC\,0315 and NGC\,4750) the LR-R and LR-V cubes were not aligned, as they were observed in different nights. To estimate the photometric centre of every cube, we created a broad band-like image by collapsing along the wavelength dimension and fitted them to ellipses. The central positions of the ellipses were taken as references for the photometric centre. When the positions differed significantly ($>$2 square-pixels, i.e. 0.8\arcsec) from the LR-V data cube to the LR-R, we shifted the cubes to set their photometric centres at the same spatial positions. Consequently, their final maps cover a smaller field of view than the rest of the targets (see Sect.~\ref{Sect:Results}). The fluxes of both data cubes where scaled in the wavelength range common to both LR-R and LR-V cubes to match their continuum levels, using as reference the LR-R cube. \\

\noindent Finally, we complemented the spectroscopic data with optical images of the galaxies observed with ALFOSC, located at the Nordic Optical Telescope (NOT) or the Hubble Space Telescope (HST). The complete description of these images can be found in HM22. As mentioned in Sect.~\ref{Sec:Intro}, we obtained H$\alpha$ images of 70 targets among which we find all the nine galaxies of this paper. These images have a spatial sampling of 0.21\arcsec\,pix$^{-1}$ and 0.05\arcsec\,pix$^{-1}$ (ALFOSC/NOT and HST, respectively), and are useful to compare with possible substructures unresolved kinematically with the MEGARA data cubes (spatial sampling of 0.62\arcsec).

\subsection{Stellar continuum modelling}
\label{SubSec:datared_stars}

\noindent We used the Penalized PiXel-Fitting (\textsc{pPXF}) software by \citet{Cappellari2003} (see \citealt{Cappellari2017} and references therein) to subtract the stellar contribution from the host galaxy. This is crucial for our LINERs, where the emission from the AGN is usually heavily diluted by the stellar continuum, especially in the Balmer lines \citep[see e.g.][]{Agostino2023}. As mentioned in Sect.~\ref{Sect:Datareduction}, we combined the LR-V and LR-R cubes (hereafter we refer to the LR-VR cube) to increase the continuum range in the spectra and improve the overall fitting by including more stellar features. For the correct modelling of the stellar population, we masked all the emission, Na\,I doublet absorption, telluric and atmospheric lines. We used the additive and multiplicative Legendre polynomials to adjust the continuum shape of the selected templates, with the lowest possible degree to obtain a proper modelling. We set the degree of the additive and multiplicative polynomials to be 8 and 4, respectively, for all the targets (C22). In the same way we fitted the stellar continuum of LR-B data cubes individually (see Sect.~\ref{Sect:Datareduction}; additive and multiplicative polynomials 8 and 4, respectively). We used the synthetic stellar library by \cite{GD2005}, which provides a total spectral coverage from 4000 to 7000\AA, enough to cover the spectral range of our data cubes with the same resolution. The 413 synthetic stellar spectra correspond to a metallicity of 0.02\,Z$_{\odot}$, and cover the whole range of other stellar parameters, such as the temperature or the gravity \citep[see ][C22 for more details]{GD2005,Martins2005}.\\

\begin{figure}
    \centering
        \includegraphics[width=.99\columnwidth]{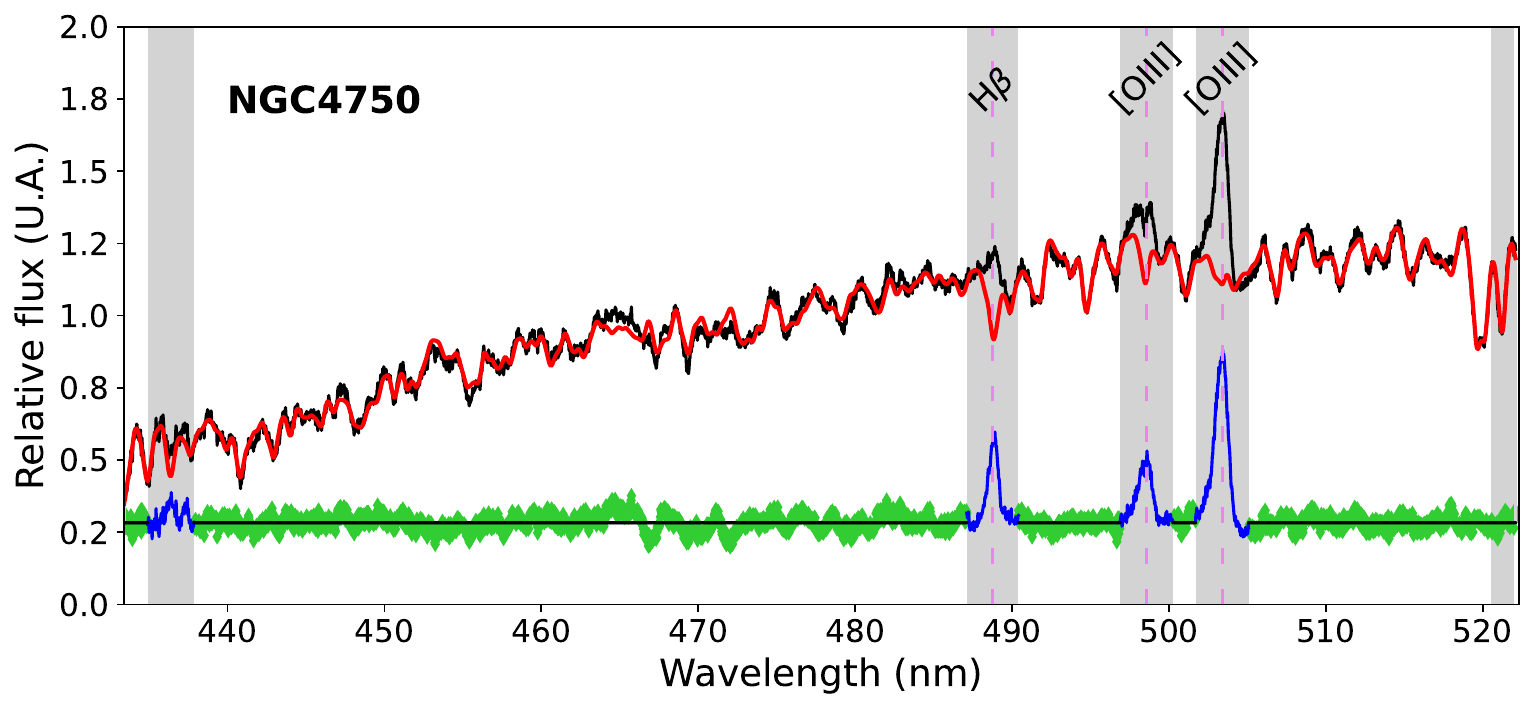}
        \includegraphics[width=\columnwidth]{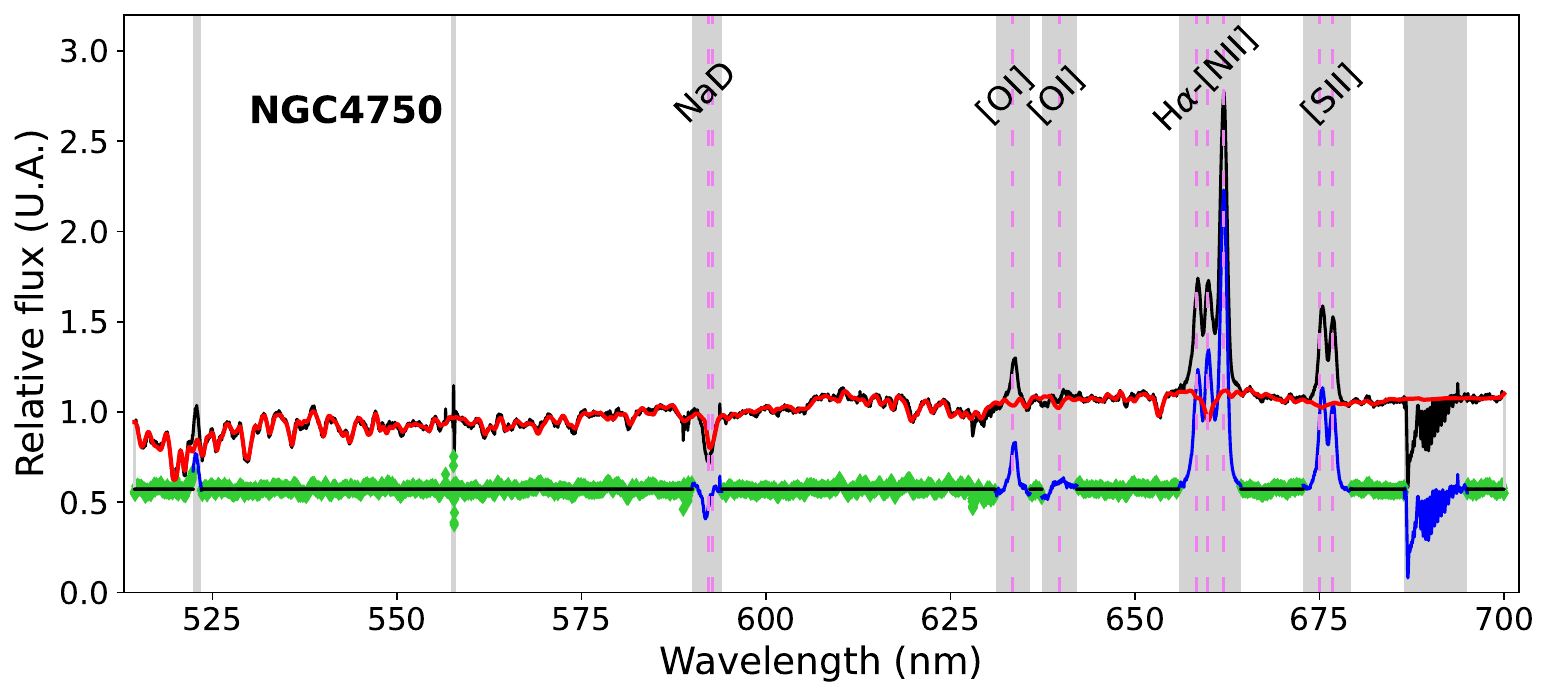}
    \caption{Stellar modelling of the LR-B (top) and LR-VR (bottom) integrated spectra in the PSF region of NGC\,4750. The red line shows the best fitting using \textsc{pPXF}, the green line shows the residuals of the fit, and the blue line is the final subtracted spectrum. The main emission and absorption lines are indicated with dashed vertical lines. Grey regions were masked during the fitting.}
    \label{Fig5:examplePPXF}
\end{figure}

\noindent In total, we fitted the combined LR-VR cubes for 7 LINERs, namely: NGC\,0266, NGC\,0315, NGC\,3226, NGC\,3245, NGC\,4438, NGC\,4750 and NGC\,5055; the LR-V cube for NGC\,4278; and the LR-B cubes for NGC\,4278 and NGC\,4750. From the modelling we obtained the main kinematical properties on a pixel by pixel basis (see Figs.~\ref{Fig5:examplePPXF} and~\ref{Fig5:exampleEmissionLineFitting}) and generated the final maps for each galaxy. The main results are found on the top panels of Figs.~\ref{Fig5:KinMaps_NGC0266} to~\ref{Fig5:KinMaps_NGC5055}, where we show the stellar properties of velocity, velocity dispersion ($\sigma$) and the galactic continuum maps of each LINER. The stellar modelling for both the LR-VR combined cube and the LR-B cube of NGC\,4750 is shown in Fig.~\ref{Fig5:examplePPXF} as an example.

\subsection{Emission line modelling}
\label{SubSec:datared_gas}

\noindent Once the stellar component was subtracted, we modelled the emission lines using the Non-Linear Least-Squares Minimization and Curve-Fitting package, \textsc{lmfit} \citep{lmfit}. We applied a similar technique as in our previous works \citep[see][C22]{Cazzoli2018,HM2020,Cazzoli2020}. The process is briefly summarised here. 

\noindent We first modelled the [S\,II] lines to use them as a reference for the remaining emission lines (i.e. the [N\,II] doublet, H$\alpha$ and [O\,I] doublet when possible), fixing their central wavelength (associated to the gas velocity) and its width (associated to the velocity dispersion). We imposed the intensity ratios between the [O\,III] (LR-B), [O\,I], and [N\,II] doublets to be 2.99, 3.13 and 2.99, respectively \citep{Osterbrock2006}. The velocity dispersion was corrected from the instrumental broadening as follows: $\sigma = \sqrt{\sigma_{obs}^{2}-\sigma_{inst}^{2}}$. To ensure a proper modelling of the profiles, we fitted the reference lines (i.e. [S\,II], [O\,III] and H$\beta$, see below) when their S/N was larger than 5.

\noindent Instead of using the [S\,II] as reference (the so-called S-method), we could also use the [O\,I]\,6300\AA\,emission line (the O-method; see more details in \citealt{Cazzoli2018} and \citealt{HM2020}). However the [O\,I] line is detected with enough S/N ($>$5) only in three galaxies from the sample (i.e. NGC\,3226, NGC\,4750 and NGC\,5055), and exclusively in the inner regions ($<$2\arcsec\,from the centre). Thus, the O-method was discarded given the generally low S/N of the line. Additionally, we found that the kinematics derived from the [S\,II] lines reproduced better the profiles of the [N\,II]-H$\alpha$ blend (i.e. with lower residuals) than the [O\,I]. Only the S-method can be applied to model the HR-R cubes, given that this cube lacks the [O\,I] lines due to the wavelength coverage. 

\noindent For the LR-B cube, we fitted the [O\,III] and H$\beta$ lines independently of the emission lines in the LR-R/HR-R cubes. Similarly to the kinematics of [O\,I] and H$\alpha$ lines, we found that [O\,III] and H$\beta$ have different kinematics (see Sect.~\ref{SubSec:ResultsGasKin}) as already reported for NGC\,1052 using MUSE data in C22. Thus the emission lines of these data cubes were modelled separately, with the intensity ratio between the two [O\,III] lines being the only fixed parameter in the fit. \\

\noindent For the line modelling, we allowed up to three Gaussian components per emission line, plus an additional very broad component (full-width at half maximum: FWHM\,$>$\,1000\,km\,s$^{-1}$) only within the point-spread function (PSF) region to reproduce the broad Balmer lines originated in the broad line region (BLR). The PSF region is determined individually for each galaxy from the FWHM of brightness distribution of the standard star used for the data flux calibration. We refer to the three Gaussian and the BLR components as the primary (P), secondary (S), tertiary (T) and broad (B), respectively (see Sect.~\ref{SubSec:datared_CalcOutfParam}). They were ordered to ensure a continuity in the velocities and widths at the final maps, which implies that the primary component usually has smaller widths than the secondary. In order to evaluate the number of components preventing overfitting, we used the $\varepsilon$ criterion \citep[][C22]{Cazzoli2018,HM2020,Cazzoli2020}. This parameter $\varepsilon$ is defined as the standard deviation of the residuals under the lines when fitted with a Gaussian component, with respect to its value in the continuum ($\varepsilon_{\rm c}$) on a 50\AA-wide region with no emission lines. We considered the inclusion of an additional component whenever $\varepsilon > 2.5$, as this means that the standard deviation of the remaining residuals could not be explained simply as continuum variations. We visually inspected the spectra for $\varepsilon$ between 2 and 2.5.

\begin{figure}
    \centering
    \includegraphics[width=\columnwidth]{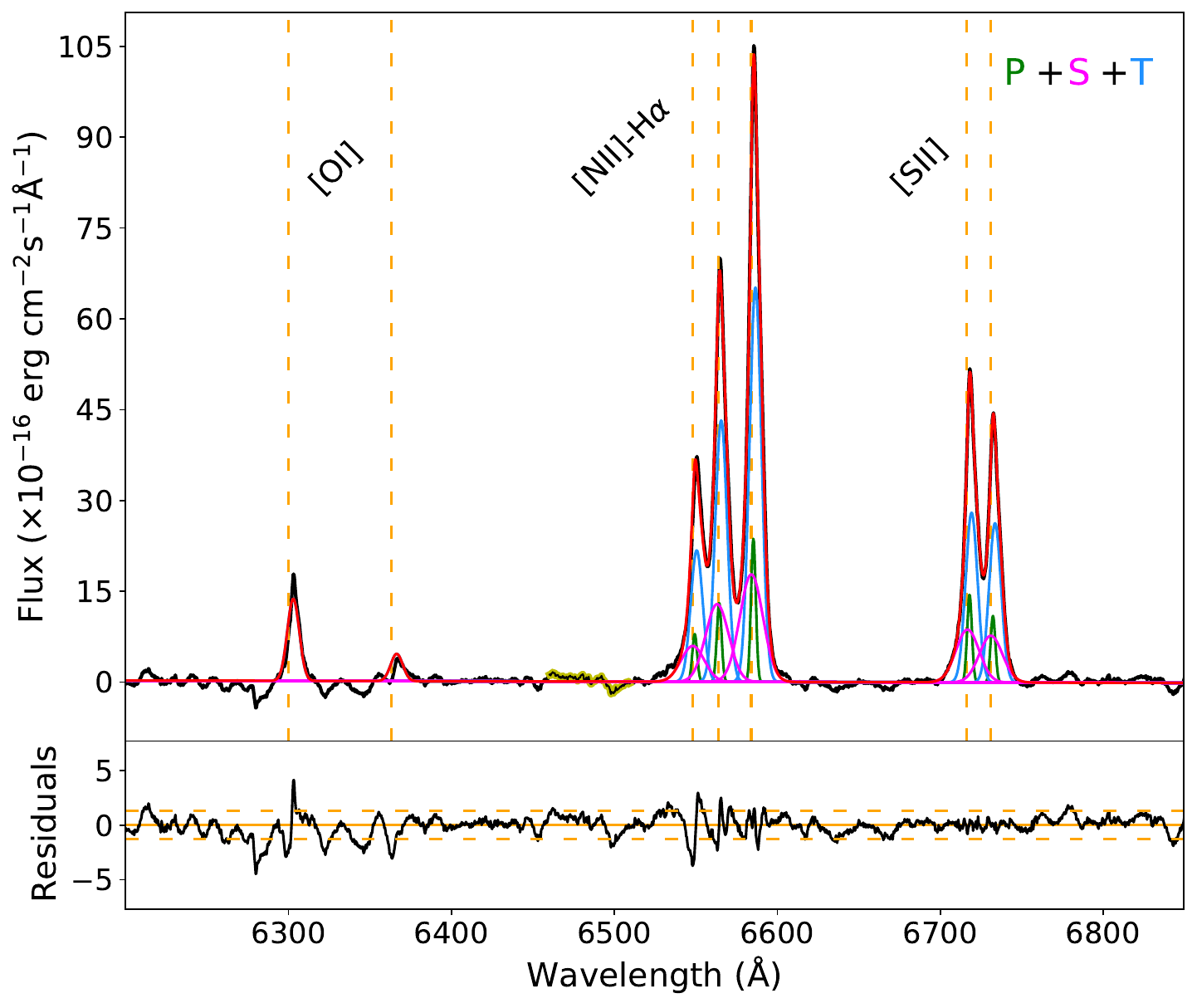}
    \caption{Ionised gas modelling of the integrated LR-R spectrum on the PSF region of NGC\,4438. In the top panel the green, pink and blue Gaussians represent the primary (P), secondary (S) and tertiary (T) kinematical components, respectively. The global fit is in red and the region where we estimated the $\varepsilon$ from the continuum ($\varepsilon_{c}$) is marked in yellow (see Sect.~\ref{SubSec:datared_gas}). In the bottom panel the residuals of the fit are shown in black, with the horizontal, dashed, yellow lines indicating 3$\varepsilon_{c}$.}
    \label{Fig5:exampleEmissionLineFitting}
\end{figure}

\noindent To test whether there was a very broad component in the Balmer lines for each galaxy, we produced a high S/N spectrum integrating all the individual spectra in the PSF region, so that the BLR component could be more easily detected. If a BLR component was required, we included it in the individual pixels corresponding to the PSF region. The ionised gas modelling of the integrated spectrum in the PSF region of NGC\,4438 is shown in Fig.~\ref{Fig5:exampleEmissionLineFitting} as an example. 

\subsection{Estimation of the ionising source through line ratios}
\label{SubSec:datared_BPTs}

\noindent By using the ratios between the fluxes of the emission lines of the individual components, we can derive their origin by using the BPT diagrams \citep{Baldwin1981,Veilleux1987,Kauffmann2003,Kewley2006}. These diagrams use log([S\,II]/H$\alpha$), log([N\,II]/H$\alpha$), log([O\,I]/H$\alpha$) and log([O\,III]/H$\beta$) to study the most probable ionising source of the gas, dividing mainly between star-forming and AGN ionisation (LINER or Seyfert). We could also include shock models in these diagrams, as they can ionise the medium producing similar ratios to that of true LINERs \citep[see e.g.][C22, and Sect.~\ref{Sec:Intro}]{Marquez2017,Cazzoli2018}.

\noindent With our data, we can only produce the BPT diagrams for two objects, namely NGC\,4278 and NGC\,4750, since only in these cases we have the LR-B data cube that allows us to estimate the log([O\,III]/H$\beta$) ratio (see Sect.~\ref{SubSec:datared_gas}). For the remaining sources we consider conservative limits for log([S\,II]/H$\alpha$), log([N\,II]/H$\alpha$) and log([O\,I]/H$\alpha$) to provide some insights into the most likely ionising source of the gas. Using the dividing lines from \cite{Kauffmann2003} and \cite{Kewley2006}, and assuming a log([O\,III]/H$\beta$) ratio $> -1.0$ (as found for the LINERs using long-slit spectroscopy in \citealt{Cazzoli2018}), we see that if log([N\,II]/H$\alpha$)\,$> 0.2$, log([S\,II]/H$\alpha$)\,$> 0.0$ and log([O\,I]/H$\alpha$)\,$> -1.0$, the regions in the galaxy lay in the AGN-photoionisation region. We note that in all cases, a shock contribution could not be discarded (see Sect.~\ref{SubSec:ResultsGasLineRatios}). 

\subsection{Association of the individual emission line components to a physical interpretation}
\label{SubSec:datared_CalcOutfParam}

\noindent Based on the emission line fitting, we can associate a physical interpretation to each kinematical component used in the modelling. The primary component of the gas is typically interpreted as the gaseous disc of the galaxy, which is expected to be correlated with the stellar component of regular galaxies\footnote{We note that this is not always the case, as some galaxies may have polar discs or may have gone through processes leading to misalignments \citep[e.g.][]{Sarzi2006}.}. The secondary and tertiary components could be associated to rotational or non-rotational motions of the gas in the galaxies, depending on the properties.

\noindent We identify possible outflows as typically done in the literature \citep[see e.g.][and references therein]{Harrison2014,Fiore2017,Cazzoli2018,HM2020,Venturi2021}, as a (normally) blueshifted and broad component, with very distinct properties to that of the primary (systemic) component of the gas. These differences do not only concern velocity and $\sigma$, but also the morphology of that component that could be filamentary or bubble-like \citep[see e.g.][HM22]{Cazzoli2018,Fluetsch2019,HM2020}. On the contrary, the ionised gas disc is typically continuous, with a rotating pattern all over the FoV.

\noindent As mentioned in Sect.~\ref{Sec:Intro}, outflows may suppress the star formation of galaxies if they are able to remove the available gas. For determining the impact that the outflows may have, we need to estimate their main properties. In particular, these are the mass of the outflow ($M_{\rm OF,ion}$), the mass outflow rate ($\dot{M}_{\rm OF,ion}$), the kinetic energy ($E_{\rm OF,ion}$) and the kinetic power ($\dot{E}_{\rm OF,ion}$) of the ionised gas. The expressions used for estimating these parameters are taken from C22. The total mass of the ionised gas, $M_{\rm OF,ion}$, using H$\alpha$ is calculated as in \cite{Venturi2021} \citep[see also][]{Carniani2015,Fiore2017}: 
    \begin{equation}
        M_{\rm OF,ion} = 3.2\times 10^{5} \times \left( \frac{L_{\rm H\alpha}}{10^{40}\,{\rm erg\,s}^{-1}} \right) \left( \frac{100\,{\rm cm}^{-3}}{\rm n_{e^{-}}}\right)   \label{Eq_MHa}
    \end{equation}
    \noindent This mass depends on the H$\alpha$ luminosity and the electron density of the gas ($n_{\rm e^{-}}$), that can be estimated from the ratio between the lines of the [S\,II] doublet \citep{Osterbrock2006}. However, for estimating the total mass using H$\beta$ or [O\,III] luminosities instead, there are several expressions assuming different conditions about the properties (i.e. geometry, density, etc) of the outflow \citep[see e.g.][]{CanoDiaz2012,Harrison2014,Carniani2015,Husemann2016,Fiore2017,Kakkad2020,Speranza2021}. In \cite{Carniani2015}, they assume T$_{\rm e} \sim$10$^{4}$\,K and $n_{\rm e^{-}} \sim$\,500\,cm$^{-3}$ to derive the expression: 
    \begin{equation}
        M_{\rm OF,ion}^{\rm H\beta} = 1.7\times 10^{9} \times C \times \left( \frac{L_{\rm H\beta}}{10^{44}\,{\rm erg\,s}^{-1}} \right)  \left( \frac{500\,{\rm cm}^{-3}}{\rm n_{e^{-}}} \right)  \label{Eq_MHb}
    \end{equation}

    \noindent where the constant C is defined as $\rm \langle n_{e^{-}}\rangle^{2} / \langle n_{e^{-}}^{2}\rangle$ \citep[see][]{Fiore2017}.
    \noindent \cite{Carniani2015} note that there is a difference in the mass determinations using H$\beta$ or [O\,III] lines, such that H$\beta$ always traces a higher mass than the [O\,III], which instead would give a lower limit to the outflow mass. 
    
    \noindent Assuming a biconical geometry for the outflow, then the mass outflow rate, $\dot{M}_{\rm OF,ion}$, is estimated as in \cite{Cresci2015} (see also \citealt{Fiore2017}): 
    \begin{equation}
       \dot{M}_{\rm OF,ion} = 3\times \frac{V_{\rm OF,ion}}{R_{\rm OF,ion}}\times M_{\rm OF,ion}    \label{eq2}
    \end{equation}
    \noindent It depends on the outflow mass, the maximum velocity ($V_{\rm OF,ion}$) and its spatial extent ($R_{\rm OF,ion}$). However, if the outflow has instead a shell-like structure, the $\dot{M}_{\rm OF,ion}$ is estimated as in \cite{Husemann2019}: 
    \begin{equation}
        \dot{M}_{\rm OF,ion} = \left( \frac{V_{\rm OF,ion}}{\rm 100\,km\,s^{-1}}\right) \left( \frac{M_{\rm OF,ion}}{10^{6}}\right) \left(\frac{\rm 100\,pc}{\rm \Delta R}\right)   \label{eqMofrate_2}
    \end{equation}
    In this case, the mass outflow rate is measured at the position of the shell structure, not accounting for the overall structure. The $V_{\rm OF,ion}$ is the maximum velocity within the shell, and $\Delta \rm R$ is the shell-thickness. With the mass and the mass outflow rate, we can estimate the kinetic energy of the outflow, $E_{\rm OF,ion}$, as in \cite{Venturi2021} and the kinetic power of the outflow, $\dot{E}_{\rm OF,ion}$, as in \cite{Rose2018} \citep[see also][]{Santoro2020}:
    \begin{equation}
        E_{\rm OF,ion} = \frac{1}{2}\times \sigma^{2}_{\rm OF,ion} \times M_{\rm OF,ion}     \label{eq3}
    \end{equation}

    \begin{equation}
        \dot{E}_{\rm OF,ion} = \frac{1}{2}\times \dot{M}_{\rm OF,ion}\times (V^{2}_{\rm OF,ion} + 3\sigma_{\rm OF,ion})    \label{eq4}
    \end{equation}

\noindent We note that there are other formulae in the literature that could be used to estimate the same parameters \citep[e.g.][]{CanoDiaz2012,Liu2013,Harrison2014}. However, the expressions mentioned here assume an electron temperature T$\sim$10$^{4}$\,K and an electronic density $n_{\rm e^{-}} \sim$500\,cm$^{-3}$, that provide a better match with what we actually found in our objects (see Sect.~\ref{SubSec_ResultOutflows}).

\section{Results}
\label{Sect:Results}

\noindent In this section we describe the main results from the stellar (see Sect.~\ref{SubSec:ResultsStars}) and ionised gas components of the galaxies (see Sect.~\ref{SubSec:ResultsGas}). We show the kinematic and flux maps for both stars and ionised gas, accounting for the different emission lines and kinematical components (see Figs.~\ref{Fig5:KinMaps_NGC0266} to~\ref{Fig5:KinMaps_NGC5055}). In all cases, the ionised gas kinematic maps are derived from the [S\,II] line (all lines share the same kinematics; see Sect.~\ref{SubSec:datared_gas}); the H$\beta$ and [O\,III]\,5007\AA\,maps are given when the LR-B cube is available (see Table~\ref{Table:obslog}). The derived parameters from the maps are given in Table~\ref{Table:resultsKin}. The line ratio maps are in Appendix~\ref{Appendix:BPTs}, and the BPT diagrams for NGC\,4278 and NGC\,4750 are in Fig.~\ref{Fig5:BPT_NGC4750}. 

\noindent The following subsections refer to eight out of the nine targets (NGC\,1052 was analysed in detail in C22; see Table~\ref{Table:obslog}). In this work we focus exclusively on the study of the ionised gas phase, thus on the emission lines analysis. \\ 

\noindent For each galaxy (see Sects.~\ref{SubSec:ResultsStars} and~\ref{SubSec:ResultsGasKin} and Table~\ref{Table:resultsKin}), we refer to: the velocity semi amplitude ($\Delta$v) defined as half the difference between the maximum and minimum velocities; and the central sigma ($\sigma^{\rm PSF}_{\star}$ and $\sigma^{\rm PSF}_{\rm gas}$) as the average velocity dispersion measured within the PSF region. The velocities are estimated subtracting the systemic velocity of the galaxy calculated from the redshifts reported in Table~\ref{Table:galaxieslisted}. The position angles (PA$_{\star}$ and PA$_{\rm gas}$) were measured using the maximum and minimum velocities of the maps, using as reference the receding side of the velocity maps from north to east\footnote{We only estimated the PA for the primary component of the ionised gas whenever it resembled to a gas disc (see Sect.~\ref{SubSec:Discussion_morphologies})}. If the components are spatially unresolved, we report exclusively the average kinematic properties. All the values for the ionised gas given per component are referred to as P, S or T (see Sect.~\ref{SubSec:datared_gas}). The reported line ratios are in logarithm scale (see Sect.~\ref{SubSec:datared_BPTs}).

\subsection{Stellar component kinematics}
   \label{SubSec:ResultsStars}
   
    \noindent The stellar component maps are shown in the upper panels of Figs.~\ref{Fig5:KinMaps_NGC0266} to~\ref{Fig5:KinMaps_NGC5055}. The main properties are listed in Table~\ref{Table:resultsKin}, and the position-velocity (PV) and position-velocity dispersion (PS) diagrams are in Appendix~\ref{AppendixA:PV-PS}. Globally, the stars are always rotating with spider-regular rotational patterns, except for NGC\,0315, where both velocity and velocity dispersion maps are homogeneous with no particular structures (see Fig.~\ref{Fig5:KinMaps_NGC0315} and Table~\ref{Table:resultsKin}). \\
    For NGC\,0266, the velocity map shows a rotating pattern (V/$\sigma \sim 1.8$; see Fig.~\ref{Fig5:KinMaps_NGC0266}), with two regions with more extreme velocities than the rest of the map (see position-velocity diagram in Fig.~\ref{Fig:PV-PS_1}). They are located from the PSF region up to $\sim$4\arcsec\,in diameter in the major axis direction (PA$\sim$121$^{\circ}$). Contrarily, the velocity dispersion is rather homogeneous throughout the FoV, peaking at the photometric centre ($\sigma$\,=\,205\,km\,s$^{-1}$), decreasing radially out to a radius of $\sim$1.2\arcsec\,down to $\sim$170\,km\,s$^{-1}$, and then it gets uniform at larger radii ($>2.5\arcsec$).

    \noindent For NGC\,3245 and NGC\,5055, the H$\alpha$ emission is highly diluted due to an important contribution from the stars in the Balmer spectral lines, specially in the nuclear region (for NGC\,5055 see \citealt{Dullo2021}). Thus to ensure a proper modelling of their stellar component within the PSF region, we used higher additive polynomials (degree 10 vs 8; see Sect.~\ref{SubSec:datared_stars}) than for the rest of the objects (see integrated PSF spectrum of NGC5055 in Fig.~\ref{Fig5:PPXF_NGC5055}). For both NGC\,3245 and NGC\,5055, the velocity maps show a clear rotational pattern (V/$\sigma$ of 1.6 and 1.4, respectively; see Sect.~\ref{SubSec:Discussion_morphologies}) along a PA of 178$^{\circ}$ and $99^{\circ}$, respectively (see Table~\ref{Table:resultsKin}). The velocity dispersion is homogeneously distributed for NGC\,5055 (median $\sigma_{\star} \sim$\,121\,km\,s$^{-1}$; see Table~\ref{Table:resultsKin}), whereas for NGC\,3245 it is centrally peaked ($\sigma_{\star}^{\rm PSF}=245$\,km\,s$^{-1}$; see Table~\ref{Table:resultsKin}). 

    \noindent For NGC\,3226 and NGC\,4278 the stars rotate with a regular pattern and similar semi-amplitude velocities (106\,km\,s$^{-1}$ and 104\,km\,s$^{-1}$ along a PA of 49$^{\circ}$ and 20$^{\circ}$, respectively; see Table~\ref{Table:resultsKin}). In both cases, the velocity dispersion is centrally peaked ($\sigma^{\rm PSF}_{\star}$\,$\sim$\,210 and 336\,km\,s$^{-1}$, for NGC\,3226 and NGC\,4278 respectively; see Table~\ref{Table:resultsKin}) and gradually drops from the centre to the outskirts (see Fig.~\ref{Fig:PV-PS_1}). For NGC\,4278, contrary to the previous galaxies, we obtained the stellar component maps by using the LR-B and LR-V data cubes individually and not the combined LR-VR cube, as we have the HR-R cube instead of the LR-R (see Table~\ref{Table:obslog} and Sect.~\ref{SubSec:datagather}). The analysis with both spectral setups (LR-B and LR-V) gave consistent results, so in Fig.~\ref{Fig5:KinMaps_NGC4278} we show only that of the LR-V cube.

    \noindent For NGC\,4438 and NGC\,4750, the velocity maps of the stars also show a rotational pattern with similar properties ($\Delta$v\,$\sim$\,111 and 97\,km\,s$^{-1}$, respectively; see Table~\ref{Table:resultsKin}), but with a rather homogeneous velocity dispersion ($\sigma^{\rm PSF}_{\star}$\,$\sim$\,148 and 134\,km\,s$^{-1}$, respectively; see Figs.~\ref{Fig5:KinMaps_NGC4438} and~\ref{Fig5:KinMaps_NGC4750}). NGC\,4438 is known to have an ionised gas bubble, traced in several works through the H$\alpha$ emission \citep{Kenney2002, Masegosa2011}. To map this feature completely with MEGARA, we centred the north-west bubble instead of the nucleus in the FoV of MEGARA (see Sect.~\ref{Sect:sample_data}). As a consequence, we detect better the redder velocities than the bluer velocities in the stellar velocity field, simply because is not centred in the FoV (see Fig.~\ref{Fig5:KinMaps_NGC4438}). We detected sightly larger values of $\sigma$ ($\sim$\,180\,km\,s$^{-1}$) on the location of the ionised gas bubble.
    
\subsection{Ionised gas component}   
\label{SubSec:ResultsGas}
   \subsubsection{Main kinematics}
   \label{SubSec:ResultsGasKin}

   \noindent We fitted exclusively one Gaussian component for the emission lines of two galaxies (NGC\,0266 \& NGC\,5055), two components for five galaxies (NGC\,0315, NGC\,3226, NGC\,3245, NGC\,4278 and NGC\,4750), and three components for NGC\,4438 and NGC\,1052 (see C22). We added a BLR component to the spectra in the PSF region for five galaxies (namely NGC\,0315, NGC\,1052, NGC\,3226, NGC\,4438 and NGC\,4750), although for NGC\,4438 the detection is not well constrained; these detections are discussed in Sect.~\ref{Sect:Discussion}. \\
   
\begin{figure*}
   \includegraphics[width=\textwidth]{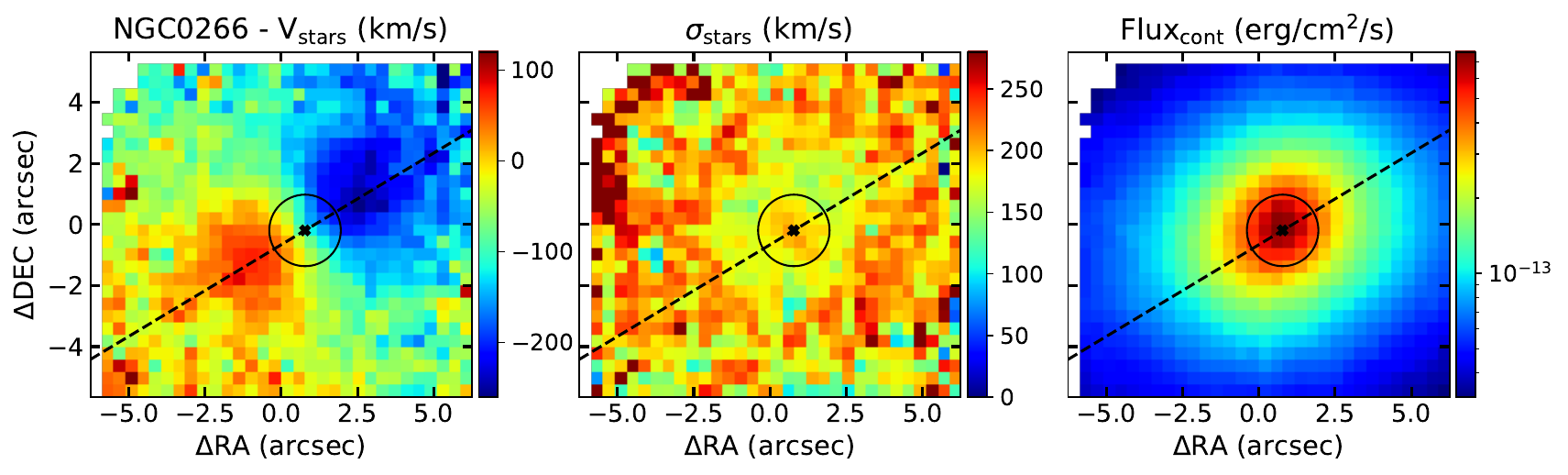}
   \includegraphics[width=\textwidth]{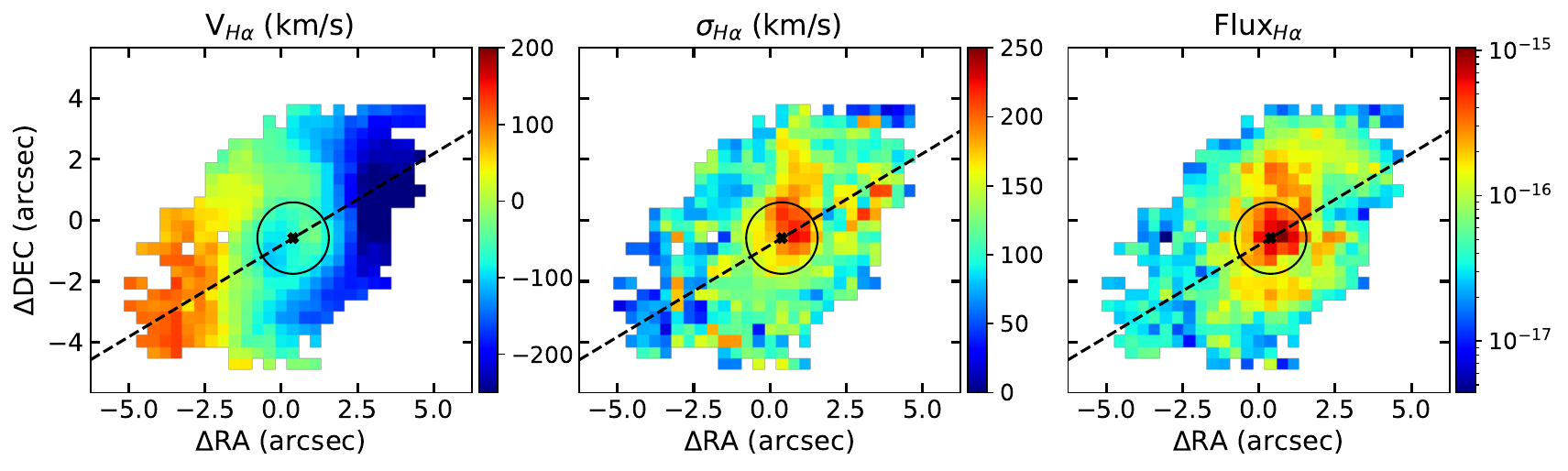}
   \caption{Kinematic maps for the stellar (top panel) and ionised gas (bottom panel) components of NGC\,0266. The stellar kinematics was calculated using the combined LR-VR data cube, whereas the ionised gas kinematics is extracted from the [S\,II] lines (all the lines share the same kinematics, see Sect.~\ref{SubSec:datared_gas}). In all panels from left to right: velocity in km\,s$^{-1}$ (corrected from the systemic velocity of the galaxy; see redshifts in Table~\ref{Table:galaxieslisted}), velocity dispersion in km\,s$^{-1}$ and integrated flux per spaxel (stars) or integrated H$\alpha$ flux (gas) in erg\,s$^{-1}$\,cm$^{-2}$. The black (grey) dashed line, when present, indicates the position angle of the stellar (primary ionised gas) component. The black cross indicates the photometric centre, and the PSF region is marked with a black circle. For all the maps, north is up and east to the left.}
   \label{Fig5:KinMaps_NGC0266}
   \end{figure*}

   \noindent \textit{NGC\,0266} (Figs.~\ref{Fig5:KinMaps_NGC0266} and~\ref{Fig:NGC0266_BPTs}): We find that $\sim$40\% of the spaxels have S/N$>$5 in the [S\,II] lines. The ionised gas is elongated in the direction of the stellar major axis (see Fig.~\ref{Fig5:KinMaps_NGC0266}). In general, the emission line profiles are narrow (median $\sigma \sim$\,116\,km\,s$^{-1}$) and thus were fitted with a single Gaussian component all over the FoV. The structure of the ionised gas velocity map clearly differs from that of the stars, with a large (i.e. $\sim$6.5\arcsec), non-rotating region in the inner parts (see bottom panel in Fig.~\ref{Fig5:KinMaps_NGC0266}) twisted towards the north direction. The outer parts of the velocity map (at radius $>$2.5\arcsec\,in the east-west direction) show a rotating-like pattern oriented along the stellar major axis, with a semi-amplitude velocity of $\sim$212\,km\,s$^{-1}$. The H$\alpha$ flux is concentrated on a spiral-like shape, spatially coincident with the twisted region of nearly flat rotation in the gas velocity map. The velocity dispersion map is centrally peaked ($\sigma^{\rm PSF}_{\rm gas,N} =$\,173$\pm$28\,km\,s$^{-1}$; see Table~\ref{Table:resultsKin}) with the largest values corresponding to the spiral-like feature seen in flux and velocity, and then dropping radially outwards from that structure. \\
   
\begin{figure*}
   \includegraphics[width=.98\textwidth]{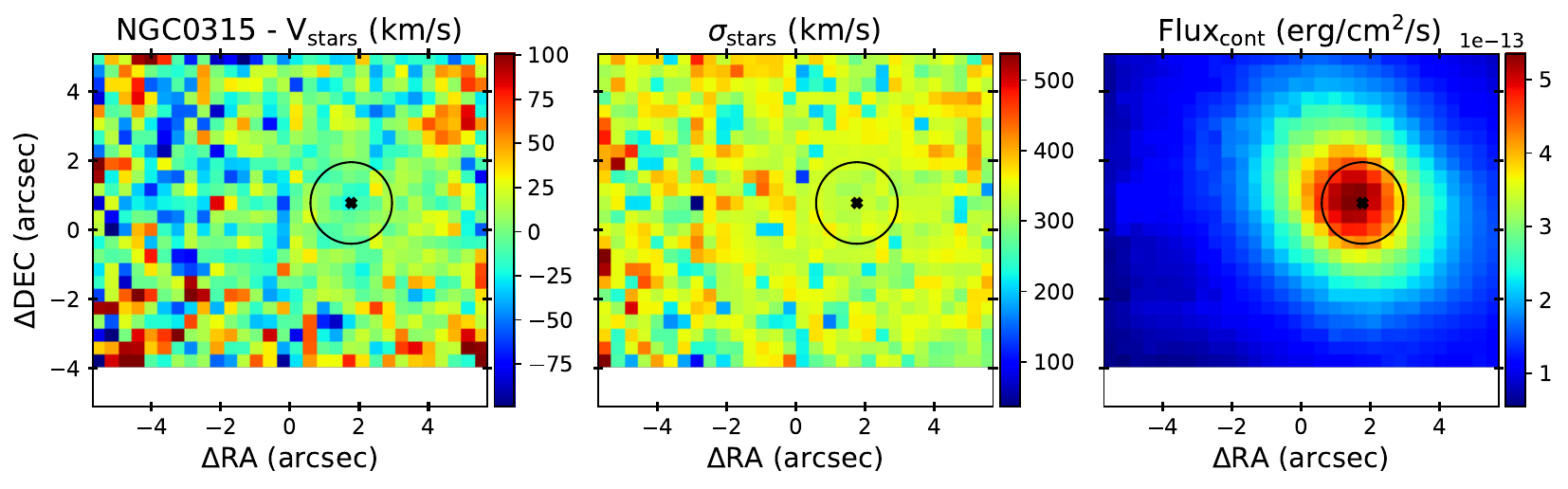}
   \includegraphics[width=\textwidth]{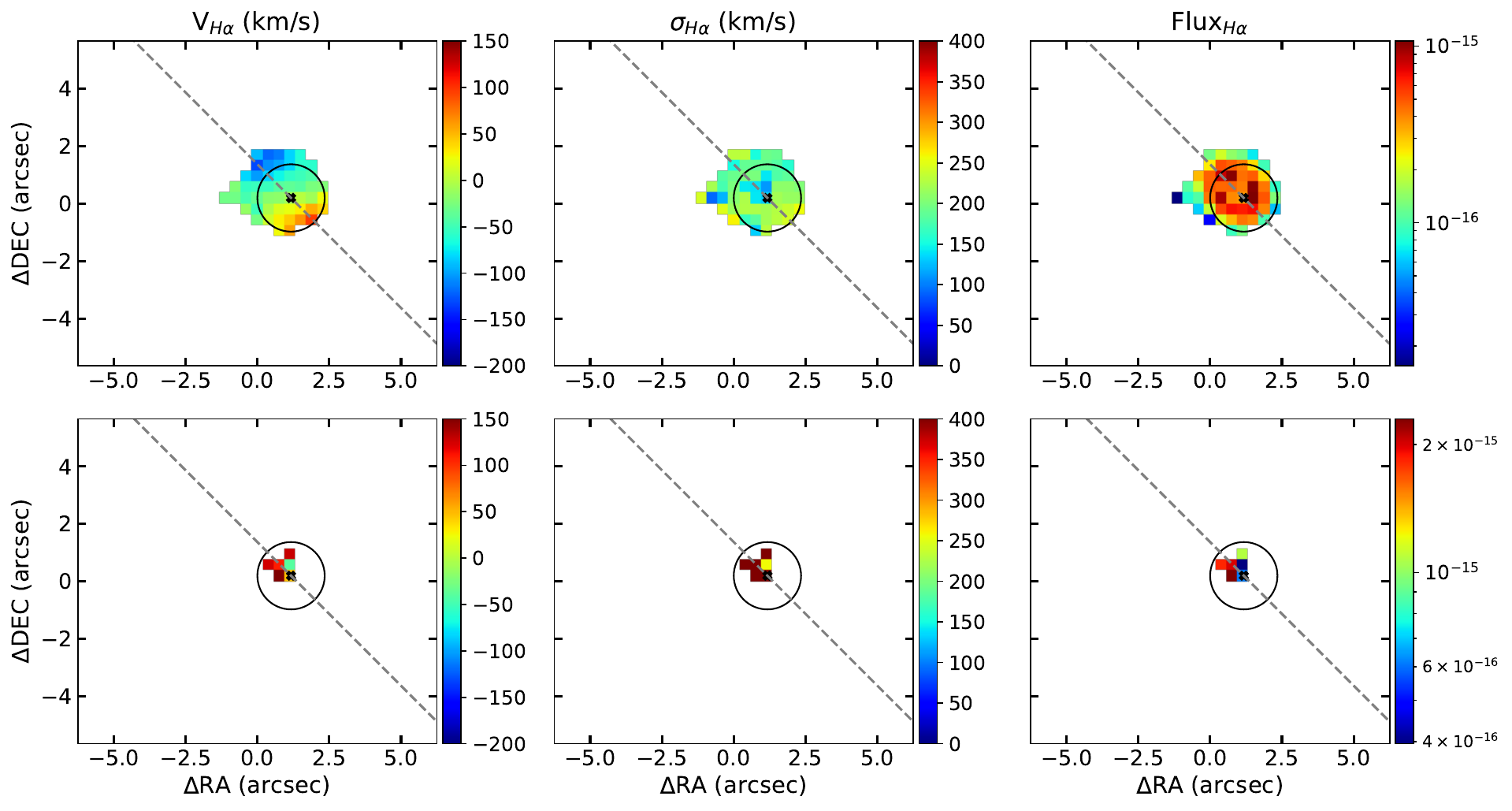}
   \caption{Kinematic maps for the stellar (top panel) and ionised gas components of NGC\,0315. The spectra of this galaxy was modelled with a primary (second panel) and an unresolved, secondary (third panel) kinematical components per emission line. A full description is provided in Fig.~\ref{Fig5:KinMaps_NGC0266}.}
   \label{Fig5:KinMaps_NGC0315}
   \end{figure*}

   \noindent \textit{NGC\,0315} (Figs.~\ref{Fig5:KinMaps_NGC0315} and~\ref{Fig:NGC0315_BPTs}):    
   \noindent The emission corresponding to the ionised gas in this galaxy, measured with the S-method, is concentrated around the nucleus, and reaches only distances up to $\sim$2\arcsec\,in radius from the photometric centre. The emission lines show complex profiles, with various kinematical components difficult to disentangle. Most of the emission ($\sim$70\%) is located inside the PSF region, hence unresolved. We modelled the spectra with two kinematic components per forbidden line plus a very broad component in H$\alpha$. The primary component is barely resolved (total extension of $\sim$3.9\arcsec, i.e. $\sim$1.4\,kpc; see Table~\ref{Table:galaxieslisted}), with a rotational pattern that differs from the velocity map of the stars (see Fig.~\ref{Fig5:KinMaps_NGC0315}), and homogeneous $\sigma$ (see Table~\ref{Table:resultsKin}). The secondary component and the broad component in H$\alpha$ are unresolved (total extension for the secondary of $<$\,1.5\arcsec, i.e. $<$\,500\,pc; the broad component is within the PSF region). The BLR component has an average velocity of -53\,km\,s$^{-1}$ and $\sigma_{\rm gas,BLR}$ = 1077\,km\,s$^{-1}$. \\
   
\begin{figure*}
   \includegraphics[width=\textwidth]{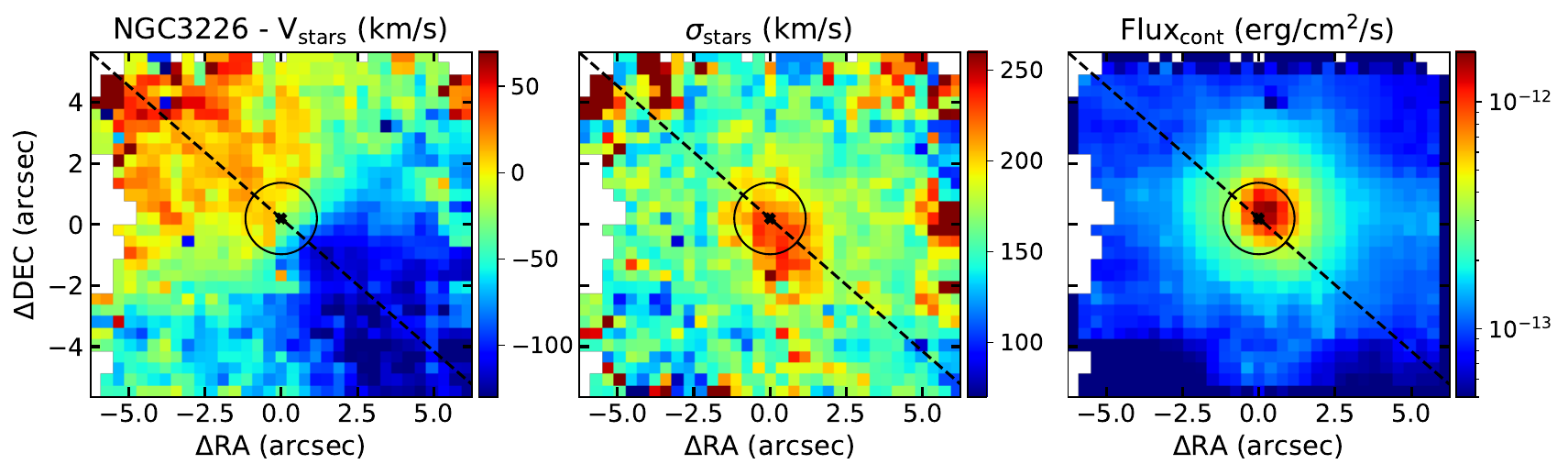}
   \includegraphics[width=\textwidth]{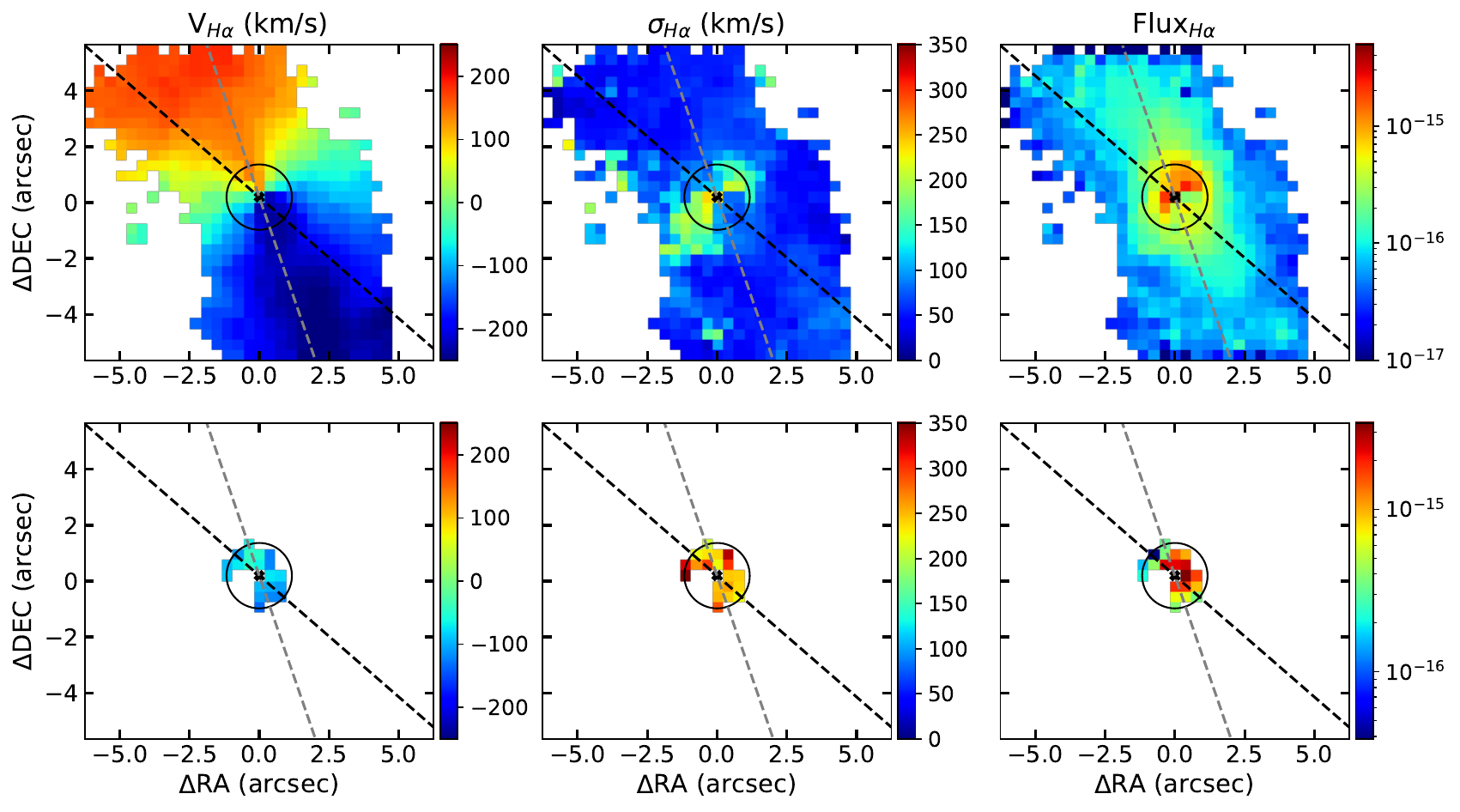}
   \caption[Kinematic maps for the stellar and ionised gas components of NGC\,3226.]{Kinematic maps for the stellar (top panel) and ionised gas components of NGC\,3226. The spectra of this galaxy was modelled with a primary (second panel) and an unresolved secondary (third panel) kinematical components per emission line. A full description is provided in Fig.~\ref{Fig5:KinMaps_NGC0266}.}
   \label{Fig5:KinMaps_NGC3226}
   \end{figure*}

   \noindent \textit{NGC\,3226} (Figs.~\ref{Fig5:KinMaps_NGC3226} and~\ref{Fig:NGC3226_BPTs}): We identified two kinematical components per emission line plus a spatially unresolved broad component in H$\alpha$ (see Sect.~\ref{Sect:Discussion}). The velocity map and the H$\alpha$ flux map of the primary component are oriented in the north-east direction, with a position angle differing in 30$^{\circ}$ from that of the stars (PA$_{\rm gas}$\,=\,19$^{\circ}$; see Table~\ref{Table:resultsKin})\footnote{We note that, as we consider the maximum and minimum velocities from the map to estimate PA$_{\star}$, the determination could be affected by S/N issues in the north-east corner of the FoV. Hence both PAs could agree.}. Moreover, this component rotates with a semi-amplitude velocity twice larger than that of the stars (see Table~\ref{Table:resultsKin}). In general, this component is very narrow ($\sigma_{\rm gas,P}\leq$\,100\,km\,s$^{-1}$) although a nuclear structure with wider profiles ($\sigma\sim$\,200\,km\,s$^{-1}$) is visible oriented almost perpendicular to the major axis of the ionised gas (extended $\sim$4.7\arcsec, i.e. $\sim$400\,pc at PA\,$\sim$107$^{\circ}$). The secondary component is unresolved and embedded in the PSF region ($\sim$2.2\arcsec, i.e. $\sim$190\,pc). It is clearly blueshifted (average velocity of -91\,km\,s$^{-1}$) with larger velocity dispersion than the primary in the same spatial region (difference of $\sim$60\,km\,s$^{-1}$; see Table~\ref{Table:resultsKin}). We detect a very broad component ($\langle \sigma_{\rm gas, BLR} \rangle$\,$=$\,979$\pm$65\,km\,s$^{-1}$) with $\langle \rm v \rangle$\,$=$\,269\,$\pm$\,20\,km\,s$^{-1}$ by modelling the integrated PSF spectrum. \\
      
\begin{figure*}
   \includegraphics[width=.98\textwidth]{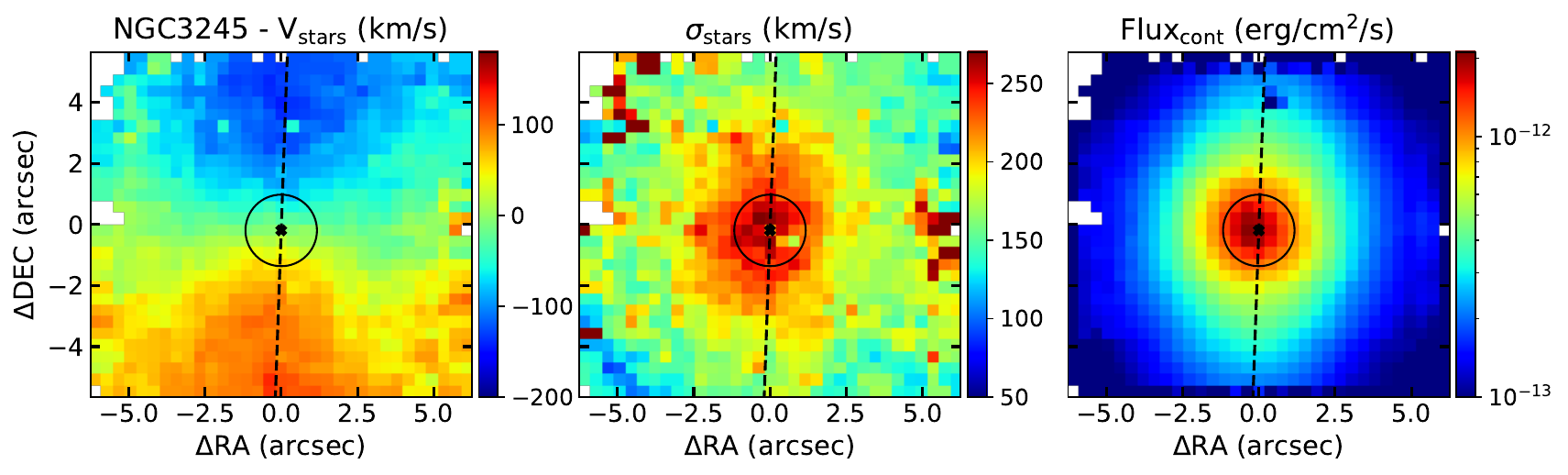}
   \includegraphics[width=\textwidth]{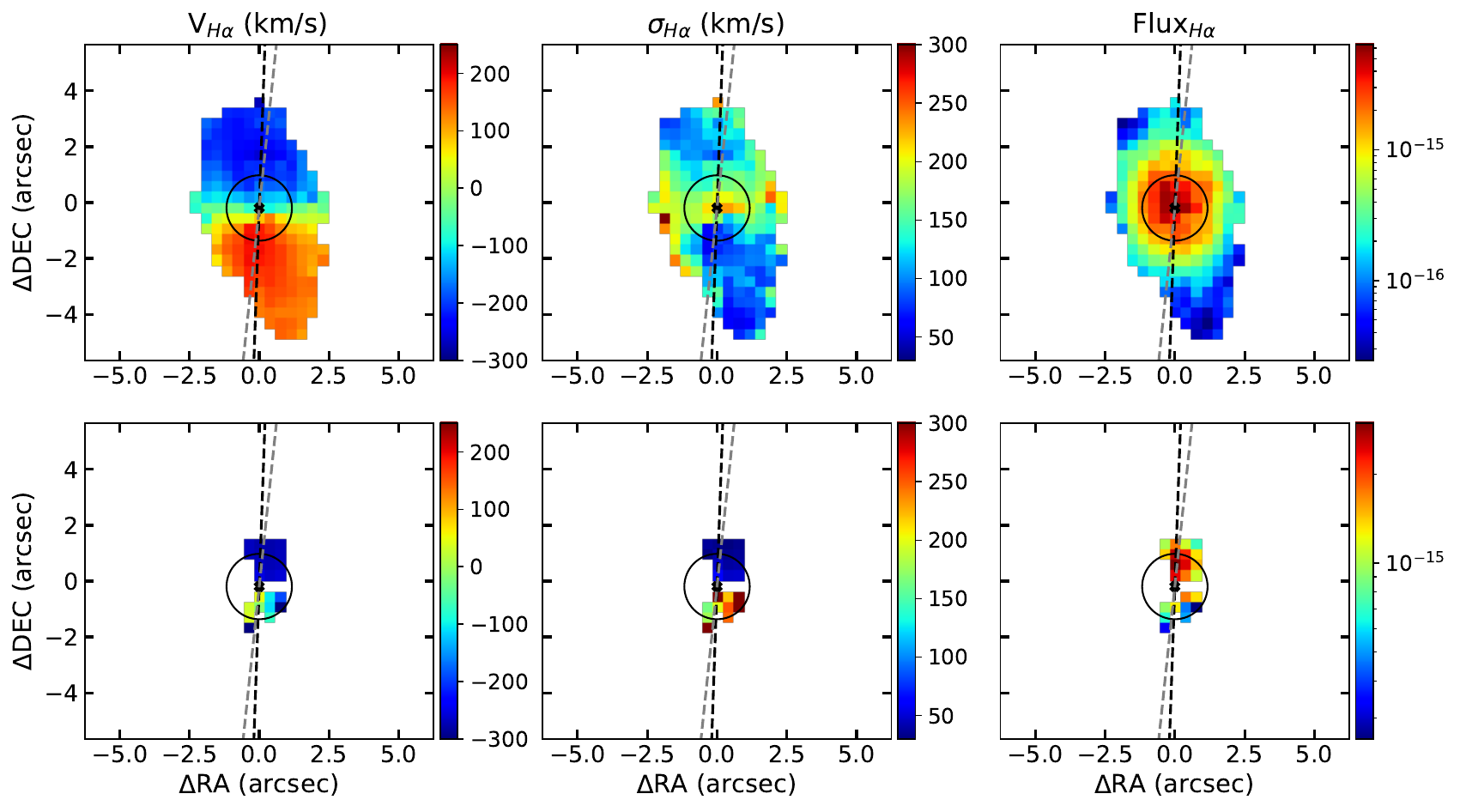}
   \caption{Kinematic maps for the stellar (top panel) and ionised gas components of NGC\,3245. The spectra of this galaxy was modelled with a primary (second panel) and a secondary (third panel) kinematical components per emission line. A full description is provided in Fig.~\ref{Fig5:KinMaps_NGC0266}.}
   \label{Fig5:KinMaps_NGC3245}
   \end{figure*}

   \noindent \textit{NGC\,3245} (Figs.~\ref{Fig5:KinMaps_NGC3245} and~\ref{Fig:NGC3245_BPTs}):    
   \noindent The ionised gas in the inner regions has a complex structure that was modelled using two kinematical components per emission line. The primary component shows a clear rotation pattern, larger than that of the stars (differences of $\sim$\,100\,km\,s$^{-1}$; see Table~\ref{Table:resultsKin}) despite its relatively low extension ($\sim$10\arcsec\,in the north-south direction and $\sim$\,5\arcsec\,in the east-west direction). The velocity dispersion map has an enhanced region ($\sigma$\,$>$\,140\,km\,s$^{-1}$) in the east-west direction resembling a bicone that does not correspond to any particular feature in the velocity or flux maps. The secondary component is unresolved ($\sim$\,1.7\arcsec\,$\times$\,3.6\arcsec, i.e. $\sim$150\,pc\,$\times$\,320\,pc), and blueshifted ($\langle \rm v \rangle$\,$\sim$\,-200\,km\,s$^{-1}$). Its average value is consistent to that of the primary component in the same spatial region (see Table~\ref{Table:resultsKin}). As for NGC\,3226, we tested the presence of a very broad component in the integrated PSF spectrum. To model the lines, we added a non-rotating ($\langle \rm v \rangle$\,$\sim$\,-30\,km\,s$^{-1}$), very broad component ($\sigma >$\,700\,km\,s$^{-1}$), with a mild contribution to the total flux of the H$\alpha$+[N\,II] blend ($<$10\%). \\
   
\begin{figure*}
   \centering
   \includegraphics[width=.89\textwidth]{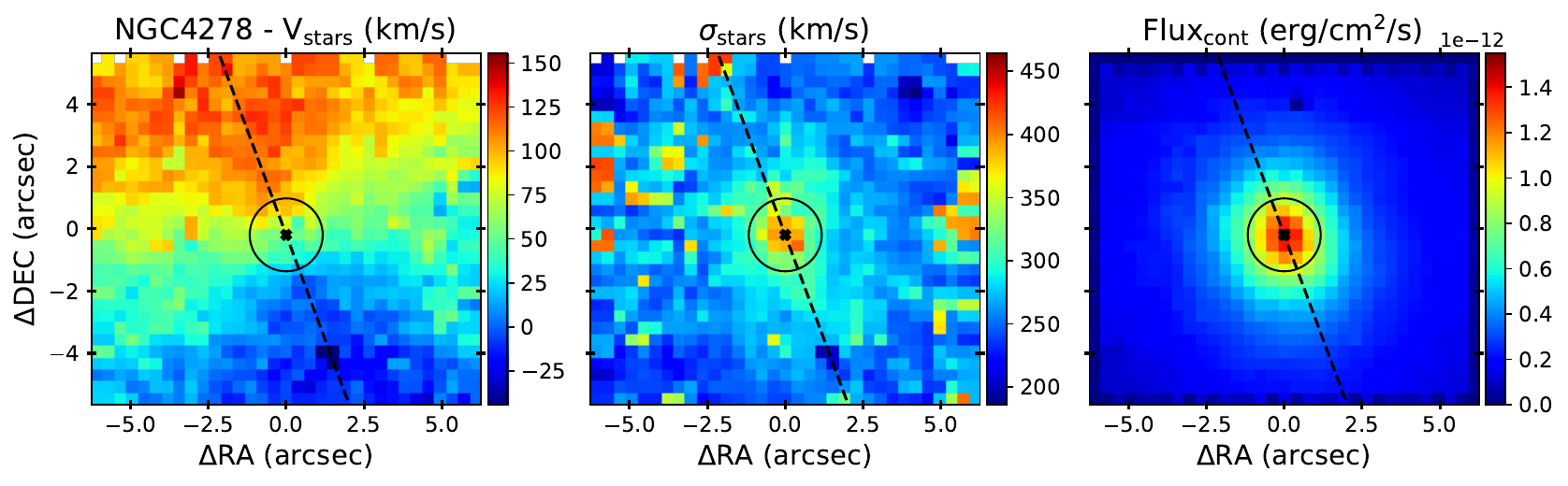}
   \includegraphics[width=.9\textwidth]{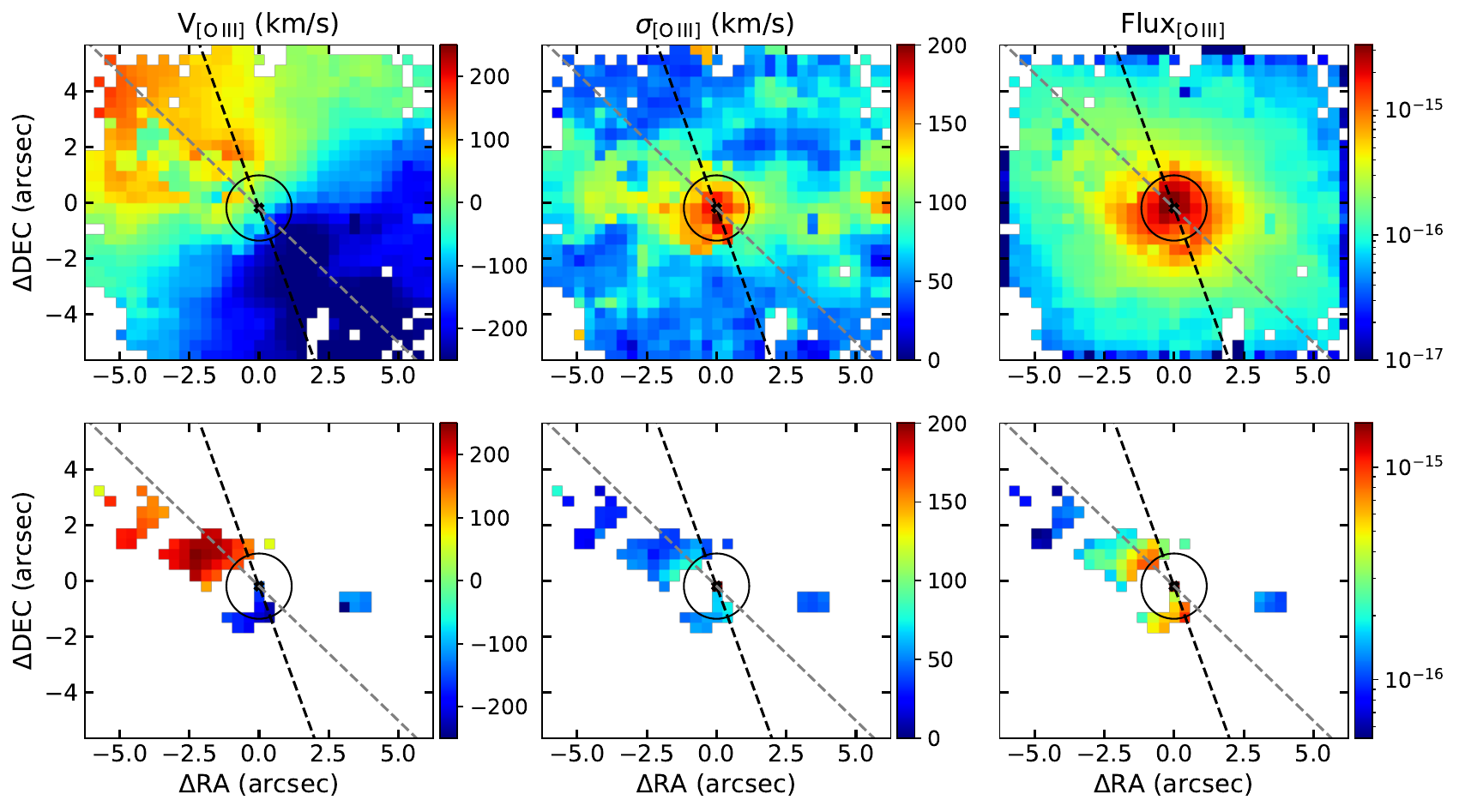}
   \includegraphics[width=.9\textwidth]{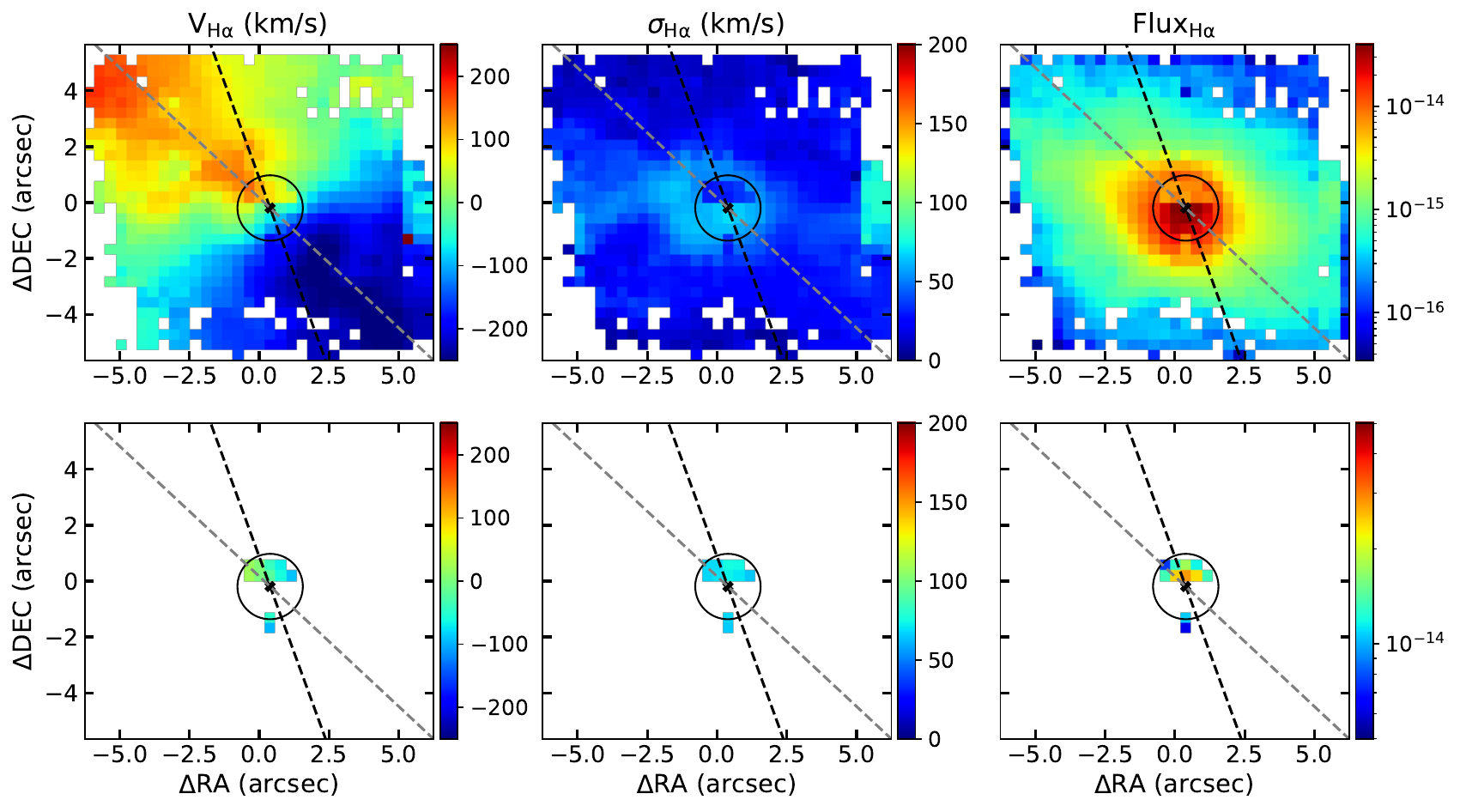}
   \caption{Kinematic maps for the stellar component of NGC\,4278. For this galaxy we modelled the stellar component using only the LR-V cube (see Sect.~\ref{Sect:Datareduction}). The second and third (fourth and fifth) rows of panels correspond to the primary and secondary component of the [O\,III] (H$\alpha$) maps, respectively. A full description is provided in Fig.~\ref{Fig5:KinMaps_NGC0266}.}
   \label{Fig5:KinMaps_NGC4278}
   \end{figure*}

   \noindent \textit{NGC\,4278} (Figs.~\ref{Fig5:KinMaps_NGC4278},~\ref{Fig5:BPT_NGC4750} and~\ref{Fig:NGC4278_BPTs}):    
   \noindent The analysis of the emission lines in the LR-B cube reveals the presence of two different kinematical components. The [O\,III] and H$\beta$ lines show very similar $\sigma$ for all the components (mean difference $<$\,10\,km\,s$^{-1}$), but they are decoupled in velocity, with a difference of $\sim$50\,km\,s (larger for H$\beta$), as it was already found in the case of NGC\,1052 (see C22). Given that the S/N of [O\,III] is larger than that of H$\beta$ (97\% vs 78\% spaxels with S/N\,$>$\,5), we focus on the [O\,III] line for the LR-B analysis. The ISM and the stellar velocity maps have different PA (20$^{\circ}$ vs 46$^{\circ}$; see Table~\ref{Table:resultsKin}). The primary component of the gas is rotating ($\rm \Delta v/2$\,$\sim$\,240\,km\,s$^{-1}$), with a receding side that presents several regions with rest-frame velocities located north-east from the nucleus. The velocity dispersion is centrally peaked ($\sigma_{\rm gas,P}^{\rm PSF}$\,$\sim$\,146\,km\,s$^{-1}$), and has an enhanced region ($\sigma$\,$\sim$\,120\,km\,s$^{-1}$) west-east from the nucleus (PA\,$\sim$\,45$^{\circ}$), extending along the whole FoV. The [O\,III] flux is distributed along the major axis of the primary gas component. The secondary component is resolved, expanding up to $\sim$6\arcsec, mainly in the north-east direction. It is oriented in a similar direction to the primary component, but with a shell-like structure. The velocities are more extreme than for the primary ($\rm \Delta v/2$\,$\sim$280\,km\,s$^{-1}$), specially in the receding side, while the velocity dispersion is smaller ($\langle \sigma\rangle$\,$\sim$\,50\,km\,s$^{-1}$). We have not detected a BLR component in H$\beta$.
   
   \noindent The H$\alpha$ and [S\,II] lines have a similar velocity map to that of [O\,III] ($\rm \Delta v/2$\,$\sim$\,222\,km\,s$^{-1}$) rotating along a similar PA (47$\pm 7 ^{\circ}$). The velocity dispersion is similar too, as we detect for H$\alpha$ a region with enhanced values along the east-west direction. However, the different spectral resolutions (50\,km\,s$^{-1}$ for LR-R vs 15\,km\,s$^{-1}$ for HR-R) are reflected in the velocity dispersion, much smaller for the HR-R data cube (average $\sigma$\,$\sim$\,30\,km\,s$^{-1}$) than for the LR-B (average $\sigma$\,$\sim$\,70\,km\,s$^{-1}$). We detect an unresolved secondary component only in a few spaxels ($\sim$1\%), mainly in the PSF region, that is not spatially coincident with the secondary in [O\,III]. This secondary component in H$\alpha$ has rest-frame velocities and a velocity dispersion of $\sim$70\,km\,s$^{-1}$. As we have not subtracted the stellar continuum for this cube given the small wavelength range (see Sect.~\ref{SubSec:datared_stars}), the detection of the secondary component could be affected. Moreover, this may also affect the detection of a BLR component, which presence we cannot constrain although there are hints of it in some individual spaxels. \\
   
\begin{figure*}
   \includegraphics[width=\textwidth]{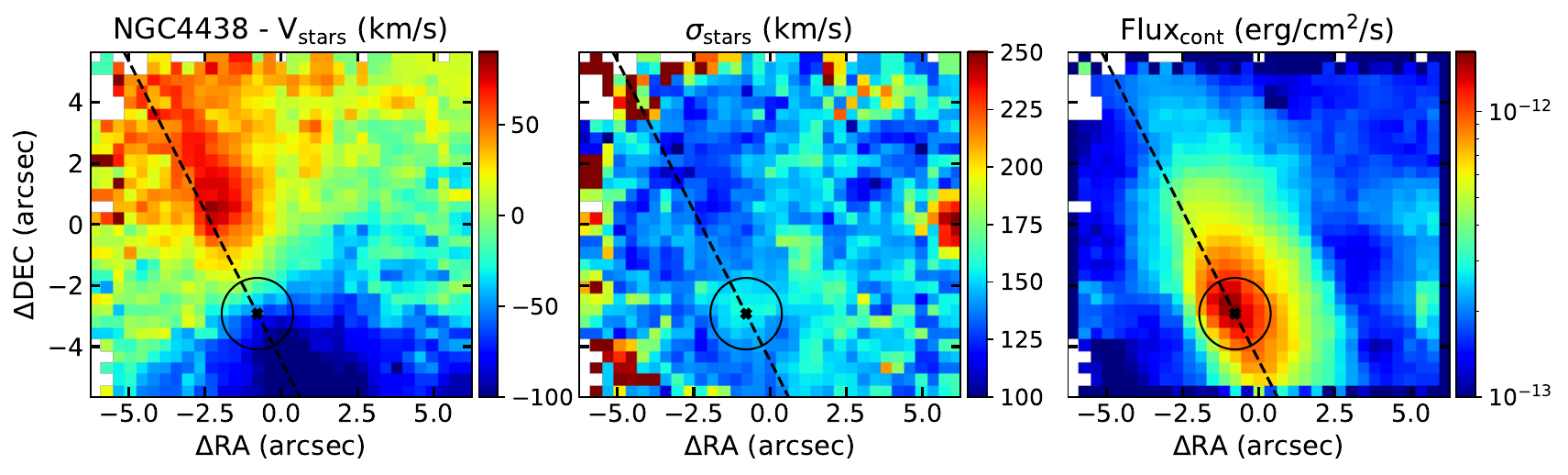}
   \includegraphics[width=\textwidth]{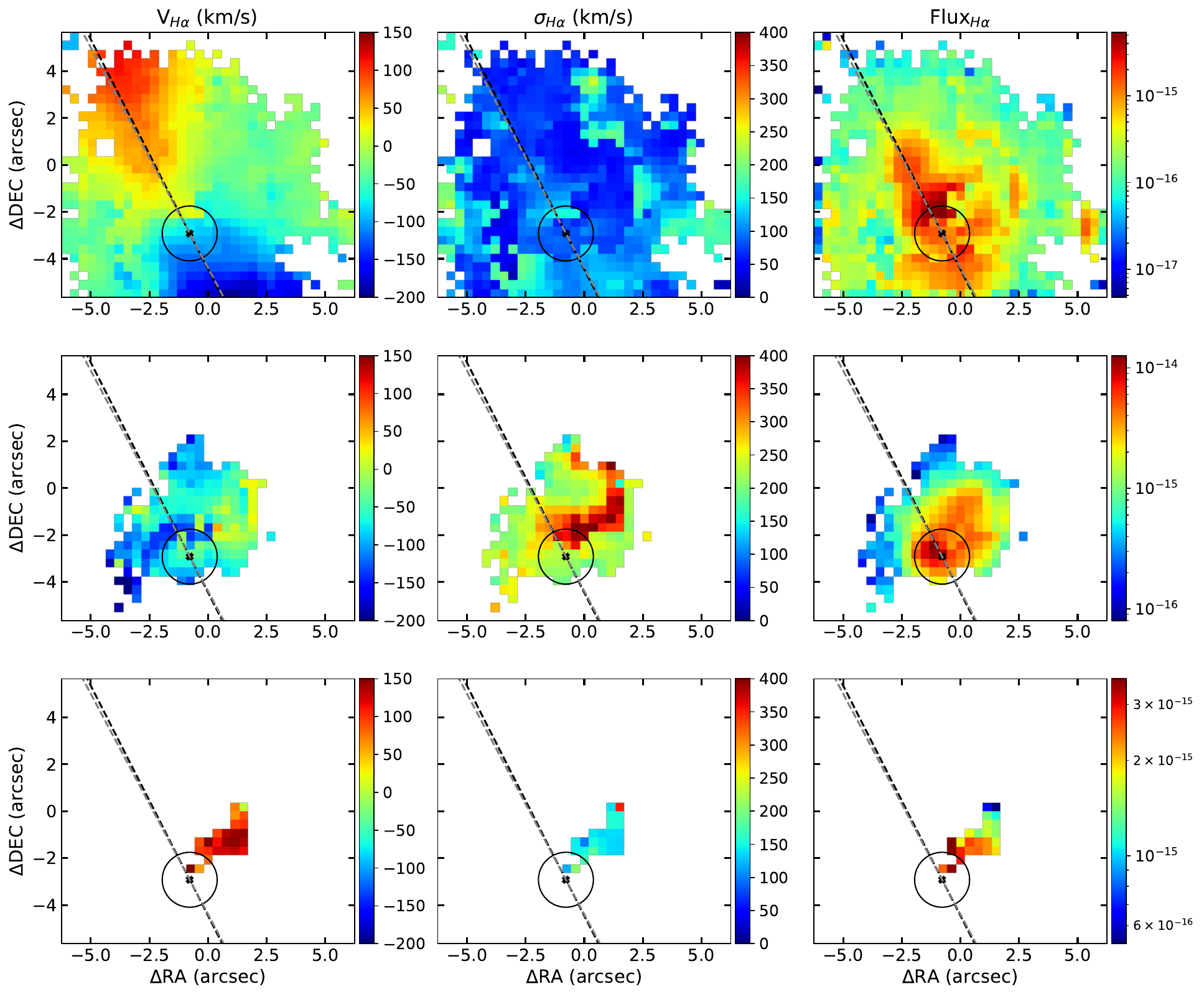}
   \caption[Kinematic maps for the stellar and ionised gas components of NGC\,4438.]{Kinematic maps for the stellar (top panel) and ionised gas components of NGC\,4438. The emission lines were modelled with a primary (second panel), secondary (third panel) and tertiary (bottom panel) kinematical components. A full description in Fig.~\ref{Fig5:KinMaps_NGC0266}.}
   \label{Fig5:KinMaps_NGC4438}
   \end{figure*}

   \noindent \textit{NGC\,4438} (Fig.~\ref{Fig5:KinMaps_NGC4438} and~\ref{Fig:NGC4438_BPTs})): We modelled the ionised gas spectra of this galaxy with three different kinematical components per forbidden emission line. This is the only galaxy in this work for which all the kinematical components are resolved out of the PSF region. The primary, narrowest component ($\sigma_{\rm gas,P}^{\rm PSF}$\,$=$\,99\,$\pm$\,28\,km\,s$^{-1}$) is detected with S/N$>$5 in almost the whole FoV ($\sim$\,75\%). Its velocity field is very similar to that of the stellar component (PA$_{\rm gas}$\,=\,28$^{\circ}$), with $\rm \Delta v/2$\,$\sim$\,154\,km\,s$^{-1}$ (see Table~\ref{Table:resultsKin}). The H$\alpha$ flux is distributed along the disc, as the stellar continuum, tracing a regular rotating gas disc. We found an enhancement in the emission line fluxes in the same direction as the bubble (north-west). We visually inspected those spaxels finding hints of a secondary component, although not statistically significant to include them ($\varepsilon$\,$<$\,2 in the [S\,II] lines). Thus the flux enhancement is probably due to the bubble being slightly more extended, but our data do not allow to further constraint its presence. The $\sigma$ map of the primary component has an homogeneous structure except in the FoV limits (due to a lower S/N), with typical values $\sigma$\,$\leq$\,200\,km\,s$^{-1}$. \\
   \noindent The H$\alpha$ flux map of the secondary component has a bubble-like structure and spatially matches the extension of the bubble detected in the images in previous works ($\sim$\,10.5\arcsec i.e. $\sim$\,300\,pc). In the velocity map (see Fig.~\ref{Fig5:KinMaps_NGC4438}) we distinguish a receding (north-west) and an approaching (south-east) side, oriented at the same PA of the bubble (PA$_{\rm gas,S}$\,$\sim$\,123$^{\circ}$). Overall the $\sigma$ is higher than for the primary component ($\sigma <$\,200\,km\,s$^{-1}$ only in 10\% of the spaxels), specially enhanced at the borders of the bubble and less intense in the inner parts (see Sect.~\ref{SubSec:Discussion_NGC4438}). \\
   \noindent Finally, the tertiary component (not found in [O\,I]) is also present in the spaxels along the north-west direction from the nucleus. This component is generally redshifted and narrow ($\langle \sigma\rangle$\,$\sim$\,135\,km\,s$^{-1}$), being $\sigma$ comparable to that of the primary component (see Table~\ref{Table:resultsKin}). However, it has a filamentary-like morphology, different to the other components, that reaches as far as the secondary component ($\sim$\,4.8\arcsec, i.e. $\sim$\,130\,pc). We discuss its possible origin in Sect.~\ref{SubSec:Discussion_NGC4438}. \\

\begin{figure*}
   \includegraphics[width=.95\textwidth]{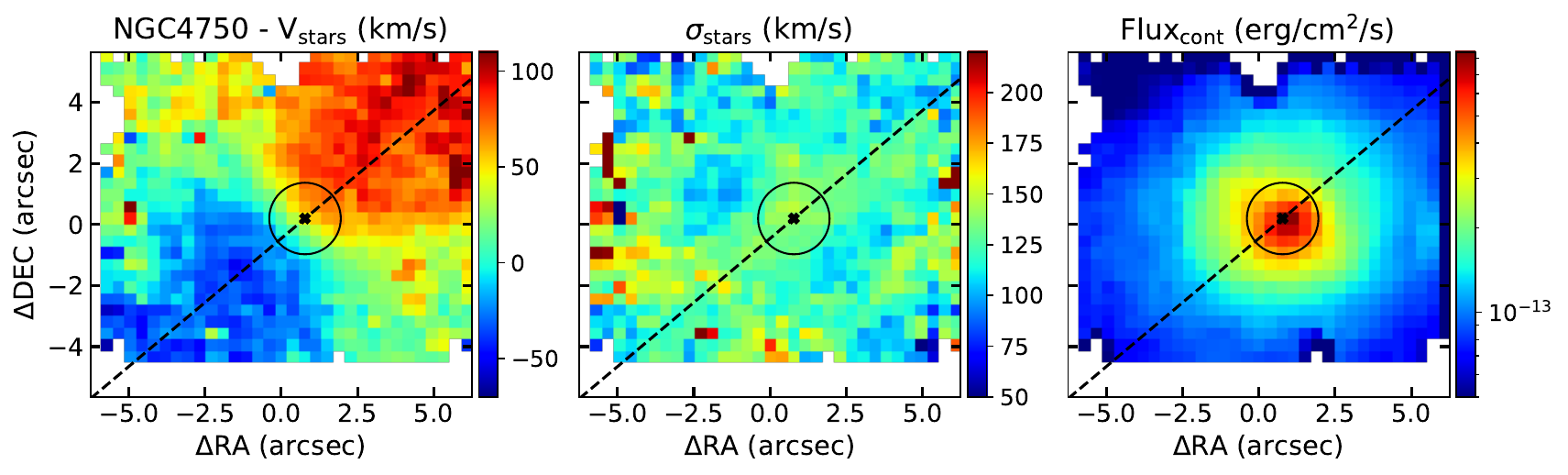}
   \includegraphics[width=.95\textwidth]{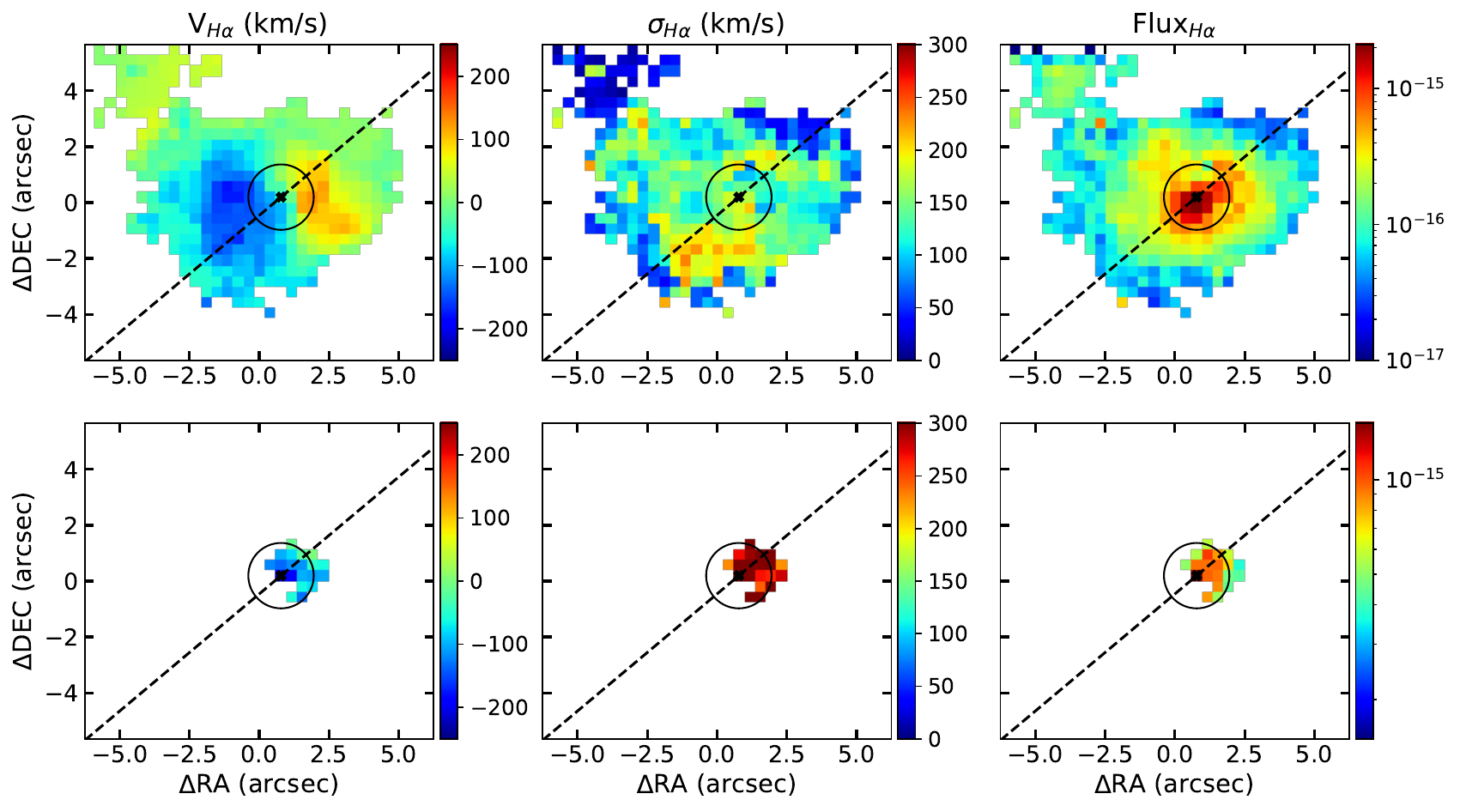}
   \includegraphics[width=.95\textwidth]{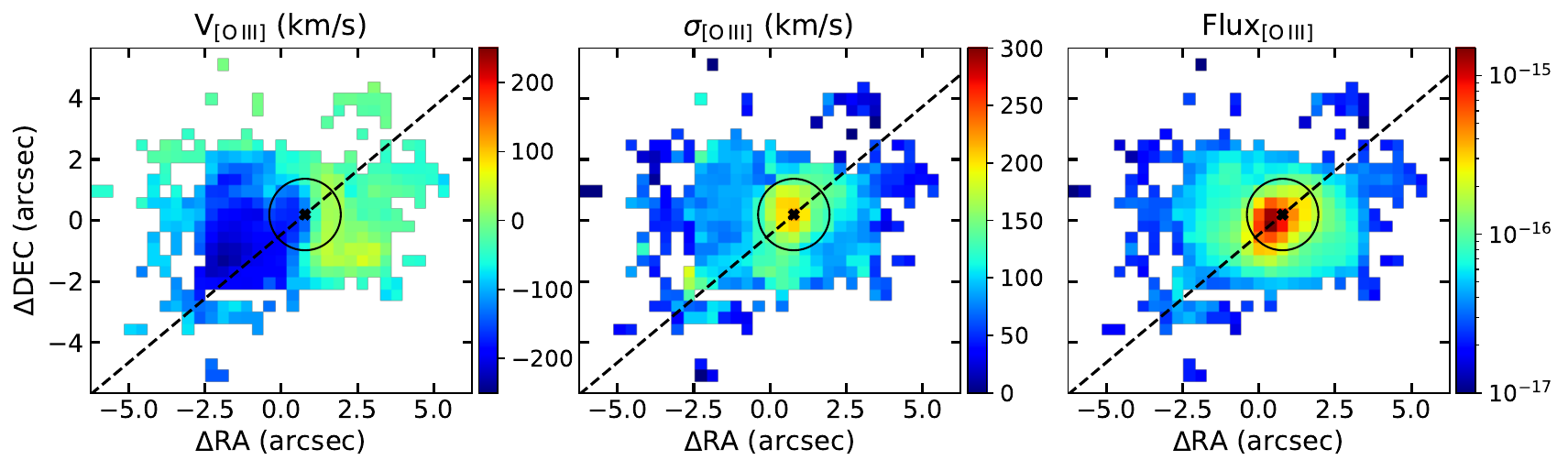}
   \caption{Kinematic maps for the stellar (top panel) and ionised gas components of NGC\,4750. The spectra of this galaxy was modelled with a primary (second panel) and a secondary (third panel) kinematical components per emission line. The kinematic maps for the primary component of the [O\,III] line are in the fourth panel. A full description is provided in Fig.~\ref{Fig5:KinMaps_NGC0266}.}
   \label{Fig5:KinMaps_NGC4750}
   \end{figure*}

   \noindent \textit{NGC\,4750} (Figs.~\ref{Fig5:KinMaps_NGC4750},~\ref{Fig5:BPT_NGC4750} and~\ref{Fig:NGC4750_BPTs}):  In contrast to the stellar component, we found peculiar kinematic maps for the ionised gas. We modelled the LR-VR combined cube with two components per emission line plus an additional very broad component associated to the BLR in H$\alpha$ (see Sect.~\ref{Sect:Discussion}). The primary component shows two blobs in the velocity map near the nucleus, with velocities differing from the rest of the map, barely rotating. These two regions located eastward (approaching side) and westward (receding side) from the nucleus do not correlate with any particular feature in the velocity dispersion map (average $\sigma$ in the whole map of $\sim$\,120\,km\,s$^{-1}$), that has a S-like structure of high $\sigma$ (i.e. $\sigma >$\,200\,km\,s$^{-1}$). The position angle of the rotating region of the ISM velocity map differs by $\sim$\,45$^{\circ}$ from that of the stellar component (PA$_{\star}$\,$\sim$\,175$^{\circ}$ vs PA$_{\rm gas}$\,$\sim$\,210$^{\circ}$; see Table~\ref{Table:resultsKin}). The flux is enhanced in a region located at $\sim$\,5\arcsec\,north-east from the centre, that corresponds to a low velocity dispersion region ($\sigma\sim$50\,km\,s$^{-1}$) of approximately 2\arcsec$\times$2\arcsec\,in size. 
   The secondary component is barely resolved out of the PSF region, slightly extended to the west from the nucleus, up to 2.5\arcsec. It is mainly blueshifted and broader than the primary component (difference in $\sigma$ of 150\,km\,s$^{-1}$; see Table~\ref{Table:resultsKin}). We found a very broad, BLR component in H$\alpha$ with $\langle v\rangle$\,$=$\,156$\pm$18\,km\,s$^{-1}$ and $\langle \sigma\rangle$\,$=$\,929$\pm$127\,km\,s$^{-1}$ (see Sect.~\ref{Sect:Discussion}).
   
   \noindent For the LR-B cube we could model both [O\,III] and H$\beta$\,lines with a single component. The S/N of the emission lines is $> 5$ only for 31\% and 29\% of the cube for [O\,III]$\lambda$5008\AA\,and H$\beta$, respectively. Contrarily to the emission lines in the LR-R cube, we do not detect a secondary component for any of the lines in this cube, probably because of the lower S/N. The primary component has a similar behaviour to the LR-VR cube, with two blobs in the velocity map. There is a shift of $\sim 25$\,km\,s$^{-1}$ between the velocity of the blobs of [O\,III] and H$\alpha$. The velocity dispersion is consistent within the errors to that of H$\alpha$\,in the PSF region ([O\,III] $\sigma_{\rm gas,P}^{\rm PSF} = 155 \pm 34$\,km\,s$^{-1}$ and H$\beta$\,$\sigma_{\rm gas,P}^{\rm PSF} = 116 \pm 32$\,km\,s$^{-1}$ vs H$\alpha$\,$\sigma_{\rm gas,P}^{\rm PSF} = 153 \pm 29$\,km\,s$^{-1}$; see Table~\ref{Table:resultsKin}). It is enhanced in the same direction of the H$\alpha$ S-like structure, but with a different morphology (just elongated SE-NW for [O\,III]). We do not detect a broad component in H$\beta$ (neither in individual spaxels nor in the integrated spectrum), which matches its classification as a type 1.9 LINER. \\
   
\begin{figure*}
   \includegraphics[width=\textwidth]{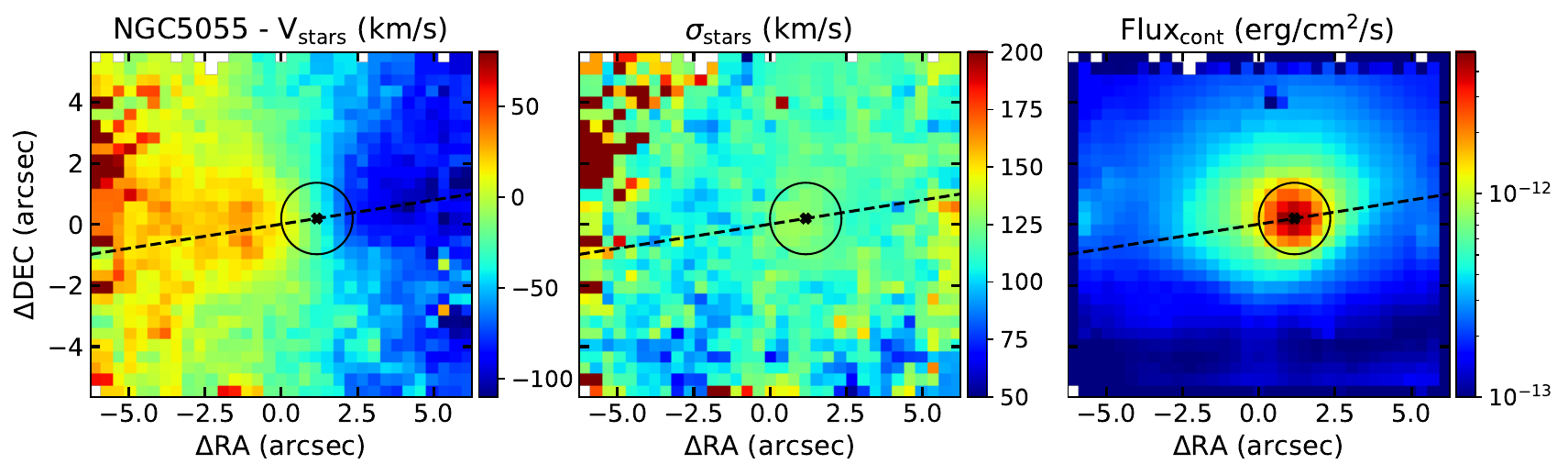}
   \includegraphics[width=\textwidth]{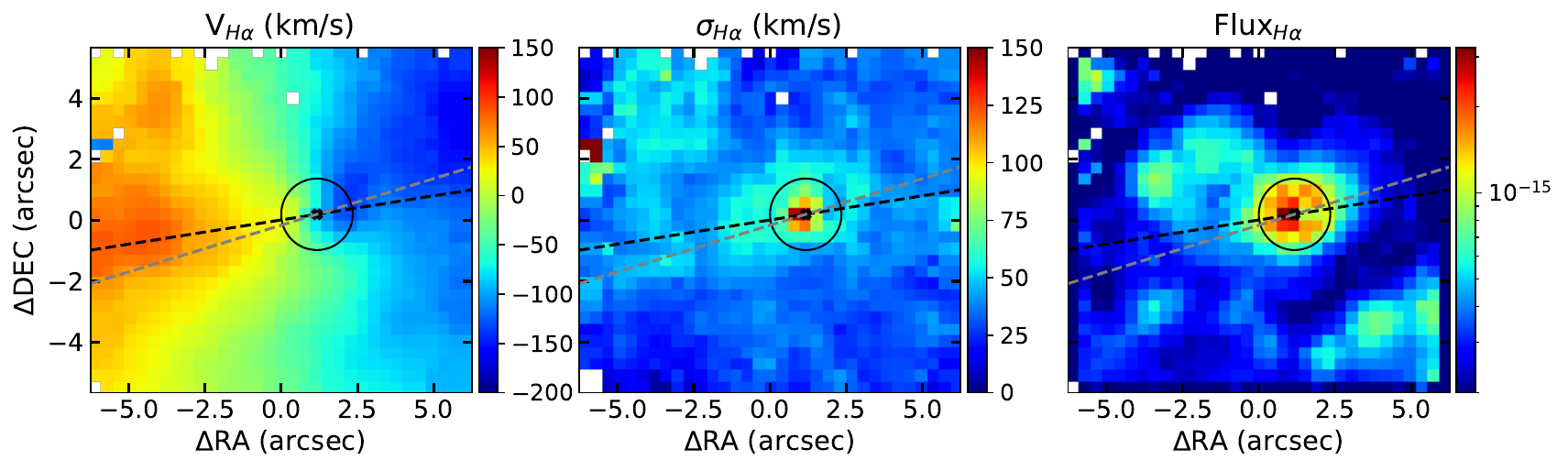}
   \caption{Kinematic maps for the stellar (top panel) and ionised gas (bottom panel) components of NGC\,5055. The spectra of this galaxy was modelled with a single narrow component on the emission lines. A full description is provided in Fig.~\ref{Fig5:KinMaps_NGC0266}. }
   \label{Fig5:KinMaps_NGC5055}
   \end{figure*}

\begin{figure}
   \includegraphics[width=\columnwidth]{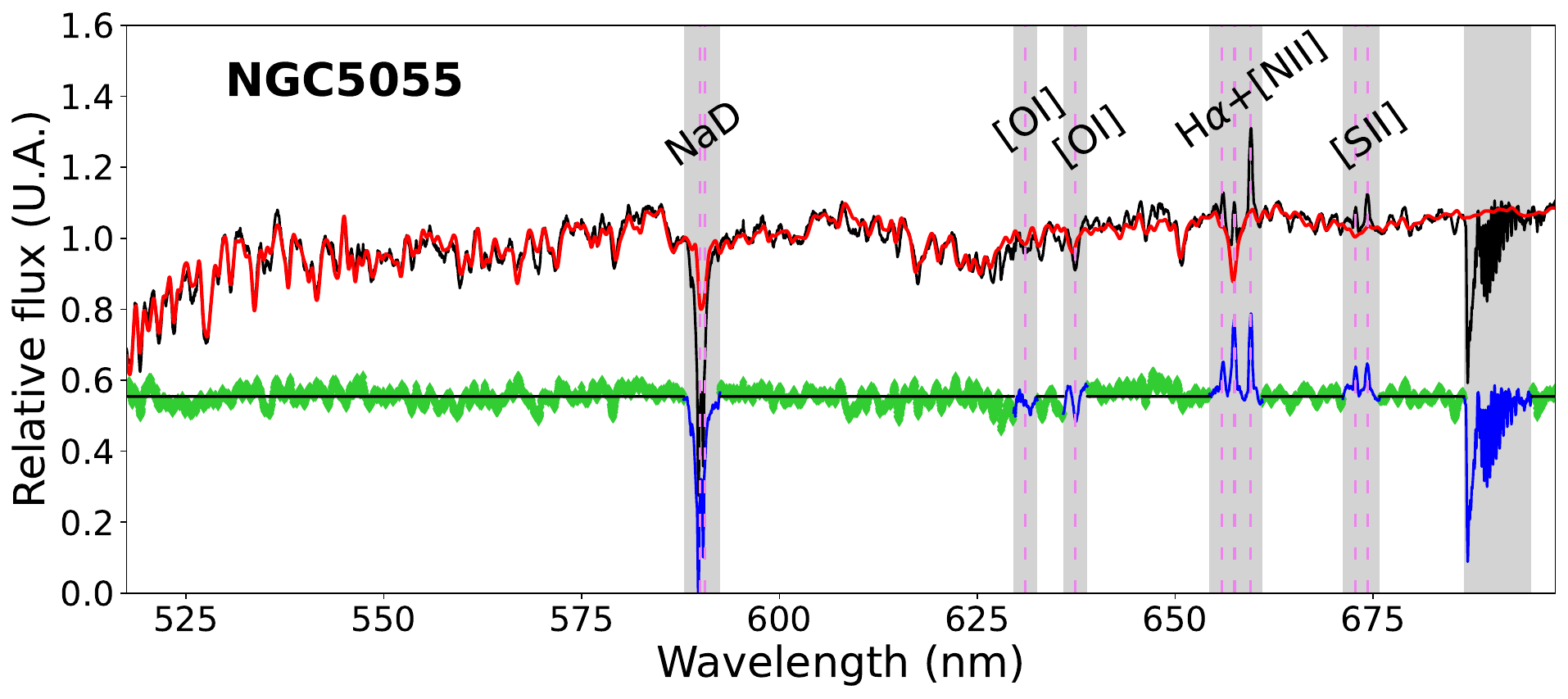}
   \caption{Stellar component modelling of the integrated PSF region for NGC\,5055. Same description as in Fig.~\ref{Fig5:examplePPXF}. }
   \label{Fig5:PPXF_NGC5055}
   \end{figure}

   \noindent \textit{NGC\,5055} (Figs.~\ref{Fig5:KinMaps_NGC5055} and~\ref{Fig:NGC5055_BPTs}):    
   \noindent In the ionised gas spectra, the [S\,II] lines have always S/N\,$>$\,5, whereas the [O\,I] is not detected in the FoV. The line profiles are quite narrow ($\sigma$\,$<$\,140\,km\,s$^{-1}$), so we could model them with a single kinematic component, including the PSF region. The median velocity dispersion of the primary component is $\sigma$\,=\,40$\pm$20\,km\,s$^{-1}$. The velocity map follows a rotating pattern with a semi-amplitude velocity of 126\,km\,s$^{-1}$, larger than that of the stars along a similar direction (PA$_{\rm gas} \sim$\,107$^{\circ}$, see Table~\ref{Table:resultsKin}). The velocity dispersion map peaks at the photometric centre, but it becomes asymmetric in the north-east side with respect to the south-west region (see Fig.~\ref{Fig5:KinMaps_NGC5055}). Finally, the H$\alpha$\,flux map is centrally peaked, with various enhanced regions in the north-east and south-west direction (see Fig.~\ref{Fig5:Compare}). The blobs in the north-east direction are anti-correlated with the regions mentioned before where $\sigma$ is more prominent.

   \begin{table*}
   \small
        \caption{Main kinematical parameters derived from the stellar modelling and for the individual components detected in the gas modelling of the H$\alpha$ line for all the cubes.}
        \label{Table:resultsKin}
    \centering
        \begin{tabular}{l|ccc|ccccc|cc}
        \hline \hline
        Galaxy & PA$_{\star}$ & $\Delta v_{\star}$ & $\sigma^{\rm PSF}_{\star}$ & PA$_{\rm gas,P}$ & $\Delta v_{\rm gas,P}$ & $\sigma^{\rm PSF}_{\rm gas,P}$ &  $\langle v_{\rm gas,S}\rangle$ & $\sigma^{\rm PSF}_{\rm gas,S}$ & Broad & FWHM \\ 
         & ($^{\circ}$) & (km\,s$^{-1}$) & (km\,s$^{-1}$) & ($^{\circ}$) & (km\,s$^{-1}$) & (km\,s$^{-1}$) & (km\,s$^{-1}$) & (km\,s$^{-1}$) & (Y/N) & (km\,s$^{-1}$) \\
        (1) & (2) & (3) & (4) & (5) & (6) & (7) & (8) & (9) & (10) & (11) \\ \hline
        NGC\,0266 & 121$\pm$7      & 172 & 182$\pm$14 & --         &  --   & 173$\pm$28 & --            &  --         &  N   & --             \\ 
        NGC\,0315 & --             & --  & 335$\pm$13 & 225$\pm$8  &  156  & 195$\pm$38 & 96$\pm$77     & 480$\pm$120 &  Y   & 2655$\pm$360   \\ 
        NGC\,3226 & 49$\pm$4       & 106 & 210$\pm$17 & 19$\pm$4   &  219  & 136$\pm$55 & $-91 \pm$27   & 262$\pm$42  &  Y   & 2305$\pm$153   \\ 
        NGC\,3245 & 178$\pm$3      & 144 & 245$\pm$18 & 174$\pm$6  &  232  & 154$\pm$46 & $-214\pm 142$ & 198$\pm$173 &  N   & --             \\ 
        NGC\,4278$^{(\dagger)}$    & 20$\pm$1   & 104 & 336$\pm$35 & 46$\pm$10  &  240  & 146$\pm$23    & 82$\pm$150  &  49$\pm$27  &      &                \\
        NGC\,4438 & 27$\pm$2       & 111 & 148$\pm$7  & 28$\pm$6   &  154  & 99$\pm$28  & $-$8$\pm$35    & 256$\pm$57  &  [Y] & [1980$\pm$350] \\ 
        NGC\,4750 & 210$\pm$10 & 97  & 134$\pm$7  & --         &  --   & 153$\pm$29 & $-$92$\pm$50   & 297$\pm$41  &  Y   & 2188$\pm$300   \\ 
        NGC\,5055 & 99$\pm$4       &  79 & 50$\pm$4   & 107$\pm$11 &  126  & 69$\pm$26  & --            &  --         &  N   & --             \\
        \hline
        \end{tabular} \\
        \tablefoot{Columns indicate: (1) object, $^{(\dagger)}$ indicates that the values correspond to the [O\,III] line; (2) major axis position angle of the stellar velocity maps using the receding sides as reference; (3) semi-amplitude velocity of the stellar maps; (4) central velocity dispersion in the PSF region of the stellar maps; (5) position angle for the major axis of the primary component of the ionised gas; (6) and (7) semi-amplitude velocity and central velocity dispersion (in the PSF region) of the primary component of the ionised gas; (8) and (9) average velocity and velocity dispersion integrated in the PSF region for the secondary component of the ionised gas; (10) indicates if there is or not a broad component associated to the BLR detected in the Balmer lines and (11) the FWHM of that component. For all columns, the reported errors are 1$\sigma$ values. [] indicates that the BLR fit is not well constrained.}
        \end{table*}

   \subsubsection{Line ratios}
   \label{SubSec:ResultsGasLineRatios}

\begin{figure*}
   \includegraphics[width=\textwidth]{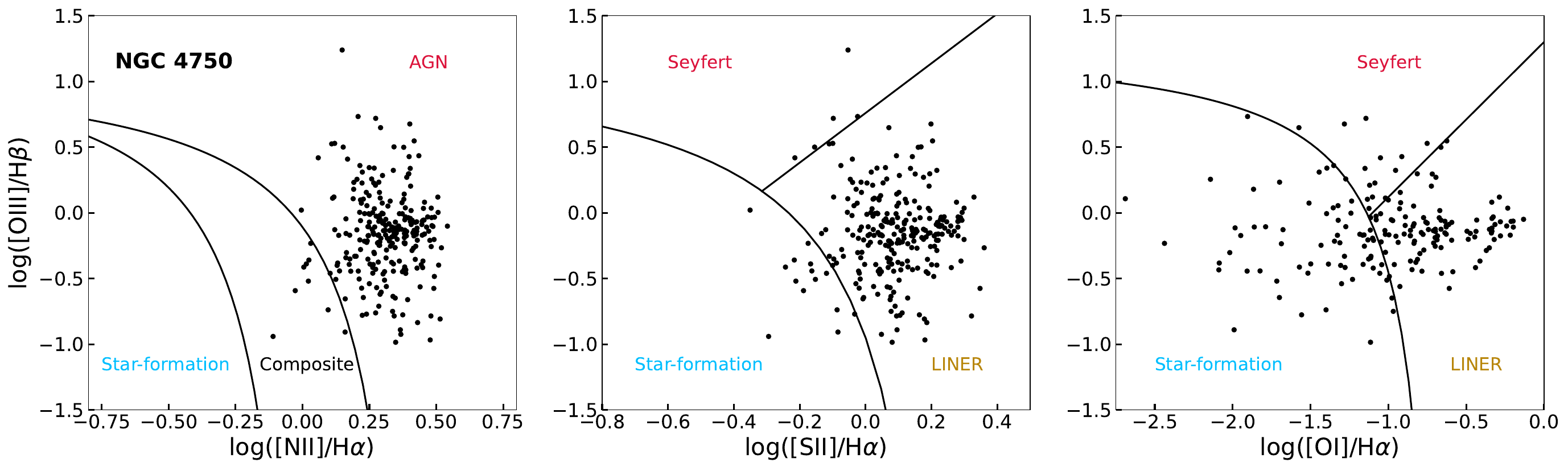}
   \includegraphics[width=\textwidth]{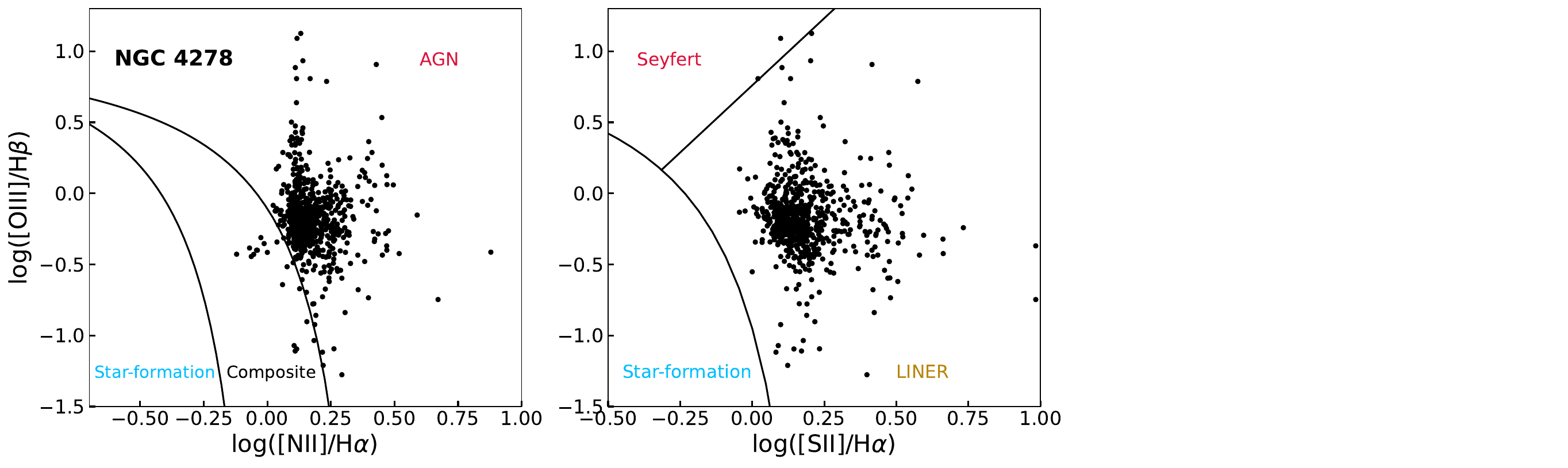}
   \caption{BPT diagrams of the primary kinematical component of the ionised gas for NGC\,4750 (upper panels) and NGC\,4278 (bottom panels), with the dividing lines from \cite{Kauffmann2003} and \cite{Kewley2006}. For NGC\,4278 the [O\,I] lines are not available in the wavelength range covered by the HR-R data cube (see Sect.~\ref{Sect:sample_data}).}
   \label{Fig5:BPT_NGC4750}
   \end{figure*}

   \noindent We derived the log([S\,II]/H$\alpha$) and log([N\,II]/H$\alpha$) ratios for all galaxies and components, log([O\,I]/H$\alpha$) for five objects (namely NGC\,0266, NGC\,0315, NGC\,3226, NGC\,4438 and NGC\,4750), and log([O\,III]/H$\beta$) for two objects (namely, NGC\,4278 and NGC\,4750), for which we could produce the BPT diagrams of their primary kinematical component (see Fig.~\ref{Fig5:BPT_NGC4750}). In general, all the line ratios for all the different components can be explained by AGN photoionisation (see our conservative limits in Sect.~\ref{SubSec:datared_BPTs}) except for NGC\,3245 and NGC\,5055, that are consistent with star-forming/composite photoionisation\footnote{We note that the line ratios in the discs could have a contribution from post-AGB stars, particularly in those objects with no detection of the [OI] line, as is the case for NGC\,5055 \citep{Ho1997,Sarzi2010,Singh2013,Ricci2015}}. 
   
   \noindent For NGC\,3245 the line ratios for both the primary and secondary components are very similar (see Fig.~\ref{Fig:NGC3245_BPTs} and Table~\ref{Table:resultsBPTs}), laying both in the star-forming/composite regimes of the BPTs (see Sect.~\ref{SubSec:datared_BPTs}). The ratios in the central region of the primary component are lower than the average (i.e. log([S\,II]/H$\alpha$)\,$= -0.36 \pm 0.08$ and log([N\,II]/H$\alpha$)\,$= -0.04 \pm 0.07$, see Table~\ref{Table:resultsBPTs}), showing an increase towards the north and south parts of the FoV (log([S\,II]/H$\alpha$)\,$\sim 0.08$ and log([N\,II]/H$\alpha$)\,$\sim 0.32$). For NGC\,5055 (see Fig.~\ref{Fig:NGC5055_BPTs} and Table~\ref{Table:resultsBPTs}), the line ratio maps show some regions with smaller ratios (log([S\,II]/H$\alpha$)\,$\sim -0.5$ and log([N\,II]/H$\alpha$)\,$\sim -0.4$) than the mean values (see Table~\ref{Table:resultsBPTs}) at the north-east and south limits of the FoV. These regions are correlated with the various blobs visible in the kinematical maps corresponding to the low-$\sigma$ regions and the H$\alpha$ flux map enhancements (see Sect.~\ref{SubSec:ResultsGasKin}). \\

   \noindent All the components used to model the spectra of NGC\,0266 and NGC\,0315 (see Figs.~\ref{Fig:NGC0266_BPTs} and~\ref{Fig:NGC0315_BPTs}) have homogeneous line ratios, with mean values consistent with AGN ionisation (see Table~\ref{Table:resultsBPTs} and Sect.~\ref{SubSec:datared_BPTs}). The line ratios of the secondary kinematical component of NGC\,3226 (see Fig.~\ref{Fig:NGC3226_BPTs}) and the secondary and tertiary (not detected in [O\,I]) components of NGC\,4438 (see Fig.\ref{Fig:NGC4438_BPTs}) are consistent with AGN photoionisation (see Sect.~\ref{SubSec:datared_BPTs} and Table~\ref{Table:resultsBPTs}), specially in the direction of the bubble for the latter galaxy. Despite that the average values of their primary component line ratios are also consistent with AGN photoionisation (AGN/composite for NGC\,3226), their maps show some substructures. On the one hand, for NGC\,3226 (see Fig.~\ref{Fig:NGC3226_BPTs}), the log([S\,II]/H$\alpha$) map is roughly homogeneous, whereas log([N\,II]/H$\alpha$) and log([O\,I]/H$\alpha$) have slightly larger ratios along the minor axis, coincident with the enhanced-$\sigma$ region (mean of $0.04$ and $-0.98$, respectively). Additionally, log([O\,I]/H$\alpha$) is enhanced in the FoV limits ($\sim -0.8$). On the other hand, for NGC\,4438 the line ratios are in general smaller along the galactic disc (log([N\,II]/H$\alpha$) and log([S\,II]/H$\alpha$) $0.05\pm 0.10$ and $-0.04\pm 0.15$, respectively; not mapped by the [O\,I] due to low S/N), thus consistent with the composite region. On the contrary, they are larger in the direction of the bubble (log([N\,II]/H$\alpha$), log([S\,II]/H$\alpha$) and log([O\,I]/H$\alpha$) $0.28 \pm 0.11$, $0.21 \pm 0.15$ and $-0.66 \pm 0.41$, respectively), falling in the AGN region (see Sect.~\ref{SubSec:datared_BPTs} and Fig.~\ref{Fig:NGC4438_BPTs}). \\

   \noindent The only two objects for which we could produce the BPT diagrams are NGC\,4278 and NGC\,4750 (see Table~\ref{Table:resultsBPTs} and Fig.~\ref{Fig5:BPT_NGC4750}). In both cases the primary component is consistent with AGN/LINER photoionisation in the diagrams. Although in the case of NGC\,4278 we detected a secondary component in all the lines, they are in different spatial positions, so we cannot represent them in the BPT diagrams. Nevertheless, the secondary component of both galaxies are also consistent with the AGN region (see Table~\ref{Table:resultsBPTs} and Figs.~\ref{Fig:NGC4278_BPTs} and~\ref{Fig:NGC4750_BPTs}). 
   
   \noindent For NGC\,4278 in general the line ratios of the primary component are quite homogeneous (standard deviation $\sim 0.1$; see Fig.~\ref{Fig:NGC4278_BPTs}), except slightly higher values near the limits of the FoV probably due to S/N effects, and in the region of the secondary component for the log([O\,III]/H$\beta$) ratio ($0.15 \pm 0.10$; see Table~\ref{Table:resultsBPTs}). We note that the spectra were not corrected from the stellar contribution from the host (see Sect.~\ref{SubSec:datared_stars}), so the ratios with H$\alpha$ are upper limits. Nevertheless, considering the small contribution of the stars in the H$\alpha$ line in \cite{Cazzoli2018}, we do not expect a drastic difference in the final line ratios.

   \noindent For NGC\,4750 the more distinctive kinematic feature in the primary component is a blob $\sim 5$\,\arcsec\,north-east from the nucleus, that also has different line ratios from the rest of the maps (log([N\,II]/H$\alpha$)\,$= -0.24 \pm 0.08$, log([S\,II]/H$\alpha$)\,$= -0.44 \pm 0.11$ and log([O\,I]/H$\alpha$)\,$=-1.84 \pm 0.35$), consistent with ionisation due to star-forming (see Fig.~\ref{Fig:NGC4750_BPTs}). Overall, the ratios of the primary component are consistent with the AGN/LINER region in the BPTs, except for the [O\,I] ratio (see Fig.~\ref{Fig5:BPT_NGC4750} and Table~\ref{Table:resultsBPTs}), with larger dispersion (standard deviation of [O\,I]\,$\sim 0.6$ vs $\sim 0.2$ for [N\,II]).

   \begin{table*}
   \small 
        \caption{Average line ratios of the ionised gas for all the derived components.}
        \label{Table:resultsBPTs}
    \centering
        \begin{tabular}{ll|ccccc}
        \hline \hline
        Galaxy & Component & log([S\,II]/H$\alpha$) & log([N\,II]/H$\alpha$) & log([O\,I]/H$\alpha$) & log([O\,III]/H$\beta$)  \\ 
        (1) & (2) & (3) & (4) & (5) & (6) \\ \hline
        NGC\,0266 & Primary    & \phantom{1}0.22$\pm$0.19 & \phantom{1}0.42$\pm$0.17 & $-0.68\pm$0.30 & --                       \\
        NGC\,0315 & Primary    & \phantom{1}0.28$\pm$0.21 & \phantom{1}0.53$\pm$0.18 & $-0.43\pm$0.33 & --                       \\
                  & Secondary  & \phantom{1}0.34$\pm$0.31 & \phantom{1}0.13$\pm$0.36 & $-0.23\pm$0.30 & --                       \\
        NGC\,3226 & Primary    & \phantom{1}0.06$\pm$0.12 & $-0.04\pm$0.09           & $-1.08\pm$0.19 & --                       \\
                  & Secondary  & \phantom{1}0.05$\pm$0.25 & \phantom{1}0.17$\pm$0.16 & $-0.60\pm$0.49 & --                       \\
        NGC\,3245 & Primary    & $-0.30\pm$0.12           & 0.01$\pm$0.01            & --             & --                       \\
                  & Secondary  & $-0.55\pm$0.24           & $-0.21\pm$0.15           & --             & --                       \\
        NGC\,4278 & Primary    & \phantom{1}0.18$\pm$0.12 & \phantom{1}0.17$\pm$0.09 & --             & $-0.18\pm$0.28           \\
                  & Secondary  & \phantom{1}0.17$\pm$0.03 & \phantom{1}0.17$\pm$0.04 & --             & \phantom{1}0.35$\pm$0.30 \\
        NGC\,4438 & Primary    & \phantom{1}0.05$\pm$0.19 & \phantom{1}0.15$\pm$0.15 & $-0.96\pm$0.49 & --                       \\
                  & Secondary  & \phantom{1}0.10$\pm$0.12 & \phantom{1}0.16$\pm$0.11 & $-1.10\pm$0.49 & --                       \\
                  & Tertiary   & \phantom{1}0.28$\pm$0.16 & \phantom{1}0.25$\pm$0.06 & --             & --                       \\
        NGC\,4750 & Primary    & \phantom{1}0.07$\pm$0.14 & \phantom{1}0.30$\pm$0.13 & $-1.08\pm$0.55 & $-0.15\pm$0.33           \\
                  & Secondary  & \phantom{1}0.23$\pm$0.16 & \phantom{1}0.42$\pm$0.18 & $-0.80\pm$0.47 & --                       \\
        NGC\,5055 & Primary    & $-0.23\pm$0.03           & $-0.09\pm$0.03           & --             & --                       \\
        \hline
        \end{tabular} \\
        \tablefoot{Columns indicate: (1) object; (2) kinematical component from the emission lines used to derived the ratios; (3), (4), (5) and (6) are the average line ratios in logarithm scale, considering the standard deviation as the 1$\sigma$ error.} 
\end{table*}

    \subsection{Estimation of the properties of the detected outflows}
    \label{SubSec_ResultOutflows}

    \noindent We have detected kinematical components consistent with an ionised outflow (i.e. mainly blueshifted secondary component) in five out of the nine galaxies in the sample (namely, NGC\,1052, NGC\,3226, NGC\,3245, NGC\,4278, NGC\,4438 and NGC\,4750; see Table~\ref{Table:discussion_comp}). In this section we characterise their main properties using physical relations (see equations in Sect.~\ref{SubSec:datared_CalcOutfParam}) that depend on observational parameters, such as the velocity, outflow extension, and fluxes, that we can estimate from the kinematical and flux maps. The results are summarised in Table~\ref{Table:resultsOutflows}. \\
    
    \noindent To estimate the mass outflow rate, $\dot{\rm M}_{\rm OF,ion}$ according to \cite{Venturi2021}, we have to calculate the intrinsic luminosity of H$\alpha$, that depends on the ratio of the H$\alpha$/H$\beta$ fluxes. We cannot estimate the ratio using MEGARA data, as we only have both H$\alpha$ and H$\beta$ for NGC\,4278 and NGC\,4750, but for the first the outflow is only detected in H$\beta$ and for the latter in H$\alpha$; for the rest of the galaxies, only H$\alpha$ is available. Thus we decided to measure the H$\alpha$/H$\beta$ ratio uniformly in the ground-based long-slit spectra from \cite{Cazzoli2018}. The long-slit nuclear spectrum of NGC\,3245 was analysed in \cite{HM2020}, although only in the H$\alpha$ line range; thus for this galaxy we used the ratio H$\alpha$/H$\beta = 4.76$ reported by \cite{Ho1997}. 

    \noindent As already mentioned, NGC\,4278 is the only galaxy for which we detect the possible outflow exclusively in the wavelength range of the LR-B data cube. This implies that the equation~\ref{Eq_MHa} mentioned in Sect.~\ref{SubSec:datared_CalcOutfParam} cannot be used to estimate the mass of the outflow. We note though that there could be differences on the mass determinations depending on the line and on the assumptions of the gas temperature. For our determination of $M_{\rm OF,ion}$ using H$\beta$, assuming a T\,$\sim 10^{4}$\,K and $n_{\rm e}$\,$\sim 500$\,cm$^{-3}$ seem more appropriate (see Sect.~\ref{SubSec:datared_CalcOutfParam}), so in Table~\ref{Table:resultsOutflows} we report the estimated parameters following \cite{Carniani2015}. The average $M_{\rm OF,ion}$ using H$\alpha$ is $7.5 \pm 1.2 \times 10^{4}$\,$M_{\sun}$, whereas for H$\beta$ using the expressions of the various works from the literature (see Sect.~\ref{SubSec:datared_CalcOutfParam}), $M_{\rm OF,ion}^{\rm H\beta}$ varies from $1.8\times 10^{3}$ to $6.6\times 10^{3}$\,$M_{\sun}$ with typical errors from $50$ to $160$\,$M_{\sun}$. The values for [O\,III] are typically one order of magnitude less than those estimated with H$\beta$. 
    
    \noindent As seen in Fig.~\ref{Fig5:KinMaps_NGC4278} and Sect.~\ref{SubSec_ResultOutflows}, the outflow in NGC\,4278 is distributed with a shell-like morphology that has to be taken into account when estimating $\dot{M}_{\rm OF,ion}$ (see Eq.~\ref{eqMofrate_2} in Sect.~\ref{SubSec:datared_CalcOutfParam}). We identified two main shells at $R \sim 3.8$\arcsec\ and $R \sim$5.4\arcsec\ in the north-east direction with respect to the photometric centre, with thicknesses ($\Delta R$) of $2.9$\arcsec$\pm 0.4$\arcsec and $1.3$\arcsec$\pm 0.4$\arcsec, respectively. Thus, we calculated $\dot{M}_{\rm OF,ion}$, $E_{\rm OF,ion}$ and $\dot{E}_{\rm OF,ion}$ in the two shells, finding similar results for both (we report in Table~\ref{Table:resultsOutflows} the values for the closest shell to the nucleus). We note that these values do not represent the overall properties of the outflow given its morphology, but only lower limits. \\

    \noindent Globally, we obtain for our ionised gas (H$\alpha$) outflows a mean outflow velocity $\sim 340$\,km\,s$^{-1}$, a mean extension of $\sim 400$\,pc, and a mean density of $\sim 600$\,cm$^{-3}$. The galaxy with the most powerful outflow is NGC\,1052, being the most extended outflow expanding at the largest velocities from the sample. On the contrary, the least powerful outflow is that of NGC\,4438, with properties typically one order of magnitude smaller than for the other objects (see Table~\ref{Table:resultsOutflows}). The mean $\dot{\rm M}_{\rm OF,ion}$ is $0.17 \pm 0.06 M_{\sun}$ with a standard deviation of $0.13$, which is translated to a mean $\dot{\rm E}_{\rm OF,ion}$ of $\sim 2 \times 10 ^{40}$\,erg\,s$^{-1}$, lower values than for other more luminous AGNs from the literature \citep[see e.g.][]{Fiore2017,Baron2019,Venturi2021}. We discuss the implications of these results in Sect.~\ref{SubSec:Discussion_outflows}. 
   
   \begin{table*}
   \small 
        \caption{Outflow parameters derived for the LINERs with a detection, including NGC\,1052 (see \citealt{Cazzoli2022}).}
        \label{Table:resultsOutflows}
    \centering
        \begin{tabular}{l|cccccccc}
        \hline \hline
        Galaxy & $v_{\rm OF,ion}$ & $R_{\rm OF,ion}$ & $M_{\rm OF,ion}$ & $\dot{M}_{\rm OF,ion}$ & $E_{\rm OF,ion}$ & $\dot{E}_{\rm OF,ion}$ & $\dot{E}_{\rm OF,ion}$/$L_{\rm bol}$ & $n_{\rm e}$ \\ 
         & (km\,s$^{-1}$) & (pc) & ($M_{\sun}$) & ($M_{\sun}$\,yr$^{-1}$) & (erg) & (erg\,s$^{-1}$) &  & (cm$^{-3}$)\\
        (1) & (2) & (3) & (4) & (5) & (6) & (7) & (8) & (9) \\ \hline
        NGC\,1052        & 655 & 792 & 1.6$\pm$0.6\,$\times$\,10$^{5}$ & 0.36$\pm$0.27     & 1.3$\pm$0.9\,$\times$\,10$^{53}$ & 8.8$\pm$3.5\,$\times$\,10$^{40}$ &  0.01   &  360  \\
        NGC\,3226        & 138 & 242 & 1.5$\pm$0.1\,$\times$\,10$^{5}$ & 0.26$\pm$0.05     & 9.0$\pm$3.1\,$\times$\,10$^{52}$ & 1.7$\pm$0.6\,$\times$\,10$^{40}$ &  0.008   &  591  \\
        NGC\,3245        & 580 & 322 & 2.1$\pm$0.1\,$\times$\,10$^{4}$ & 0.11$\pm$0.03     & 0.5$\pm$0.4\,$\times$\,10$^{51}$ & 1.2$\pm$0.7\,$\times$\,10$^{40}$ &  0.006  &  787  \\
        NGC\,4278$^{(*)}$ & 241 & 122 & 2.0$\pm$0.5\,$\times$\,10$^{3}$  & 0.004$\pm$0.001   & 4.5$\pm$2.6\,$\times$\,10$^{49}$ & 7.2$\pm$1.7\,$\times$\,10$^{37}$ &  0.00002   &  314  \\
        NGC\,4438         & 213 & 307 & 1.90$\pm$0.01\,$\times$\,10$^{3}$ & 0.004$\pm$0.001 & 1.0$\pm$0.6\,$\times$\,10$^{51}$ & 2.7$\pm$1.4\,$\times$\,10$^{38}$ &  0.0001 &  710  \\
        NGC\,4750         & 191 & 318 & 4.2$\pm$0.1\,$\times$\,10$^{4}$  & 0.08$\pm$0.02   & 3.8$\pm$1.0\,$\times$\,10$^{52}$ & 7.5$\pm$2.7\,$\times$\,10$^{39}$ &  0.02   &  686  \\
        \hline
        \end{tabular} \\
        \tablefoot{Columns indicate: (1) object; (2) maximum velocity of the component associated to the outflow; (3) extension of the outflow; (4) mass of the ionised outflow; (5) mass outflow rate; (6) kinetic energy; (7) kinetic power of the outflow; (8) energy rate and (9) mean electronic density. $^{(*)}$ indicates that the $M_{\rm OF,ion}$ was estimated with the H$\beta$ luminosity instead of H$\alpha$. See Sect.~\ref{SubSec_ResultOutflows} for more details on the calculations.}
\end{table*}

\section{Discussion}
\label{Sect:Discussion}

   \noindent The discussion is focused on the interpretation of the kinematical components used to model the emission lines, including NGC\,1052 (see C22). All the targets, except for NGC\,5055, were analysed in our previous imaging \citep[][HM22]{Masegosa2011} and long-slit spectroscopic\footnote{These works analysed both ground-based and space-based nuclear spectra of the galaxies. As they detected varying results due to the different resolutions of the instruments, we will conservatively use only the ground-based results to do a fair comparison with the MEGARA data, also ground-based.} works \citep[][]{Cazzoli2018,HM2020}. The interpretations given to the components in each work are summarised in Table~\ref{Table:discussion_comp}. For NGC\,0266 and NGC\,4750 there were signatures of outflows with both long-slit spectroscopic and imaging data. For NGC\,0315 and NGC\,3226 there were hints of an outflow with the long-slit spectra but not with the imaging data. For NGC\,3245 and NGC\,4438, the kinematical components from the long-slit spectroscopic data were explained with rotational movements in \cite{HM2020} and \cite{Cazzoli2018}, respectively, whereas both were classified as outflow candidates based on the imaging morphology (\lq Bubble\rq\,candidate in HM22). NGC\,4278 was classified as candidate for an inflow in \cite{Cazzoli2018}, and as \lq Core-halo\rq\,in \cite{Masegosa2011}. For NGC\,5055, we have no prior kinematical information in our works, but we have a morphological classification of the H$\alpha$ gas as \lq Core-halo\rq\,in \cite{Masegosa2011}. In Appendix~\ref{Appendix:IndComm} we report a more detailed comparison between all these works for each object individually. \\

   \noindent In general, with the MEGARA data we find extended emission for all galaxies except for NGC\,0315, which is mainly unresolved (see Sect.~\ref{Sect:Results}). We modelled with at least two different kinematical components the emission lines in seven out of the nine objects. The secondary component traces an unresolved kinematical component (i.e. contained in the PSF region) in three cases (namely NGC\,0315, NGC\,3226 and NGC\,3245), whereas it is extended for four objects (namely NGC\,1052, NGC\,4278, NGC\,4438 and NGC\,4750). We needed a third component to model the emission lines only in two objects: unresolved in NGC\,1052 (see C22) and extended in NGC\,4438 (see Fig.~\ref{Fig5:KinMaps_NGC4438} and Sect.~\ref{SubSec:Discussion_NGC4438}). We have interpreted the secondary component as a possible signature of outflows for NGC\,1052 (extensively discussed in C22), NGC\,3226, NGC\,3245, NGC\,4278, NGC\,4438 and NGC\,4750. Additionally, we modelled exclusively the H$\alpha$ line with a very broad component associated to the BLR for four LINERs, namely NGC\,0315, NGC\,1052, NGC\,3226 and NGC\,4750, and we cannot constrain its presence for NGC\,4438 (see Table~\ref{Table:resultsKin}). These results are consistent with the previous works by \cite{Cazzoli2018} and \cite{HM2020}, using long-slit spectroscopic data. The non-detections with the ground-based data of a BLR component in H$\alpha$ for type-1.9 LINERs, as mentioned in \cite{Cazzoli2018}, could be a consequence of the low luminosity of these AGNs, whose BLR can be diluted if seen with relatively low spatial resolution (due to the long-slit aperture and the spaxel size). As mentioned in Sect.~\ref{SubSec:ResultsGasKin}, we did not detect any BLR component for the H$\beta$ line, when available (see Table~\ref{Table:resultsKin}), thus confirming the nature of the galaxies as either type 1.9 or type 2 LINERs.
   
   \noindent The flux maps of the primary component for all galaxies, except for NGC\,0266, NGC\,4438 (see Sect.~\ref{SubSec:Discussion_NGC4438}) and NGC\,4750, are centrally peaked and show no particular morphologies. There is, however, a particular feature that is common for five targets (NGC\,1052, NGC\,3226, NGC\,3245, NGC\,4278 and NGC\,4750) in the ionised gas velocity dispersion maps: an enhanced $\sigma$ region with a \lq butterfly\rq-like shape. For NGC\,1052 and NGC\,4750 this feature is found along the stellar major axis; whereas for NGC\,3226, NGC\,3245 and NGC\,4278, it is almost perpendicular to the major axis. Given that these galaxies have a secondary component, as NGC\,1052 for which it was interpreted as an outflow, it is likely that we are in a similar case. Thus the high-$\sigma$ region would be turbulent gas produced by an outflow, probed by the secondary component of the gas. We discuss this feature in more detail in Sect.~\ref{SubSec:Discussion_highsigma}, and the properties and origin of the outflows in Sect.~\ref{SubSec:Discussion_outflows}. \\ 

   \noindent There are five targets with radio emission detected in the sample (namely NGC\,0266, NGC\,0315, NGC\,1052, NGC\,4278 and NGC\,4438; see \citealt{Nagar2005,Giroletti2005,Hota2007,Baldi2018}). There are three radio jets: in NGC\,1052, already discussed in C22; in NGC\,0315, for which the ionised gas emission is barely resolved in our data, so a possible correlation with the jet can not be explored; and in NGC\,4278 (see Sect.~\ref{SubSec:Discussion_outflows}); and two compact detections in radio continuum in NGC\,0266 and NGC\,4438. We discuss a possible connection between the radio and the ionised gas emission in Sects.~\ref{SubSec:Discussion_highsigma} and~\ref{SubSec:Discussion_NGC4438}. 

   \begin{table}
   \small 
        \caption{Interpretation of the secondary component fitted to the ionised gas emission lines for each galaxy in this and previous works.}.
        \label{Table:discussion_comp}
    \centering
        \begin{tabular}{l|cccc}
        \hline \hline
        Galaxy & C18 & HM20 & HM22 & This work \\ 
        (1) & (2) & (3) & (4) & (5) \\ \hline
        NGC\,0266         & Outflow     & --       &  Bubble     & Rotation   \\ 
        NGC\,0315         & Outflow     & --       &  Core-halo  & Unresolved \\ 
        NGC\,1052         & Outflow     & --       &  Bubble     & Outflow    \\ 
        NGC\,3226         & Outflow     & --       &  Dusty      & Outflow    \\ 
        NGC\,3245         & --          & Rotation &  Bubble     & Outflow    \\ 
        NGC\,4278         & Inflow      & --       &  Core-halo  & Outflow    \\ 
        NGC\,4438         & Rotation    & --       &  Bubble     & Outflow    \\ 
        NGC\,4750         & Candidate   & --       &  Bubble     & Outflow    \\ 
        NGC\,5055         & --          & --       &  Core-halo  & Rotation   \\ 
        \hline
        \end{tabular}  \\
        \tablefoot{Columns indicate: (2) \cite{Cazzoli2018}, (3) \cite{HM2020} and (4) HM22, and (5) this work (C18, HM20, HM22 and This work, respectively). \lq In/Outflow\rq\,indicates candidates of hosting an in/outflow, \lq Rotation\rq\,indicates those for which the kinematical components were consistent with being produced by rotational motions, \lq Candidate\rq\,indicates those candidates for non-rotational motions. \lq Bubble\rq\,refers to bubble-like, filamentary emission; \lq Core-halo\rq\,to compact emission in the circumnuclear region; and \lq Dusty\rq\,when dust lanes covered the central parts of the galaxy. A complete definition of the latter three categories can be found in HM22.}
\end{table}

   \begin{figure*}
   \includegraphics[width=.7\columnwidth]{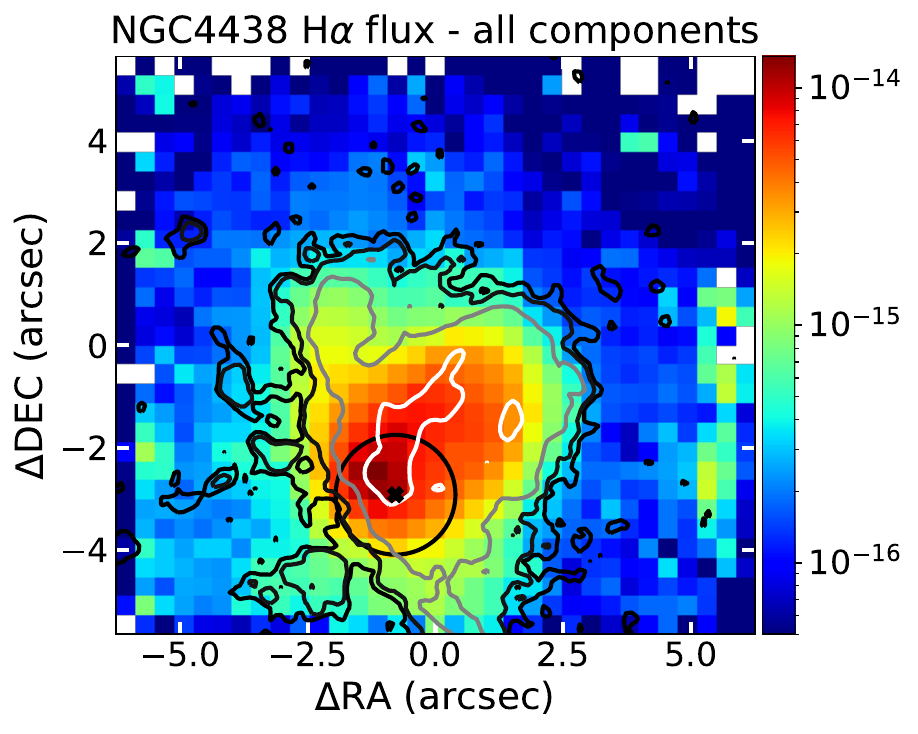}
   \includegraphics[width=.7\columnwidth]{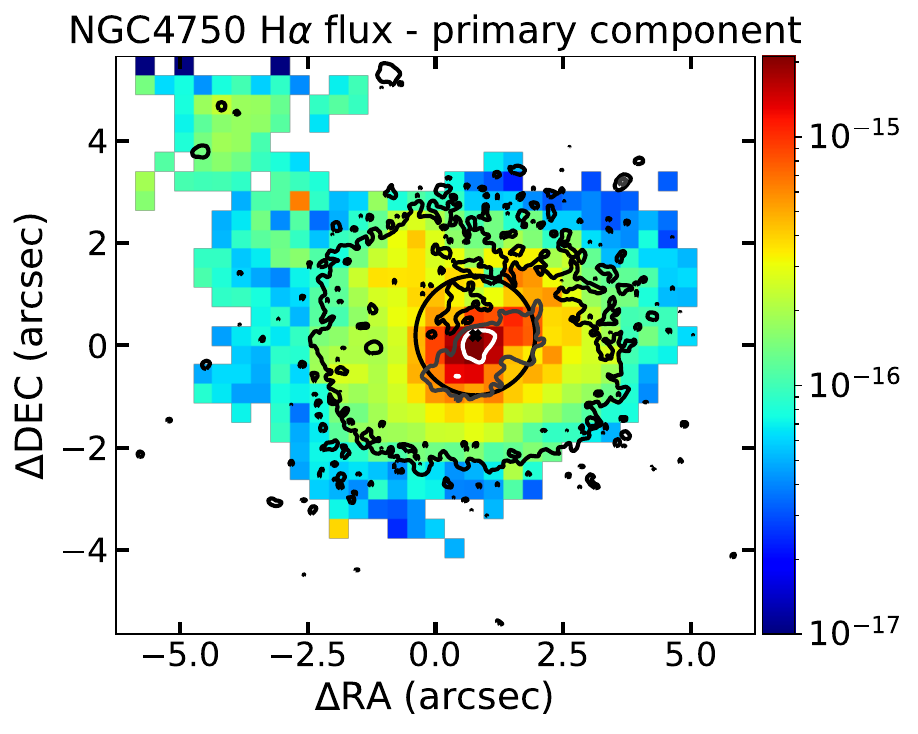}
   \includegraphics[width=.7\columnwidth]{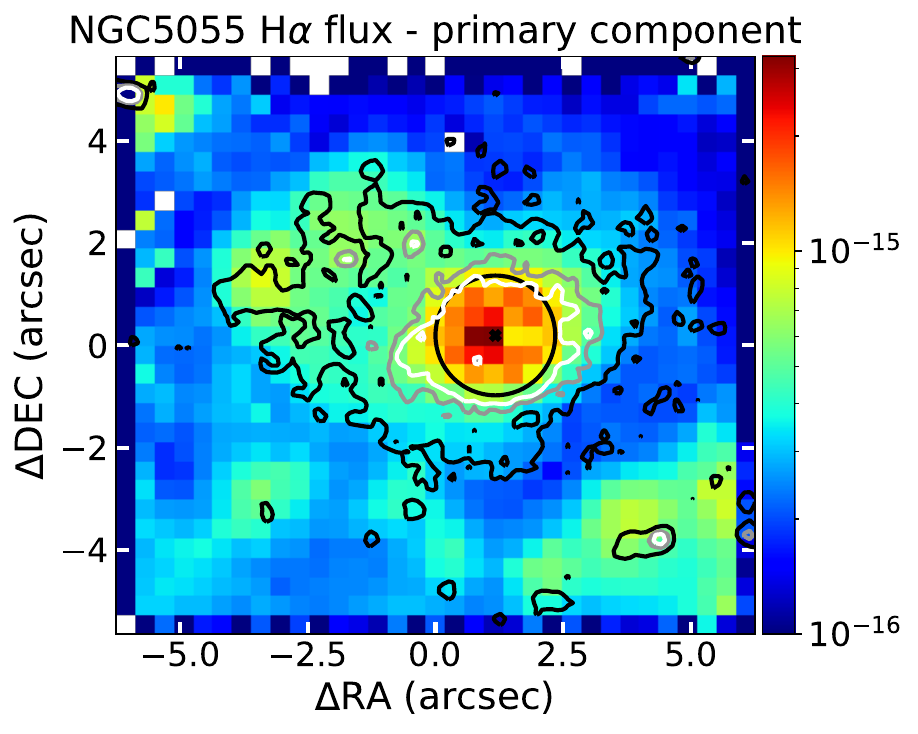}
   \caption{H$\alpha$ flux derived from the MEGARA data superimposed with the contours of the narrow band HST images of the H$\alpha$ emission from \cite{Masegosa2011} and HM22 for NGC\,4438 (total flux accounting for all components), NGC\,4750 and NGC\,5055 (only primary component of the emission lines). The inner black circle and the black crosses indicate the PSF region and the photometric centre. The contours for NGC\,4750 and NGC\,4750 represent the 3$\sigma$, 5$\sigma$ and 10$\sigma$ (black, grey and white) flux levels of the narrow band image, with an extra contour at 100$\sigma$ for NGC\,4438 (black, black, grey, white).}
   \label{Fig5:Compare}
   \end{figure*}

\subsection{Dynamical support and galaxy morphologies}
\label{SubSec:Discussion_morphologies}

   \noindent We can estimate the dynamical support of the galaxies through the V/$\sigma$ ratio, estimated from the amplitude of the velocity maps and the mean value of $\sigma$ in the stellar maps over the FoV of MEGARA. We note that none of the position-velocity diagrams (see Figs.~\ref{Fig:PV-PS_1} and~\ref{Fig:PV-PS_2}) reach to the flattening region of the curves, so the values to estimate the velocity amplitude and the mean velocity dispersion correspond only to the inner parts of the stellar discs, except for NGC\,1052, which was completely sampled with MUSE data (see C22). 

   \noindent NGC\,0266, NGC\,3245, NGC\,4438, NGC\,4750 and NGC\,5055 are classified as spiral galaxies, and thus we should expect for them a velocity map with rotational patterns and a centrally peaked velocity dispersion. In fact, the V/$\sigma$ values for these objects are the largest (1.8, 1.6, 1.4, 1.5 and 1.4, respectively), consistent with being dynamically cold discs. For NGC\,0266 there is a bar in the inner regions disturbing the velocity map in the FoV of MEGARA, with two kinematically distinct blobs in velocity and a depression in the velocity dispersion (see Figs.~\ref{Fig5:KinMaps_NGC0266} and~\ref{Fig:PV-PS_1}) already reported in the literature in previous works \citep{Cappellari2011,Foster2016}. NGC\,4438, NGC\,4750 and NGC\,5055 show a strong rotational pattern but $\sigma$ is not centrally peaked (see Figs.~\ref{Fig5:KinMaps_NGC4438}, ~\ref{Fig5:KinMaps_NGC4750},~\ref{Fig5:KinMaps_NGC5055}, and~\ref{Fig:PV-PS_2}). NGC\,3245 has one of the strongest rotational patterns and a centrally peaked velocity dispersion, consistent with an spiral object (see Figs.~\ref{Fig5:KinMaps_NGC3245} and~\ref{Fig:PV-PS_1}). 

   \noindent NGC\,0315 is the only galaxy classified as elliptical that indeed shows a non-rotating map with V/$\sigma$ of 0.7, thus dominated by random motions. However, as discussed in C22, NGC\,1052 was classified as an elliptical galaxy (see Table~\ref{Table:galaxieslisted}), but the stellar kinematics suggest that is rather a lenticular galaxy. This is also the case for NGC\,3226 and NGC\,4278, classified as ellipticals but with a strong rotational pattern and centrally peaked $\sigma$ (see Figs.~\ref{Fig5:KinMaps_NGC3226},~\ref{Fig5:KinMaps_NGC4278} and~\ref{Fig:PV-PS_1}). Their V/$\sigma$ are 1.2 and 1.0, pretty similar to that found for NGC\,1052 in the MUSE data (see C22). Moreover, the stellar and ionised gas components of NGC\,3226 and NGC\,4278 are misaligned by $\sim$30$^{\circ}$ (see Table~\ref{Table:resultsKin}). In the case of NGC\,3226, it is probably as a consequence of its interaction with the companion galaxy NGC\,3227. Thus NGC\,3226 and NGC\,4278 have very similar characteristics to NGC\,1052, which was interpreted in C22 as probably a lenticular galaxy given its strong rotational patterns. Nevertheless, some elliptical galaxies may have strong rotational patterns that would classify them as fast rotators \citep{Cappellari2016}. We note though that for NGC\,3226 and NGC\,4278 we are mapping exclusively the inner regions, so other options such as the presence of a small nuclear disc could also be causing the rotational patterns. This is probably not the case for NGC\,1052, where C22 mapped almost completely the whole galaxy, detecting rotation out to the FoV limits ($\sim$80\arcsec; i.e. 8.8\,kpc).
   
   \noindent The difference in amplitude between the radial velocities measured in the gas and the stars in NGC\,3226, NGC\,3245, and NGC\,5055, which range between $v_{\mathrm{gas}}-v_{\star}$$\sim$\,60\,km\,s$^{-1}$ in NGC\,5055 and  $\sim$\,100\,km\,s$^{-1}$ in NGC~3245, is likely due to the stars not being fully supported by rotation in these regions, which leads to an (asymmetric) drift between the two components. This is supported by the relatively large V$/\sigma$ ratios found, with line-of-sight velocity and velocity dispersion values measured for the stars in these objects of $v_{\star} \sim$\,70-100\,km\,s$^{-1}$ and $\sigma_{\star} \sim$\,100-200\,km\,s$^{-1}$, respectively (see Figs.~\ref{Fig5:KinMaps_NGC3226},~\ref{Fig5:KinMaps_NGC3245}, and~\ref{Fig5:KinMaps_NGC5055}). In fact, adopting a slope for the Stromberg linear equation of 5.4, similar to that of the Solar neighbourhood\footnote{Note that in these regions neither the value adopted might be valid nor the assumptions behind the Stromberg linear equation applicable to infer the drift values found \citep[see][]{Binney2008}.}, the expected drift would be a factor $\times$2-3 larger than the measured differences (see more on the calculations on Mario Chamorro-Cazorla's PhD thesis). 

\subsection{Feedback effects on the gas disc}
\label{SubSec:Discussion_highsigma}

     \noindent As mentioned in Sect.~\ref{SubSec_ResultOutflows}, we identified outflows as a secondary, blueshifted, broad component in the emission lines. We find what may be a correlation between the existence of an outflow probed by the secondary component, and the presence of an enhanced $\sigma$ region in the primary component. This region is found to be co-spatial with the secondary component, ascribed to an outflow for five objects (namely NGC\,1052, NGC\,3226, NGC\,3245, NGC\,4278, NGC\,4750), in most cases also co-spatial with a region of near flat rotation (NGC\,1052, NGC\,3226, NGC\,3245 and NGC\,4750). 
     Moreover, in the case of NGC\,1052 (see C22) this region was consistent with being ionised by a mixture of LINER-like and shocks photoionisation, thus it was interpreted as a cocoon most likely formed as a consequence of the launch of the outflow. For the other four objects we can only obtain the BPT diagrams for NGC\,4278 and NGC\,4750 (we have the LR-B data cubes to estimate the [O\,III]/H$\beta$ ratio), with similar ratios to that of NGC\,1052 (see Sect.~\ref{SubSec:ResultsGasLineRatios} and figures in Appendix~\ref{Appendix:BPTs}).  
     
     \noindent In NGC\,1052 (see C22) there is a radio jet oriented in the same direction of the ionised gas outflow (consistent with being triggered by the jet) and perpendicular to the high-$\sigma$ region. There is a similar situation for NGC\,4278, as there is a double-sided, parsec-scale jet detected with VLBA data \citep{Giroletti2005,Pellegrini2012}, perpendicular to the enhanced-$\sigma$ region of the primary component and also to the secondary component in MEGARA. The BPT diagrams for NGC\,4750 indicate LINER photoionisation for the secondary component and for the high-$\sigma$ region, with similar ratios to NGC\,1052 (see Sect.~\ref{SubSec:ResultsGasLineRatios} and figures in Appendix~\ref{Appendix:BPTs}). The two sides of the radio jet have different PAs and the emission was detected in previous works as a milliarcsecond or even compact radio source \citep[see e.g.][and references therein]{Schilizzi1983,Nagar2005}. \cite{Pellegrini2012} suggested that the jet may be confined inside the galaxy due to both the low-power of the AGN, that cannot launch the jet to larger distances, and the interaction with the ISM in the inner regions. The enhanced-$\sigma$ region reaches large distances in the galaxy out of the MEGARA FoV \citep[$>$\,50\arcsec, i.e. $>$\,2\,kpc;][]{Sarzi2006,Pellegrini2012}. 
     Given that this galaxy is almost face-on (inclination of 16$^{\circ}$; see Table~\ref{Table:galaxieslisted}), the difference with respect to the outflow direction in NGC\,1052 may be due to inclination effects. There are no reported detections of radio jets for NGC\,3226, NGC\,3245 and NGC\,4750. 

    \noindent Enhanced-$\sigma$ regions in the ionised gas related to a radio jet have already been reported in works as \cite{Venturi2021} with data of the MAGNUM survey, modelling the [O\,III] line for four Seyfert galaxies. In all the cases, these enhanced-$\sigma$ regions were perpendicular to low power radio jets ($<$10$^{44}$\,erg\,s$^{-1}$). \\ 

    \noindent The enhanced-$\sigma$ region of NGC\,3226 and NGC\,3245 are very similar (see Figs.~\ref{Fig5:KinMaps_NGC3226} and~\ref{Fig5:KinMaps_NGC3245}), both in size ($\sim 500$\,pc\footnote{We cannot rule out that the enhanced-$\sigma$ region is extended further out of the S/N limits of the [S\,II] lines in our MEGARA data.}), in width ($\sigma > 150$\,km\,s$^{-1}$; see Sect.~\ref{SubSec:ResultsGasKin}) and in velocity (almost flat-rotation). As already mentioned, they are both coincident with the secondary component, which is unresolved, blueshifted (average values for NGC\,3226 of $-91$\,km\,s$^{-1}$ and NGC\,3245 of $-214$\,km\,s$^{-1}$; see Table~\ref{Table:resultsKin}) and broader than the primary component of the emission lines in the same spatial region. The log([S\,II]/H$\alpha$), log([N\,II]/H$\alpha$) and log([O\,I]/H$\alpha$) ratios for both NGC\,3226 and NGC\,3245 are consistent with AGN ionisation in the primary and secondary components. Thus the $\sigma$ regions are interpreted as turbulence produced by the expansion of the outflow, in a similar case to NGC\,1052, except that there is no radio jet driving the gas outwards. 
    
    \noindent The high-$\sigma$ region in NGC\,4750 is distributed in a spiral-like shape, bordering southward and northward the regions of lowest and highest velocities, respectively, and located at the base of the secondary component, interpreted as the outflow. This morphology of the primary component may have been produced by strong dust lanes from the spiral arms \citep[see e.g.][]{Carollo2002}, and the enhanced-$\sigma$ region most likely is a perturbation produced by the outflow, as for NGC\,1052. Considering the line ratios, this region is consistent with both LINER photoionisation or shocks, thus a mixture of the AGN ionisation and the collision of the ISM with the outflowing gas may be triggering this enhanced-$\sigma$ region. 

\subsection{Characterisation of the ionised outflows}
\label{SubSec:Discussion_outflows}

    \begin{figure}
        \centering
        \includegraphics[width=\columnwidth]{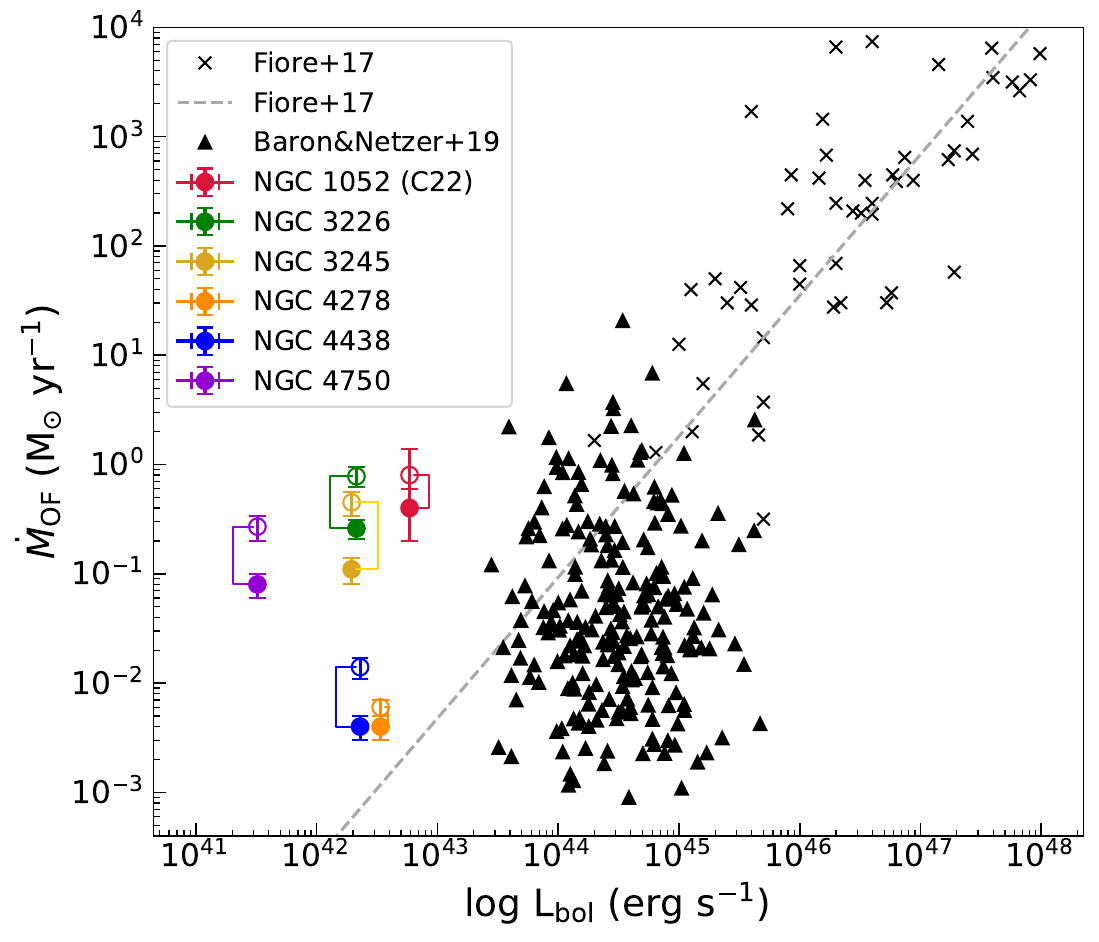}
        \caption{Mass outflow rate versus bolometric luminosity for ionised gas outflows from different objects. Black crosses represent data  obtained from \cite{Fiore2017}, and the grey, dashed line is the linear fit they report in their paper; black triangles are the data from \cite{Baron2019}; the filled circles with their respective errorbars represent the values obtained for the outflows in this work, except NGC\,1052 (red circle), estimated in C22. The empty circles are the same objects, but assuming an electron density of 200\,cm$^{-3}$, as \cite{Fiore2017} in their analysis.}
        \label{Fig5:OutflowMassRate_Fiore17}
     \end{figure}

    \noindent We can estimate the main properties of the detected outflows following the expressions in Sect.~\ref{SubSec:datared_CalcOutfParam}. We represent in Fig.~\ref{Fig5:OutflowMassRate_Fiore17} a comparison between the mass outflow rates of our sources compact to the work by \cite{Fiore2017} and \cite{Baron2019}. \cite{Fiore2017} put together data from many different works to estimate the properties of outflows depending on the gas phase, deriving various scaling relations between the outflow and the AGN properties, such as the mass outflow rate with the bolometric luminosity. This work misses the low luminosity end of the AGN family, and thus we have extrapolated the linear relation that they report down to the luminosity of our targets ($L_{\rm bol}$: $10^{41}$ - $10^{43.5}$\,erg\,s$^{-1}$, see Table~\ref{Table:galaxieslisted}). We have also added the targets from \cite{Baron2019}, which are objects at lower luminosities than \cite{Fiore2017}, but higher than our targets ($L_{\rm bol}$: $10^{43.5}$ - $10^{45.5}$\,erg\,s$^{-1}$). \cite{Baron2019} measured the properties of ionised gas outflows with SDSS DR7 spectra of 234 type-II AGNs, mainly Seyferts, in a range of z\,$=$\,0.005-0.15. They used the ionisation parameter to derive the densities, a method independent on the [S\,II] lines \citep[see e.g.][]{Davies2020,Revalski2022}. They found electron densities of $\sim 10^{4.5}$\,cm$^{-3}$, two orders of magnitude larger than those in \cite{Fiore2017}, that instead assumes a constant value of 200\,cm$^{-3}$ for all targets. We cannot compare the densities that we obtained with \cite{Baron2019}, as the ionisation parameter depends on the [O\,III]/H$\beta$ ratio, that we cannot estimate for the majority of our targets.
    
    \noindent We plotted our values in Fig~\ref{Fig5:OutflowMassRate_Fiore17}, and we can see that they tend to be located above the linear relation reported by \cite{Fiore2017} for ionised gas outflows. We note that if we consider the electron density to be 200\,cm$^{-3}$ as in \cite{Fiore2017}, instead of our derived values (see Table~\ref{Table:resultsOutflows}), the $\dot{M}_{\rm OF,ion}$ increases, hence these points located even further away from the reported linear relation in that work (see Fig~\ref{Fig5:OutflowMassRate_Fiore17}). \cite{Baron2019} also found that their data points do not follow the linear relation by \cite{Fiore2017}, even accounting for the uncertainties on both the mass outflow rate (1 dex) and bolometric luminosities (0.3-0.4 dex). They state that in their case, different outflow geometries and projection effects may play a role, given that they analyse type-II AGNs. If we were to assume densities similar to those reported in \cite{Baron2019} (i.e. $n_{e^{-}} \sim 10^{4.5}$\,cm$^{-3}$), the mass outflow rates (and consequently the energy) would be two orders of magnitude lower than what we obtain, thus closer to the linear relation although still above it (mean $\dot{M}_{\rm OF,ion}$ would go from 0.17 to 0.003).
    We note that our $\dot{M}_{\rm OF,ion}$ values are not corrected for projection effects. If we consider the outflows to be perpendicular to the galactic plane, then the deprojected mass outflow rates would depend exclusively on the inclination of the host. Given the inclinations (see Table~\ref{Table:galaxieslisted}), we estimate that the reported values in Table~\ref{Table:resultsOutflows} would diminish a 45\% up to 75\% in the galaxies inclined more than 45$^{\circ}$, whereas it would increase up to 63\% (NGC\,4750) and 250\% (for NGC\,4278) in those inclined less than 45$^{\circ}$. Despite this, the order of magnitude of the $\dot{M}_{\rm OF,ion}$ is maintained in all cases except for NGC\,3245 (where it diminish from $0.11\pm0.03$ to $0.06\pm0.02$), NGC\,4750 (from 0.08$\pm$0.02 to 0.13$\pm$0.10) and NGC\,4278 (from 0.004$\pm$0.001 to 0.014$\pm$0.004), and they would still be out of the linear relation in Fig~\ref{Fig5:OutflowMassRate_Fiore17}. Given that we cannot fully account for the projection effects in the outflows due to the uncertainty of their orientations, and that the galaxy inclination does not heavily impact the results, we prefer to use the projected values for the discussion. \\
    
    \noindent Although we only have a few points, our results indicate that it may be a flattening in the curve towards lower luminosities, such that the detected outflows are not as weak as expected. We note that these results have to be considered carefully as every work assumes different conditions, specially with respect to the electron density; thus a unified treatment, also accounting for projection effects, would be needed to extract robust conclusions. \\

    \noindent The mass outflow rates and kinetic powers of our sources are equivalent to those found by \cite{Kukreti2023} with optical and radio data for 129 radio-AGNs in a slightly larger redshift range (z$<$0.25; mean 0.12) and luminosities than our sources (14\% detection of outflows; $\dot{M}_{\rm OF,ion}$\,=\,0.09 - 0.41 $M_{\sun}$\,yr$^{-1}$ and $E_{\rm OF, ion}$\,=\,0.1-1.8$\times$\,10$^{41}$\,erg\,s$^{-1}$). These reported mass outflow rates and those in our work are small to efficiently quench the galaxies. In fact, the kinetic coupling efficiency ($\dot{E}_{\rm OF,ion}$/L$_{\rm bol}$; see Table~\ref{Table:resultsOutflows}) for our sources is always $<$\,0.01. This means that there is little energy transferred from the outflow to the medium, and so these outflows alone are not expected to heavily impact and/or quench their host galaxies \citep{Harrison2018}. However, if the turbulence of the medium is high as a consequence of the outflow and/or due to an interaction with the jet (specially if it is young), then the gas could still be unable to form stars \citep{Kukreti2023}. Further exploration of the quenching phenomena is out of the scope of this paper. 

\subsection{The bubble in NGC 4438 vs the prototypical LINER NGC 1052}
\label{SubSec:Discussion_NGC4438}

    \noindent NGC\,4438 together with NGC\,1052 (see C22) are the only galaxies for which we detected three different kinematical components in all the analysed emission lines, but spatially resolved only for NGC\,4438. Both are lenticular galaxies with different kinematical properties (see Sect.~\ref{SubSec:Discussion_morphologies}). The primary component of NGC\,4438 is likely tracing the gas disc. The velocity map is rotating regularly along the same position angle and with slightly larger velocities (difference in the semi-amplitude velocity of $\sim$40\,km\,s$^{-1}$) to the stellar component (see Table~\ref{Table:resultsKin}). This is in contrast to the case of NGC\,1052, for which the primary component was rotating along the minor axis of the stars, and it was not consistent with a regular gas disc, showing spiral-like flux maps with a non-regular morphology. The velocity dispersion of NGC\,4438 is relatively homogeneous $\sigma$, enhanced in the borders of the bubble (see Sect.~\ref{SubSec:ResultsGasKin}, whereas for NGC\,1052 we detected an enhanced-$\sigma$ region within the MEGARA FoV as already mentioned (see C22 for more details). 
    
    \noindent The secondary and tertiary components of NGC\,4438 are most likely related to the H$\alpha$ bubble detected in this galaxy, as hinted by its morphology and kinematics. \cite{Kenney2002} studied HST H$\alpha$ narrow band images of this LINER, as so did afterwards \cite{Masegosa2011}, characterising a prominent ionised gas bubble emerging north-west from the nucleus. Its projected length was estimated to be 1\,kpc and, despite it was classified as a bipolar outflow, one bubble is more visible north-west of the nucleus, probably due to intrinsic differences in the ISM in the two sides of the galaxy \citep{Kenney2002}. 

    \noindent The H$\alpha$ emission derived from the MEGARA data clearly maps the bubble emerging from the nucleus (see Sect.~\ref{SubSec:ResultsGasKin} and Fig.~\ref{Fig5:Compare}). The velocity map of the secondary component indicates that we are looking at the receding (i.e. redshifted) part of the bubble, whereas the approaching (i.e. blueshifted) part is on the south-east direction. The velocity dispersion in the borders of the bubble reaches larger values than for the inner parts (average $\sigma^{\rm MEGARA}_{\rm gas,S} \sim$340 vs $\sim$220\,km\,s$^{-1}$; see Fig.~\ref{Fig5:KinMaps_NGC4438}). These kinematical properties and the overall morphology of this component are consistent with an outflow. The previous spectroscopic analysis of this object with long-slit spectroscopy did not find strong evidences of an outflow \citep{Cazzoli2018}. This can be understood since the slit was oriented perpendicular to the outflow, so it could not be traced (PA$_{\rm slit}$\,=\,39$^{\circ}$ vs PA$_{\rm Bubble} \sim 141^{\circ}$). 
    
    \noindent The secondary component detected in NGC\,1052 had more extreme kinematics than the primary component, being broader and aligned with the radio jet. It was also interpreted as an outflow, in this case consistent with being triggered by the radio jet (see C22). 

    \noindent Given that the tertiary component in the MEGARA data is oriented in the direction of the outflow (PA$\sim$127$^\circ$), its origin may be connected to it. In fact, it is coincident with the north-western side of the bubble, traced with high-$\sigma$ by the secondary component. The tertiary component is mainly redshifted (v$^{\rm MEGARA}_{\rm gas,T} >$50\,km\,s$^{-1}$), broader than the primary but narrower than the secondary (average $\sigma^{\rm MEGARA}_{\rm gas,T}$ of 135\,km\,s$^{-1}$; see Table~\ref{Table:resultsKin}). 
    This galaxy has a double-lobed radio continuum source oriented in a similar direction to that of the bubble (PA\,$\sim$\,125$^{\circ}$; \citealt{Hota2007}). The extension of this double-lobed, asymmetric radio emission was estimated to be $\sim$250\,pc west and $\sim$730\,pc east from the nucleus \citep{Hota2007}. This is consistent with the estimated size of the outflow in the west direction ($\sim$300\,pc), as extended as the tertiary component in MEGARA data. In fact \cite{Kenney2002} confirmed the co-spatiality between this radio emission and the outer ends of the H$\alpha$ bubble, initially associating the borders of the ionised gas bubble to shocks, which \cite{Hota2007} associated to the interaction of the radio emission with the ISM. For NGC\,1052 we also reported the existence of perturbed emission in the ionised gas due to the ISM-jet interaction, through the turbulence in the enhanced-$\sigma$ region, as discussed in Sect.~\ref{SubSec:Discussion_highsigma}. However, for NGC\,1052 the tertiary component was unresolved, and thus no further interpretations were given. 
    For NGC\,4438, considering the kinematics, line ratios and spatial extension of the tertiary component derived with MEGARA, it may be tracing the optical counter-part of the radio emission. \\

    \noindent Summarising, the two galaxies are prototypical LINERs inside lenticular galaxies, with a detected outflow in ionised gas consistent with being powered by a radio jet. Both show a complex kinematical structure that needed to be modelled with multiple components, highlighting the importance of spatially resolved studies for obtaining a complete characterisation of the physics modelling the ISM in these objects. 
    Although our sample is small, the results support the idea that the jet-mode might be play a very important role in driving outflows for low luminosity AGNs.

\section{Summary and conclusions}
\label{Sect:Future}

  \noindent In this work, we analyse the optical IFS data of nine LINERs observed with MEGARA/GTC with its low and high resolution modes (R$\sim$6000 and R$\sim$20000) in different wavelength bands (LR-R, LR-V, LR-B and HR-R). Prior to the study of the ionised gas properties, a subtraction of the stellar component of the galaxies is required. For that we applied a \textsc{pPXF} modelling \citep{Cappellari2003,Cappellari2017} to the combined data cubes of LR-R and LR-V when both were available (seven LINERs, see Sect.~\ref{Sect:Datareduction}). After the stellar subtraction we obtained the spectra of the ISM and applied a multicomponent fitting of the optical ionised emission lines available in the wavelength range of each VPH. We used the [S\,II] lines as reference for modelling the rest of the emission lines as in previous works \citep{Cazzoli2018,HM2020}, and modelled [O\,III] and H$\beta$ independently given that they have different kinematics, as was found for NGC\,1052 (C22). These are the main conclusions of this work: 

  \begin{itemize}
      \item \textit{Stellar kinematics}: The stars are generally rotating following a regular pattern, with a centrally peaked or homogeneous $\sigma$ map, except for NGC\,0315, that shows no rotation. Morphologically, four galaxies were ellipticals, two spirals and two lenticulars (see Table~\ref{Table:galaxieslisted}), but our results indicate that all the ellipticals except for NGC\,0315 show a rotational pattern.
      
      \item \textit{Ionised gas kinematics}: In contrast to the stars, the ionised gas has a complex behaviour, with at least two kinematical components needed to model the emission lines in the majority of the LINERs (7), three components for two cases, and an additional very broad component in H$\alpha$ associated to the BLR in four of them. The two objects modelled with a single component in the emission lines are NGC\,0266 and NGC\,5055. For NGC\,0266, the observed disturbed kinematics can be explained by the large-scale bar, whereas for NGC\,5055 the regular velocity and $\sigma$ maps are tracing the ionised gas disc, with distinguishable star forming regions. 

      \item \textit{Enhanced-$\sigma$ regions:} For five objects, namely NGC\,1052 (see C22), NGC\,3226, NGC\,3245, NGC\,4278 and NGC\,4750, the primary component has a rotational pattern with a peculiar $\sigma$. They have an enhanced-$\sigma$ region perpendicular (parallel) to the stellar major axis in NGC\,3226, NGC\,3245 and NGC\,4278 (NGC\,1052 and NGC\,4750) with a \lq Butterfly\rq\,-like shape. This region is probably triggered by a mixture of turbulence caused by the interaction of the secondary component and/or the radio emission. The secondary component is co-spatial with the enhanced-$\sigma$ region but with differing kinematics, typically blueshifted and broad ($\langle \sigma_{\rm S}\rangle \sim$\,200\,km\,s$^{-1}$), explained here as an outflow, compact and unresolved for NGC\,3226 and NGC\,3245. 

      \item \textit{Outflow detections:} All the galaxies in our sample were candidates of hosting an outflow from our previous analyses using both imaging and long-slit spectroscopic data \citep[see][HM22, C22]{Cazzoli2018,HM2020}. The MEGARA data has higher spectral resolution than our previous analysis of the optical spectra, and traces spatially the central regions of our targets. This new analysis allowed us to confirm the presence of outflows for the majority of the sample (six objects, namely NGC\,1052, NGC\,3226, NGC\,3245, NGC\,4278, NGC\,4438 and NGC\,4750). For NGC\,4438 and NGC\,4750, the outflow is detected with a bubble-like shape emerging north-west and west from the centre, respectively, and for NGC\,4278 it has a shell-like structure, for all cases traced by the secondary component. For the remaining objects, the outflow is resolved only for NGC\,1052 (see C22). 

      \item \textit{Energetic of the outflows}: The detected outflows in H$\alpha$ show mass outflow rates ranging from 0.004 to 0.4 $M_{\sun}$\,yr$^{-1}$ (average 0.17$\pm$0.06\,$M_{\sun}$\,yr$^{-1}$) and energy rates from $\sim$\,10$^{38}$ to $\sim$\,10$^{40}$\,erg\,s$^{-1}$ (average 2\,$\times$\,10$^{40}$\,erg\,s$^{-1}$). On average, the outflows are extended to a distance of $\sim$\,400\,pc at an (absolute) velocity of $\sim$\,340\,km\,s$^{-1}$, with an electronic density of 600\,cm$^{-3}$. 

      \item \textit{Tertiary kinematical component}: For NGC\,1052 and NGC\,4438 a tertiary component is needed to model the emission lines. For NGC\,1052 this component is unresolved, whereas for NGC\,4438 it is extended, and co-spatial with both the outflow and a double-sided radio continuum emission. Thus it is most likely that these emissions are related, and we suggest that it could be the optical counterpart of the radio emission.

      \item \textit{Line ratios and BPT diagrams}: The average line ratios for the primary components of the galaxies indicate star-forming or composite photoionisation for NGC\,3245, NGC\,4438 and NGC\,5055, and AGN photoionisation for NGC\,0266, NGC\,0315, NGC\,3226, NGC\,4278 and NGC\,4750. The nuclear regions of all the galaxies are consistent with AGN photionisation. We were able to distinguish several star forming blobs in galaxies such as NGC\,4750 and NGC\,5055. The secondary and tertiary components of the emission lines are always consistent with AGN-photoionisation. We cannot exclude a possible contribution from shocks, specially regarding the [O\,I] ratios, that tend to be more affected by these processes. 

      \item \textit{Broad Line Region component}: We detected a BLR component exclusively in H$\alpha$ for five galaxies (namely NGC\,0315, NGC\,1052, NGC\,3226, NGC\,4438 and NGC\,4750), although for NGC\,4438 it is not well constrained. We did not detect it in H$\beta$, which confirms the nature of the galaxies as either type 1.9 or type 2 LINERs.
      
  \end{itemize}
    
  \noindent Our results have allowed us to characterise and unify all the results coming from various techniques, finding the existence of outflows for the majority of the LINERs. This reinforces the results from HM22 about outflows being common also in LINERs, and that although they can be detected with less time consuming techniques than IFS, only spatially resolved spectroscopy allows us to fully understand the complex kinematics of the ionised gas in these galaxies.

\begin{acknowledgements}
We thank the referee for his/her comments that have helped to improve the manuscript. We thank B. Ag{\'i}s-Gonz{\'a}lez and M. Puig for useful discussions. LHM, SC, IM and JM acknowledge financial support from the Spanish Ministerio de Ciencia, Innovaci{\'o}n y Universidades (MCIU) under the grant PID2019-106027GB-C41 and from the State Agency for Research of the Spanish MCIU through the \lq Centre of Excellence Severo Ochoa\rq\ awards to the Instituto de Astrof{\'i}sica de Andaluc{\'i}a SEV-2017-0709, and CEX2021-001131-S funded by MCIN/AEI/10.13039/501100011033. LHM acknowledges financial support from the Spanish MCIU under the grants BES-2017-082471 and PID2021-124665NB-I00. MCC, AGdP, ACM, JG, SP and NC acknowledge financial support from the Spanish MCIU under the grants RTI2018-096188-B-I00 and PID2021-123417OB-I00. MCC acknowledges the support from the "Tecnolog{\'i}as avanzadas para la exploraci{\'o}n de universo y sus componentes" (PR47/21 TAU) project funded by Comunidad de Madrid, by the Recovery, Transformation and Resilience Plan from the Spanish State, and by NextGenerationEU from the European Union through the Recovery and Resilience Facility.  \\
This work is based on data obtained with MEGARA instrument, funded by European Regional Development Funds (ERDF), through Programa Operativo Canarias FEDER 2014-2020, and based on observations made with the Gran Telescopio Canarias (GTC), installed in the Spanish Observatorio del Roque de los Muchachos of the Instituto de Astrofísica de Canarias, in the island of La Palma. The data presented here were obtained in part with ALFOSC, which is provided by the Instituto de Astrof{\'i}sica de Andaluc{\'i}a (IAA) under a joint agreement with the University of Copenhagen and NOTSA. 
This research has made use of the NASA/IPAC Extragalactic Database (NED), which is operated by the Jet Propulsion Laboratory, California Institute of Technology, under contract with the National Aeronautics and Space Administration.
We acknowledge the usage of the HyperLeda database (http://leda.univ-lyon1.fr).\\
This work has made extensive use of IRAF (v2.16) and Python (v3.8.10), particularly with \textsc{astropy} \citep[v4.2, \nolinkurl{http://www.astropy.org};][]{astropy:2013, astropy:2018}, \textsc{lmfit}, \textsc{matplotlib} \citep[v3.4.1;][]{Hunter:2007}, \textsc{photutils} \citep[v1.1.0;][]{LarryBradley_2020} and \textsc{numpy} \citep[v1.19.4;][]{Harris2020}.\\
\end{acknowledgements}

%
   \bibliographystyle{aa} 
   \bibliography{bibliography.bib} 

\begin{thebibliography}{86}
\expandafter\ifx\csname natexlab\endcsname\relax\def\natexlab#1{#1}\fi

\bibitem[{{Agostino} {et~al.}(2023){Agostino}, {Salim}, {Ellison}, {Bickley},
  \& {Faber}}]{Agostino2023}
{Agostino}, C.~J., {Salim}, S., {Ellison}, S.~L., {Bickley}, R.~W., \& {Faber},
  S.~M. 2023, \apj, 943, 174

\bibitem[{{Astropy Collaboration} {et~al.}(2018){Astropy Collaboration},
  {Price-Whelan}, {Sip{\H{o}}cz}, {G{\"u}nther}, {Lim}, {Crawford}, {Conseil},
  {Shupe}, {Craig}, {Dencheva}, {Ginsburg}, {VanderPlas}, {Bradley},
  {P{\'e}rez-Su{\'a}rez}, {de Val-Borro}, {Paper Contributors}, {Aldcroft},
  {Cruz}, {Robitaille}, {Tollerud}, {Coordination Committee}, {Ardelean},
  {Babej}, {Bach}, {Bachetti}, {Bakanov}, {Bamford}, {Barentsen}, {Barmby},
  {Baumbach}, {Berry}, {Biscani}, {Boquien}, {Bostroem}, {Bouma}, {Brammer},
  {Bray}, {Breytenbach}, {Buddelmeijer}, {Burke}, {Calderone}, {Cano
  Rodr{\'\i}guez}, {Cara}, {Cardoso}, {Cheedella}, {Copin}, {Corrales},
  {Crichton}, {D{\textquoteright}Avella}, {Deil}, {Depagne}, {Dietrich},
  {Donath}, {Droettboom}, {Earl}, {Erben}, {Fabbro}, {Ferreira}, {Finethy},
  {Fox}, {Garrison}, {Gibbons}, {Goldstein}, {Gommers}, {Greco}, {Greenfield},
  {Groener}, {Grollier}, {Hagen}, {Hirst}, {Homeier}, {Horton}, {Hosseinzadeh},
  {Hu}, {Hunkeler}, {Ivezi{\'c}}, {Jain}, {Jenness}, {Kanarek}, {Kendrew},
  {Kern}, {Kerzendorf}, {Khvalko}, {King}, {Kirkby}, {Kulkarni}, {Kumar},
  {Lee}, {Lenz}, {Littlefair}, {Ma}, {Macleod}, {Mastropietro}, {McCully},
  {Montagnac}, {Morris}, {Mueller}, {Mumford}, {Muna}, {Murphy}, {Nelson},
  {Nguyen}, {Ninan}, {N{\"o}the}, {Ogaz}, {Oh}, {Parejko}, {Parley}, {Pascual},
  {Patil}, {Patil}, {Plunkett}, {Prochaska}, {Rastogi}, {Reddy Janga},
  {Sabater}, {Sakurikar}, {Seifert}, {Sherbert}, {Sherwood-Taylor}, {Shih},
  {Sick}, {Silbiger}, {Singanamalla}, {Singer}, {Sladen}, {Sooley},
  {Sornarajah}, {Streicher}, {Teuben}, {Thomas}, {Tremblay}, {Turner},
  {Terr{\'o}n}, {van Kerkwijk}, {de la Vega}, {Watkins}, {Weaver}, {Whitmore},
  {Woillez}, {Zabalza}, \& {Contributors}}]{astropy:2018}
{Astropy Collaboration}, {Price-Whelan}, A.~M., {Sip{\H{o}}cz}, B.~M., {et~al.}
  2018, \aj, 156, 123

\bibitem[{{Astropy Collaboration} {et~al.}(2013){Astropy Collaboration},
  {Robitaille}, {Tollerud}, {Greenfield}, {Droettboom}, {Bray}, {Aldcroft},
  {Davis}, {Ginsburg}, {Price-Whelan}, {Kerzendorf}, {Conley}, {Crighton},
  {Barbary}, {Muna}, {Ferguson}, {Grollier}, {Parikh}, {Nair}, {Unther},
  {Deil}, {Woillez}, {Conseil}, {Kramer}, {Turner}, {Singer}, {Fox}, {Weaver},
  {Zabalza}, {Edwards}, {Azalee Bostroem}, {Burke}, {Casey}, {Crawford},
  {Dencheva}, {Ely}, {Jenness}, {Labrie}, {Lim}, {Pierfederici}, {Pontzen},
  {Ptak}, {Refsdal}, {Servillat}, \& {Streicher}}]{astropy:2013}
{Astropy Collaboration}, {Robitaille}, T.~P., {Tollerud}, E.~J., {et~al.} 2013,
  \aap, 558, A33

\bibitem[{{Baldi} {et~al.}(2018){Baldi}, {Williams}, {McHardy}, {Beswick},
  {Argo}, {Dullo}, {Knapen}, {Brinks}, {Muxlow}, {Aalto}, {Alberdi}, {Bendo},
  {Corbel}, {Evans}, {Fenech}, {Green}, {Kl{\"o}ckner}, {K{\"o}rding}, {Kharb},
  {Maccarone}, {Mart{\'\i}-Vidal}, {Mundell}, {Panessa}, {Peck},
  {P{\'e}rez-Torres}, {Saikia}, {Saikia}, {Shankar}, {Spencer}, {Stevens},
  {Uttley}, \& {Westcott}}]{Baldi2018}
{Baldi}, R.~D., {Williams}, D.~R.~A., {McHardy}, I.~M., {et~al.} 2018, \mnras,
  476, 3478

\bibitem[{{Baldwin} {et~al.}(1981){Baldwin}, {Phillips}, \&
  {Terlevich}}]{Baldwin1981}
{Baldwin}, J.~A., {Phillips}, M.~M., \& {Terlevich}, R. 1981, \pasp, 93, 5

\bibitem[{{Baron} \& {Netzer}(2019)}]{Baron2019}
{Baron}, D. \& {Netzer}, H. 2019, \mnras, 486, 4290

\bibitem[{{Binney} \& {Tremaine}(2008)}]{Binney2008}
{Binney}, J. \& {Tremaine}, S. 2008, {Galactic Dynamics: Second Edition}

\bibitem[{Bradley {et~al.}(2020)Bradley, Sip{\H o}cz, Robitaille, Tollerud,
  Vin{\'{\i}}cius, Deil, Barbary, Wilson, Busko, G{\"u}nther, Cara, Conseil,
  Bostroem, Droettboom, Bray, Bratholm, Lim, Barentsen, Craig, Pascual, Perren,
  Greco, Donath, de~Val-Borro, Kerzendorf, Bach, Weaver, D'Eugenio, Souchereau,
  \& Ferreira}]{LarryBradley_2020}
Bradley, L., Sip{\H o}cz, B., Robitaille, T., {et~al.} 2020, astropy/photutils:
  1.0.0

\bibitem[{{Cano-D{\'\i}az} {et~al.}(2012){Cano-D{\'\i}az}, {Maiolino},
  {Marconi}, {Netzer}, {Shemmer}, \& {Cresci}}]{CanoDiaz2012}
{Cano-D{\'\i}az}, M., {Maiolino}, R., {Marconi}, A., {et~al.} 2012, \aap, 537,
  L8

\bibitem[{{Cappellari}(2016)}]{Cappellari2016}
{Cappellari}, M. 2016, \araa, 54, 597

\bibitem[{{Cappellari}(2017)}]{Cappellari2017}
{Cappellari}, M. 2017, \mnras, 466, 798

\bibitem[{{Cappellari} \& {Copin}(2003)}]{Cappellari2003}
{Cappellari}, M. \& {Copin}, Y. 2003, \mnras, 342, 345

\bibitem[{{Cappellari} {et~al.}(2011){Cappellari}, {Emsellem}, {Krajnovi{\'c}},
  {McDermid}, {Scott}, {Verdoes Kleijn}, {Young}, {Alatalo}, {Bacon}, {Blitz},
  {Bois}, {Bournaud}, {Bureau}, {Davies}, {Davis}, {de Zeeuw}, {Duc},
  {Khochfar}, {Kuntschner}, {Lablanche}, {Morganti}, {Naab}, {Oosterloo},
  {Sarzi}, {Serra}, \& {Weijmans}}]{Cappellari2011}
{Cappellari}, M., {Emsellem}, E., {Krajnovi{\'c}}, D., {et~al.} 2011, \mnras,
  413, 813

\bibitem[{{Carniani} {et~al.}(2015){Carniani}, {Marconi}, {Maiolino},
  {Balmaverde}, {Brusa}, {Cano-D{\'\i}az}, {Cicone}, {Comastri}, {Cresci},
  {Fiore}, {Feruglio}, {La Franca}, {Mainieri}, {Mannucci}, {Nagao}, {Netzer},
  {Piconcelli}, {Risaliti}, {Schneider}, \& {Shemmer}}]{Carniani2015}
{Carniani}, S., {Marconi}, A., {Maiolino}, R., {et~al.} 2015, \aap, 580, A102

\bibitem[{{Carollo} {et~al.}(2002){Carollo}, {Stiavelli}, {Seigar}, {de Zeeuw},
  \& {Dejonghe}}]{Carollo2002}
{Carollo}, C.~M., {Stiavelli}, M., {Seigar}, M., {de Zeeuw}, P.~T., \&
  {Dejonghe}, H. 2002, \aj, 123, 159

\bibitem[{{Carrasco} {et~al.}(2018){Carrasco}, {Gil de Paz}, {Gallego},
  {Iglesias-P{\'a}ramo}, {Cedazo}, {Garc{\'\i}a Vargas}, {Arrillaga},
  {Avil{\'e}s}, {Bouquin}, {Carbajo}, {Cardiel}, {Carrera}, {Castillo Morales},
  {Castillo-Dom{\'\i}nguez}, {Esteban San Rom{\'a}n}, {Ferrusca},
  {G{\'o}mez-{\'A}lvarez}, {Izazaga-P{\'e}rez}, {Lefort}, {L{\'o}pez Orozco},
  {Maldonado}, {Mart{\'\i}nez Delgado}, {Morales Dur{\'a}n}, {M{\'u}jica},
  {Ortiz}, {P{\'a}ez}, {Pascual}, {P{\'e}rez-Calpena}, {Picazo},
  {S{\'a}nchez-Penim}, {S{\'a}nchez-Blanco}, {Tulloch}, {Vel{\'a}zquez},
  {V{\'\i}lchez}, {Zamorano}, {Aguerri}, {Barrado}, {Bertone}, {Cava},
  {Catal{\'a}n-Torrecilla}, {Cenarro}, {Ch{\'a}vez}, {Dullo}, {Eliche},
  {Garc{\'\i}a}, {Garc{\'\i}a-Rojas}, {Guichard}, {Gonz{\'a}lez-Delgado},
  {Guzm{\'a}n}, {Herrero}, {Hu{\'e}lamo}, {Hughes}, {Jim{\'e}nez-Vicente},
  {Kehrig}, {Marino}, {M{\'a}rquez}, {Masegosa}, {Mayya}, {M{\'e}ndez-Abreu},
  {Moll{\'a}}, {Mu{\~n}oz-Tu{\~n}{\'o}n}, {Peimbert}, {P{\'e}rez-Gonz{\'a}lez},
  {P{\'e}rez-Montero}, {Roca-F{\`a}brega}, {Rodr{\'\i}guez},
  {Rodr{\'\i}guez-Espinosa}, {Rodr{\'\i}guez-Merino},
  {Rodr{\'\i}guez-Mu{\~n}oz}, {Rosa-Gonz{\'a}lez}, {S{\'a}nchez-Almeida},
  {S{\'a}nchez Contreras}, {S{\'a}nchez-Bl{\'a}zquez}, {S{\'a}nchez},
  {Sarajedini}, {Silich}, {Sim{\'o}n-D{\'\i}az}, {Tenorio-Tagle}, {Terlevich},
  {Terlevich}, {Torres-Peimbert}, {Trujillo}, {Tsamis}, \&
  {Vega}}]{Carrasco2018}
{Carrasco}, E., {Gil de Paz}, A., {Gallego}, J., {et~al.} 2018, in Society of
  Photo-Optical Instrumentation Engineers (SPIE) Conference Series, Vol. 10702,
  Ground-based and Airborne Instrumentation for Astronomy VII, ed. C.~J.
  {Evans}, L.~{Simard}, \& H.~{Takami}, 1070216

\bibitem[{{Cazzoli} {et~al.}(2020){Cazzoli}, {Gil de Paz}, {M{\'a}rquez},
  {Masegosa}, {Iglesias}, {Gallego}, {Carrasco}, {Cedazo},
  {Garc{\'\i}a-Vargas}, {Castillo-Morales}, {Pascual}, {Cardiel},
  {P{\'e}rez-Calpena}, {G{\'o}mez-Alvarez}, {Mart{\'\i}nez-Delgado}, \&
  {Hermosa-Mu{\~n}oz}}]{Cazzoli2020}
{Cazzoli}, S., {Gil de Paz}, A., {M{\'a}rquez}, I., {et~al.} 2020, \mnras, 493,
  3656

\bibitem[{{Cazzoli} {et~al.}(2022){Cazzoli}, {Hermosa Mu{\~n}oz},
  {M{\'a}rquez}, {Masegosa}, {Castillo-Morales}, {Gil de Paz},
  {Hern{\'a}ndez-Garc{\'\i}a}, {La Franca}, \& {Ramos Almeida}}]{Cazzoli2022}
{Cazzoli}, S., {Hermosa Mu{\~n}oz}, L., {M{\'a}rquez}, I., {et~al.} 2022, \aap,
  664, A135

\bibitem[{{Cazzoli} {et~al.}(2018){Cazzoli}, {M{\'a}rquez}, {Masegosa}, {del
  Olmo}, {Povi{\'c}}, {Gonz{\'a}lez-Mart{\'{\i}}n}, {Balmaverde},
  {Hern{\'a}ndez-Garc{\'{\i}}a}, \& {Garc{\'{\i}}a-Burillo}}]{Cazzoli2018}
{Cazzoli}, S., {M{\'a}rquez}, I., {Masegosa}, J., {et~al.} 2018, \mnras, 480,
  1106

\bibitem[{{Chamorro-Cazorla} {et~al.}(2023){Chamorro-Cazorla}, {Gil de Paz},
  {Castillo-Morales}, {Gallego}, {Carrasco}, {Iglesias-P{\'a}ramo},
  {Garc{\'\i}a-Vargas}, {Pascual}, {Cardiel}, {Catal{\'a}n-Torrecilla},
  {Zamorano}, {S{\'a}nchez-Bl{\'a}zquez}, {P{\'e}rez-Calpena},
  {G{\'o}mez-{\'A}lvarez}, \& {Jim{\'e}nez-Vicente}}]{Chamorro-Cazorla2022}
{Chamorro-Cazorla}, M., {Gil de Paz}, A., {Castillo-Morales}, {\'A}., {et~al.}
  2023, \aap, 670, A117

\bibitem[{{Cresci} {et~al.}(2015){Cresci}, {Marconi}, {Zibetti}, {Risaliti},
  {Carniani}, {Mannucci}, {Gallazzi}, {Maiolino}, {Balmaverde}, {Brusa},
  {Capetti}, {Cicone}, {Feruglio}, {Bland-Hawthorn}, {Nagao}, {Oliva},
  {Salvato}, {Sani}, {Tozzi}, {Urrutia}, \& {Venturi}}]{Cresci2015}
{Cresci}, G., {Marconi}, A., {Zibetti}, S., {et~al.} 2015, \aap, 582, A63

\bibitem[{{Davies} {et~al.}(2020){Davies}, {Baron}, {Shimizu}, {Netzer},
  {Burtscher}, {de Zeeuw}, {Genzel}, {Hicks}, {Koss}, {Lin}, {Lutz},
  {Maciejewski}, {M{\"u}ller-S{\'a}nchez}, {Orban de Xivry}, {Ricci}, {Riffel},
  {Riffel}, {Rosario}, {Schartmann}, {Schnorr-M{\"u}ller}, {Shangguan},
  {Sternberg}, {Sturm}, {Storchi-Bergmann}, {Tacconi}, \&
  {Veilleux}}]{Davies2020}
{Davies}, R., {Baron}, D., {Shimizu}, T., {et~al.} 2020, \mnras, 498, 4150

\bibitem[{{Davies} {et~al.}(2014){Davies}, {Maciejewski}, {Hicks}, {Emsellem},
  {Erwin}, {Burtscher}, {Dumas}, {Lin}, {Malkan}, {M{\"u}ller-S{\'a}nchez},
  {Orban de Xivry}, {Rosario}, {Schnorr-M{\"u}ller}, \& {Tran}}]{Davies2014}
{Davies}, R.~I., {Maciejewski}, W., {Hicks}, E.~K.~S., {et~al.} 2014, \apj,
  792, 101

\bibitem[{{Dopita} {et~al.}(2015){Dopita}, {Ho}, {Dressel}, {Sutherland},
  {Kewley}, {Davies}, {Hampton}, {Shastri}, {Kharb}, {Jose}, {Bhatt}, {Ramya},
  {Scharw{\"a}chter}, {Jin}, {Banfield}, {Zaw}, {James}, {Juneau}, \&
  {Srivastava}}]{Dopita2015}
{Dopita}, M.~A., {Ho}, I.-T., {Dressel}, L.~L., {et~al.} 2015, \apj, 801, 42

\bibitem[{{Dullo} {et~al.}(2021){Dullo}, {Gil de Paz}, \& {Knapen}}]{Dullo2021}
{Dullo}, B.~T., {Gil de Paz}, A., \& {Knapen}, J.~H. 2021, \apj, 908, 134

\bibitem[{{Epinat} {et~al.}(2008){Epinat}, {Amram}, \& {Marcelin}}]{Epinat2008}
{Epinat}, B., {Amram}, P., \& {Marcelin}, M. 2008, \mnras, 390, 466

\bibitem[{{Fabian}(2012)}]{Fabian2012}
{Fabian}, A.~C. 2012, \araa, 50, 455

\bibitem[{{Fiore} {et~al.}(2017){Fiore}, {Feruglio}, {Shankar}, {Bischetti},
  {Bongiorno}, {Brusa}, {Carniani}, {Cicone}, {Duras}, {Lamastra}, {Mainieri},
  {Marconi}, {Menci}, {Maiolino}, {Piconcelli}, {Vietri}, \&
  {Zappacosta}}]{Fiore2017}
{Fiore}, F., {Feruglio}, C., {Shankar}, F., {et~al.} 2017, \aap, 601, A143

\bibitem[{{Fluetsch} {et~al.}(2019){Fluetsch}, {Maiolino}, {Carniani},
  {Marconi}, {Cicone}, {Bourne}, {Costa}, {Fabian}, {Ishibashi}, \&
  {Venturi}}]{Fluetsch2019}
{Fluetsch}, A., {Maiolino}, R., {Carniani}, S., {et~al.} 2019, \mnras, 483,
  4586

\bibitem[{{Foster} {et~al.}(2016){Foster}, {Pastorello}, {Roediger}, {Brodie},
  {Forbes}, {Kartha}, {Pota}, {Romanowsky}, {Spitler}, {Strader}, {Usher}, \&
  {Arnold}}]{Foster2016}
{Foster}, C., {Pastorello}, N., {Roediger}, J., {et~al.} 2016, \mnras, 457, 147

\bibitem[{{Gil de Paz} {et~al.}(2018){Gil de Paz}, {Carrasco}, {Gallego},
  {Iglesias-P{\'a}ramo}, {Cedazo}, {Garc{\'\i}a-Vargas}, {Arrillaga},
  {Avil{\'e}s}, {Bouquin}, {Carbajo}, {Cardiel}, {Carrera}, {Castillo-Morales},
  {Castillo-Dom{\'\i}nguez}, {Esteban San Rom{\'a}n}, {Ferrusca},
  {G{\'o}mez-{\'A}lvarez}, {Izazaga-P{\'e}rez}, {Lefort}, {L{\'o}pez-Orozco},
  {Maldonado}, {Mart{\'\i}nez-Delgado}, {Morales-Dur{\'a}n}, {Mujica},
  {P{\'a}ez}, {Pascual}, {P{\'e}rez-Calpena}, {Picazo}, {S{\'a}nchez-Penim},
  {S{\'a}nchez-Blanco}, {Tulloch}, {Vel{\'a}zquez}, {V{\'\i}lchez}, {Zamorano},
  {Aguerri}, {Barrado y Navascues}, {Berlanas}, {Bertone}, {Cava},
  {Catal{\'a}n-Torrecilla}, {Cenarro}, {Ch{\'a}vez}, {Dullo}, {Garc{\'\i}a},
  {Garc{\'\i}a-Rojas}, {Guichard}, {Gonz{\'a}lez-Delgado}, {Guzm{\'a}n},
  {Herrero}, {Hu{\'e}lamo}, {Hughes}, {Jim{\'e}nez-Vicente}, {Kehrig},
  {Marino}, {M{\'a}rquez}, {Masegosa}, {Mayya}, {M{\'e}ndez-Abreu},
  {Moll{\'a}}, {Mu{\~n}oz-Tu{\~n}{\'o}n}, {Peimbert}, {P{\'e}rez-Gonz{\'a}lez},
  {P{\'e}rez-Montero}, {Rodr{\'\i}guez}, {Rodr{\'\i}guez-Espinosa},
  {Rodr{\'\i}guez Merino}, {Rodr{\'\i}guez-Mu{\~n}oz}, {Rosa-Gonz{\'a}lez},
  {S{\'a}nchez-Almeida}, {S{\'a}nchez-Contreras}, {S{\'a}nchez-Bl{\'a}zquez},
  {S{\'a}nchez}, {Sarajedini}, {Silich}, {Sim{\'o}n-D{\'\i}az},
  {Tenorio-Tagle}, {Terlevich}, {Terlevich}, {Torres-Peimbert}, {Trujillo},
  {Tsamis}, \& {Vega}}]{GildePaz2018}
{Gil de Paz}, A., {Carrasco}, E., {Gallego}, J., {et~al.} 2018, in Society of
  Photo-Optical Instrumentation Engineers (SPIE) Conference Series, Vol. 10702,
  Ground-based and Airborne Instrumentation for Astronomy VII, ed. C.~J.
  {Evans}, L.~{Simard}, \& H.~{Takami}, 1070217

\bibitem[{{Gil de Paz} {et~al.}(2016){Gil de Paz}, {Carrasco}, {Gallego},
  {Iglesias-P{\'a}ramo}, {Cedazo}, {Garc{\'\i}a Vargas}, {Arrillaga},
  {Avil{\'e}s}, {Cardiel}, {Carrera}, {Castillo-Morales},
  {Castillo-Dom{\'\i}nguez}, {de la Cruz Garc{\'\i}a}, {Esteban San Rom{\'a}n},
  {Ferrusca}, {G{\'o}mez-{\'A}lvarez}, {Izazaga-P{\'e}rez}, {Lefort},
  {L{\'o}pez-Orozco}, {Maldonado}, {Mart{\'\i}nez-Delgado}, {Morales
  Dur{\'a}n}, {Mujica}, {P{\'a}ez}, {Pascual}, {P{\'e}rez-Calpena}, {Picazo},
  {S{\'a}nchez-Penim}, {S{\'a}nchez-Blanco}, {Tulloch}, {Vel{\'a}zquez},
  {V{\'\i}lchez}, {Zamorano}, {Aguerri}, {Barrado y Nav{\'a}scues}, {Bertone},
  {Cava}, {Cenarro}, {Ch{\'a}vez}, {Garc{\'\i}a}, {Garc{\'\i}a-Rojas},
  {Guichard}, {Gonz{\'a}lez-Delgado}, {Guzm{\'a}n}, {Herrero}, {Hu{\'e}lamo},
  {Hughes}, {Jim{\'e}nez-Vicente}, {Kehrig}, {Marino}, {M{\'a}rquez},
  {Masegosa}, {Mayya}, {M{\'e}ndez-Abreu}, {Moll{\'a}},
  {Mu{\~n}oz-Tu{\~n}{\'o}n}, {Peimbert}, {P{\'e}rez-Gonz{\'a}lez}, {P{\'e}rez
  Montero}, {Rodr{\'\i}guez}, {Rodr{\'\i}guez-Espinosa},
  {Rodr{\'\i}guez-Merino}, {Rodr{\'\i}guez-Mu{\~n}oz}, {Rosa-Gonz{\'a}lez},
  {S{\'a}nchez-Almeida}, {S{\'a}nchez Contreras}, {S{\'a}nchez-Bl{\'a}zquez},
  {S{\'a}nchez Moreno}, {S{\'a}nchez}, {Sarajedini}, {Silich},
  {Sim{\'o}n-D{\'\i}az}, {Tenorio-Tagle}, {Terlevich}, {Terlevich},
  {Torres-Peimbert}, {Trujillo}, {Tsamis}, \& {Vega}}]{GildePaz2016}
{Gil de Paz}, A., {Carrasco}, E., {Gallego}, J., {et~al.} 2016, in Society of
  Photo-Optical Instrumentation Engineers (SPIE) Conference Series, Vol. 9908,
  Ground-based and Airborne Instrumentation for Astronomy VI, ed. C.~J.
  {Evans}, L.~{Simard}, \& H.~{Takami}, 99081K

\bibitem[{{Gil de Paz} {et~al.}(2020){Gil de Paz}, {Pascual}, \&
  {Chamorro-Cazorla}}]{megaratools2020}
{Gil de Paz}, A., {Pascual}, S., \& {Chamorro-Cazorla}, M. 2020,
  guaix-ucm/megara-tools: v0.1.0

\bibitem[{{Giroletti} {et~al.}(2005){Giroletti}, {Taylor}, \&
  {Giovannini}}]{Giroletti2005}
{Giroletti}, M., {Taylor}, G.~B., \& {Giovannini}, G. 2005, \apj, 622, 178

\bibitem[{{Gonz{\'a}lez Delgado} {et~al.}(2005){Gonz{\'a}lez Delgado},
  {Cervi{\~n}o}, {Martins}, {Leitherer}, \& {Hauschildt}}]{GD2005}
{Gonz{\'a}lez Delgado}, R.~M., {Cervi{\~n}o}, M., {Martins}, L.~P.,
  {Leitherer}, C., \& {Hauschildt}, P.~H. 2005, \mnras, 357, 945

\bibitem[{{Gonz{\'a}lez-Mart{\'\i}n} {et~al.}(2009){Gonz{\'a}lez-Mart{\'\i}n},
  {Masegosa}, {M{\'a}rquez}, \& {Guainazzi}}]{GM2009b}
{Gonz{\'a}lez-Mart{\'\i}n}, O., {Masegosa}, J., {M{\'a}rquez}, I., \&
  {Guainazzi}, M. 2009, \apj, 704, 1570

\bibitem[{{Gonz{\'a}lez-Mart{\'{\i}}n}
  {et~al.}(2009){Gonz{\'a}lez-Mart{\'{\i}}n}, {Masegosa}, {M{\'a}rquez},
  {Guainazzi}, \& {Jim{\'e}nez-Bail{\'o}n}}]{GM2009}
{Gonz{\'a}lez-Mart{\'{\i}}n}, O., {Masegosa}, J., {M{\'a}rquez}, I.,
  {Guainazzi}, M., \& {Jim{\'e}nez-Bail{\'o}n}, E. 2009, \aap, 506, 1107

\bibitem[{Harris {et~al.}(2020)Harris, Millman, van~der Walt, Gommers,
  Virtanen, Cournapeau, Wieser, Taylor, Berg, Smith, Kern, Picus, Hoyer, van
  Kerkwijk, Brett, Haldane, del R{\'{i}}o, Wiebe, Peterson,
  G{\'{e}}rard-Marchant, Sheppard, Reddy, Weckesser, Abbasi, Gohlke, \&
  Oliphant}]{Harris2020}
Harris, C.~R., Millman, K.~J., van~der Walt, S.~J., {et~al.} 2020, Nature, 585,
  357

\bibitem[{{Harrison} {et~al.}(2016){Harrison}, {Alexander}, {Mullaney},
  {Stott}, {Swinbank}, {Arumugam}, {Bauer}, {Bower}, {Bunker}, \&
  {Sharples}}]{Harrison2016}
{Harrison}, C.~M., {Alexander}, D.~M., {Mullaney}, J.~R., {et~al.} 2016,
  \mnras, 456, 1195

\bibitem[{{Harrison} {et~al.}(2014){Harrison}, {Alexander}, {Mullaney}, \&
  {Swinbank}}]{Harrison2014}
{Harrison}, C.~M., {Alexander}, D.~M., {Mullaney}, J.~R., \& {Swinbank}, A.~M.
  2014, \mnras, 441, 3306

\bibitem[{{Harrison} {et~al.}(2018){Harrison}, {Costa}, {Tadhunter},
  {Fl{\"u}tsch}, {Kakkad}, {Perna}, \& {Vietri}}]{Harrison2018}
{Harrison}, C.~M., {Costa}, T., {Tadhunter}, C.~N., {et~al.} 2018, Nature
  Astronomy, 2, 198

\bibitem[{{Heckman}(1980)}]{Heckman1980}
{Heckman}, T.~M. 1980, \aap, 87, 152

\bibitem[{{Heckman} \& {Best}(2014)}]{Heckman2014}
{Heckman}, T.~M. \& {Best}, P.~N. 2014, \araa, 52, 589

\bibitem[{{Hermosa Mu{\~n}oz} {et~al.}(2020){Hermosa Mu{\~n}oz}, {Cazzoli},
  {M{\'a}rquez}, \& {Masegosa}}]{HM2020}
{Hermosa Mu{\~n}oz}, L., {Cazzoli}, S., {M{\'a}rquez}, I., \& {Masegosa}, J.
  2020, \aap, 635, A50

\bibitem[{{Hermosa Mu{\~n}oz} {et~al.}(2022){Hermosa Mu{\~n}oz}, {M{\'a}rquez},
  {Cazzoli}, {Masegosa}, \& {Ag{\'\i}s-Gonz{\'a}lez}}]{HM2022}
{Hermosa Mu{\~n}oz}, L., {M{\'a}rquez}, I., {Cazzoli}, S., {Masegosa}, J., \&
  {Ag{\'\i}s-Gonz{\'a}lez}, B. 2022, \aap, 660, A133

\bibitem[{{Ho} {et~al.}(1997){Ho}, {Filippenko}, \& {Sargent}}]{Ho1997}
{Ho}, L.~C., {Filippenko}, A.~V., \& {Sargent}, W.~L.~W. 1997, \apjs, 112, 315

\bibitem[{{Hota} {et~al.}(2007){Hota}, {Saikia}, \& {Irwin}}]{Hota2007}
{Hota}, A., {Saikia}, D.~J., \& {Irwin}, J.~A. 2007, \mnras, 380, 1009

\bibitem[{Hunter(2007)}]{Hunter:2007}
Hunter, J.~D. 2007, Computing in Science \& Engineering, 9, 90

\bibitem[{{Husemann} {et~al.}(2016){Husemann}, {Scharw{\"a}chter}, {Bennert},
  {Mainieri}, {Woo}, \& {Kakkad}}]{Husemann2016}
{Husemann}, B., {Scharw{\"a}chter}, J., {Bennert}, V.~N., {et~al.} 2016, \aap,
  594, A44

\bibitem[{{Husemann} {et~al.}(2019){Husemann}, {Scharw{\"a}chter}, {Davis},
  {P{\'e}rez-Torres}, {Smirnova-Pinchukova}, {Tremblay}, {Krumpe}, {Combes},
  {Baum}, {Busch}, {Connor}, {Croom}, {Gaspari}, {Kraft}, {O'Dea}, {Powell},
  {Singha}, \& {Urrutia}}]{Husemann2019}
{Husemann}, B., {Scharw{\"a}chter}, J., {Davis}, T.~A., {et~al.} 2019, \aap,
  627, A53

\bibitem[{{Kakkad} {et~al.}(2020){Kakkad}, {Mainieri}, {Vietri}, {Carniani},
  {Harrison}, {Perna}, {Scholtz}, {Circosta}, {Cresci}, {Husemann},
  {Bischetti}, {Feruglio}, {Fiore}, {Marconi}, {Padovani}, {Brusa}, {Cicone},
  {Comastri}, {Lanzuisi}, {Mannucci}, {Menci}, {Netzer}, {Piconcelli},
  {Puglisi}, {Salvato}, {Schramm}, {Silverman}, {Vignali}, {Zamorani}, \&
  {Zappacosta}}]{Kakkad2020}
{Kakkad}, D., {Mainieri}, V., {Vietri}, G., {et~al.} 2020, \aap, 642, A147

\bibitem[{{Kauffmann} {et~al.}(2003){Kauffmann}, {Heckman}, {Tremonti},
  {Brinchmann}, {Charlot}, {White}, {Ridgway}, {Brinkmann}, {Fukugita}, {Hall},
  {Ivezi{\'c}}, {Richards}, \& {Schneider}}]{Kauffmann2003}
{Kauffmann}, G., {Heckman}, T.~M., {Tremonti}, C., {et~al.} 2003, \mnras, 346,
  1055

\bibitem[{{Kenney} \& {Yale}(2002)}]{Kenney2002}
{Kenney}, J. D.~P. \& {Yale}, E.~E. 2002, \apj, 567, 865

\bibitem[{{Kewley} {et~al.}(2006){Kewley}, {Groves}, {Kauffmann}, \&
  {Heckman}}]{Kewley2006}
{Kewley}, L.~J., {Groves}, B., {Kauffmann}, G., \& {Heckman}, T. 2006, \mnras,
  372, 961

\bibitem[{{Kormendy} \& {Ho}(2013)}]{Kormendy2013}
{Kormendy}, J. \& {Ho}, L.~C. 2013, \araa, 51, 511

\bibitem[{{Kukreti} {et~al.}(2023){Kukreti}, {Morganti}, {Tadhunter}, \&
  {Santoro}}]{Kukreti2023}
{Kukreti}, P., {Morganti}, R., {Tadhunter}, C., \& {Santoro}, F. 2023, arXiv
  e-prints, arXiv:2305.03725

\bibitem[{{Liu} {et~al.}(2013){Liu}, {Zakamska}, {Greene}, {Nesvadba}, \&
  {Liu}}]{Liu2013}
{Liu}, G., {Zakamska}, N.~L., {Greene}, J.~E., {Nesvadba}, N. P.~H., \& {Liu},
  X. 2013, \mnras, 436, 2576

\bibitem[{M{\'a}rquez {et~al.}(2017)M{\'a}rquez, Masegosa, Gonz{\'a}lez-Martin,
  Hern{\'a}ndez-Garcia, Povic, Netzer, Cazzoli, \& Olmo}]{Marquez2017}
M{\'a}rquez, I., Masegosa, J., Gonz{\'a}lez-Martin, O., {et~al.} 2017,
  Frontiers in Astronomy and Space Sciences, 4, 34

\bibitem[{{Martins} {et~al.}(2005){Martins}, {Gonz{\'a}lez Delgado},
  {Leitherer}, {Cervi{\~n}o}, \& {Hauschildt}}]{Martins2005}
{Martins}, L.~P., {Gonz{\'a}lez Delgado}, R.~M., {Leitherer}, C.,
  {Cervi{\~n}o}, M., \& {Hauschildt}, P. 2005, \mnras, 358, 49

\bibitem[{{Masegosa} {et~al.}(2011){Masegosa}, {M{\'a}rquez}, {Ramirez}, \&
  {Gonz{\'a}lez-Mart{\'{\i}}n}}]{Masegosa2011}
{Masegosa}, J., {M{\'a}rquez}, I., {Ramirez}, A., \&
  {Gonz{\'a}lez-Mart{\'{\i}}n}, O. 2011, \aap, 527, A23

\bibitem[{{Mingozzi} {et~al.}(2019){Mingozzi}, {Cresci}, {Venturi}, {Marconi},
  {Mannucci}, {Perna}, {Belfiore}, {Carniani}, {Balmaverde}, {Brusa}, {Cicone},
  {Feruglio}, {Gallazzi}, {Mainieri}, {Maiolino}, {Nagao}, {Nardini}, {Sani},
  {Tozzi}, \& {Zibetti}}]{Mingozzi2019}
{Mingozzi}, M., {Cresci}, G., {Venturi}, G., {et~al.} 2019, \aap, 622, A146

\bibitem[{{Morganti}(2017)}]{Morganti2017}
{Morganti}, R. 2017, Frontiers in Astronomy and Space Sciences, 4, 42

\bibitem[{{Nagar} {et~al.}(2005){Nagar}, {Falcke}, \& {Wilson}}]{Nagar2005}
{Nagar}, N.~M., {Falcke}, H., \& {Wilson}, A.~S. 2005, \aap, 435, 521

\bibitem[{Newville {et~al.}(2014)Newville, Stensitzki, Allen, \&
  Ingargiola}]{lmfit}
Newville, M., Stensitzki, T., Allen, D.~B., \& Ingargiola, A. 2014, {LMFIT:
  Non-Linear Least-Square Minimization and Curve-Fitting for Python}

\bibitem[{{Osterbrock} \& {Ferland}(2006)}]{Osterbrock2006}
{Osterbrock}, D.~E. \& {Ferland}, G.~J. 2006, {Astrophysics of gaseous nebulae
  and active galactic nuclei, 2nd.~ed.~by D.E.~Osterbrock and
  G.J.~Ferland.~Sausalito, CA: University Science Books, 2006}

\bibitem[{{Pascual} {et~al.}(2019){Pascual}, {Cardiel}, {Gil de Paz},
  {Carasco}, {Gallego}, {Iglesias-P{\'a}ramo}, \& {Cedazo}}]{Pascual2019}
{Pascual}, S., {Cardiel}, N., {Gil de Paz}, A., {et~al.} 2019, in Highlights on
  Spanish Astrophysics X, ed. B.~{Montesinos}, A.~{Asensio Ramos},
  F.~{Buitrago}, R.~{Sch{\"o}del}, E.~{Villaver}, S.~{P{\'e}rez-Hoyos}, \&
  I.~{Ord{\'o}{\~n}ez-Etxeberria}, 227--227

\bibitem[{{Pascual} {et~al.}(2021){Pascual}, {Cardiel}, {Picazo-Sanchez},
  {Castillo-Morales}, \& {Gil De Paz}}]{MegaraDRP2020ACM}
{Pascual}, S., {Cardiel}, N., {Picazo-Sanchez}, P., {Castillo-Morales}, A., \&
  {Gil De Paz}, A. 2021, {guaix-ucm/megaradrp: v0.11}

\bibitem[{{Pellegrini} {et~al.}(2012){Pellegrini}, {Wang}, {Fabbiano}, {Kim},
  {Brassington}, {Gallagher}, {Trinchieri}, \& {Zezas}}]{Pellegrini2012}
{Pellegrini}, S., {Wang}, J., {Fabbiano}, G., {et~al.} 2012, \apj, 758, 94

\bibitem[{{Planck Collaboration} {et~al.}(2020){Planck Collaboration},
  {Aghanim, N.}, {Akrami, Y.}, {Ashdown, M.}, {Aumont, J.}, {Baccigalupi, C.},
  {Ballardini, M.}, {Banday, A. J.}, {Barreiro, R. B.}, {Bartolo, N.}, {Basak,
  S.}, {Battye, R.}, {Benabed, K.}, {Bernard, J.-P.}, {Bersanelli, M.},
  {Bielewicz, P.}, {Bock, J. J.}, {Bond, J. R.}, {Borrill, J.}, {Bouchet, F.
  R.}, {Boulanger, F.}, {Bucher, M.}, {Burigana, C.}, {Butler, R. C.},
  {Calabrese, E.}, {Cardoso, J.-F.}, {Carron, J.}, {Challinor, A.}, {Chiang, H.
  C.}, {Chluba, J.}, {Colombo, L. P. L.}, {Combet, C.}, {Contreras, D.},
  {Crill, B. P.}, {Cuttaia, F.}, {de Bernardis, P.}, {de Zotti, G.},
  {Delabrouille, J.}, {Delouis, J.-M.}, {Di Valentino, E.}, {Diego, J. M.},
  {Dor\'e, O.}, {Douspis, M.}, {Ducout, A.}, {Dupac, X.}, {Dusini, S.},
  {Efstathiou, G.}, {Elsner, F.}, {En\ss{}lin, T. A.}, {Eriksen, H. K.},
  {Fantaye, Y.}, {Farhang, M.}, {Fergusson, J.}, {Fernandez-Cobos, R.},
  {Finelli, F.}, {Forastieri, F.}, {Frailis, M.}, {Fraisse, A. A.},
  {Franceschi, E.}, {Frolov, A.}, {Galeotta, S.}, {Galli, S.}, {Ganga, K.},
  {G\'enova-Santos, R. T.}, {Gerbino, M.}, {Ghosh, T.}, {Gonz\'alez-Nuevo, J.},
  {G\'orski, K. M.}, {Gratton, S.}, {Gruppuso, A.}, {Gudmundsson, J. E.},
  {Hamann, J.}, {Handley, W.}, {Hansen, F. K.}, {Herranz, D.}, {Hildebrandt, S.
  R.}, {Hivon, E.}, {Huang, Z.}, {Jaffe, A. H.}, {Jones, W. C.}, {Karakci, A.},
  {Keih\"anen, E.}, {Keskitalo, R.}, {Kiiveri, K.}, {Kim, J.}, {Kisner, T. S.},
  {Knox, L.}, {Krachmalnicoff, N.}, {Kunz, M.}, {Kurki-Suonio, H.}, {Lagache,
  G.}, {Lamarre, J.-M.}, {Lasenby, A.}, {Lattanzi, M.}, {Lawrence, C. R.}, {Le
  Jeune, M.}, {Lemos, P.}, {Lesgourgues, J.}, {Levrier, F.}, {Lewis, A.},
  {Liguori, M.}, {Lilje, P. B.}, {Lilley, M.}, {Lindholm, V.},
  {L\'opez-Caniego, M.}, {Lubin, P. M.}, {Ma, Y.-Z.}, {Mac\'{\i}as-P\'erez, J.
  F.}, {Maggio, G.}, {Maino, D.}, {Mandolesi, N.}, {Mangilli, A.},
  {Marcos-Caballero, A.}, {Maris, M.}, {Martin, P. G.}, {Martinelli, M.},
  {Mart\'{\i}nez-Gonz\'alez, E.}, {Matarrese, S.}, {Mauri, N.}, {McEwen, J.
  D.}, {Meinhold, P. R.}, {Melchiorri, A.}, {Mennella, A.}, {Migliaccio, M.},
  {Millea, M.}, {Mitra, S.}, {Miville-Desch\^enes, M.-A.}, {Molinari, D.},
  {Montier, L.}, {Morgante, G.}, {Moss, A.}, {Natoli, P.},
  {N\o{}rgaard-Nielsen, H. U.}, {Pagano, L.}, {Paoletti, D.}, {Partridge, B.},
  {Patanchon, G.}, {Peiris, H. V.}, {Perrotta, F.}, {Pettorino, V.},
  {Piacentini, F.}, {Polastri, L.}, {Polenta, G.}, {Puget, J.-L.}, {Rachen, J.
  P.}, {Reinecke, M.}, {Remazeilles, M.}, {Renzi, A.}, {Rocha, G.}, {Rosset,
  C.}, {Roudier, G.}, {Rubi\~no-Mart\'{\i}n, J. A.}, {Ruiz-Granados, B.},
  {Salvati, L.}, {Sandri, M.}, {Savelainen, M.}, {Scott, D.}, {Shellard, E. P.
  S.}, {Sirignano, C.}, {Sirri, G.}, {Spencer, L. D.}, {Sunyaev, R.},
  {Suur-Uski, A.-S.}, {Tauber, J. A.}, {Tavagnacco, D.}, {Tenti, M.},
  {Toffolatti, L.}, {Tomasi, M.}, {Trombetti, T.}, {Valenziano, L.},
  {Valiviita, J.}, {Van Tent, B.}, {Vibert, L.}, {Vielva, P.}, {Villa, F.},
  {Vittorio, N.}, {Wandelt, B. D.}, {Wehus, I. K.}, {White, M.}, {White, S. D.
  M.}, {Zacchei, A.}, \& {Zonca, A.}}]{Planck2018}
{Planck Collaboration}, {Aghanim, N.}, {Akrami, Y.}, {et~al.} 2020, A\&A, 641,
  A6

\bibitem[{{Pogge} {et~al.}(2000){Pogge}, {Maoz}, {Ho}, \&
  {Eracleous}}]{Pogge2000}
{Pogge}, R.~W., {Maoz}, D., {Ho}, L.~C., \& {Eracleous}, M. 2000, \apj, 532,
  323

\bibitem[{{Raimundo}(2021)}]{Raimundo2021}
{Raimundo}, S.~I. 2021, \aap, 650, A34

\bibitem[{{Revalski} {et~al.}(2022){Revalski}, {Crenshaw}, {Rafelski},
  {Kraemer}, {Polack}, {Falc{\~a}o}, {Fischer}, {Meena}, {Martinez}, {Schmitt},
  {Collins}, \& {Falcone}}]{Revalski2022}
{Revalski}, M., {Crenshaw}, D.~M., {Rafelski}, M., {et~al.} 2022, \apj, 930, 14

\bibitem[{{Ricci} {et~al.}(2015){Ricci}, {Steiner}, \& {Menezes}}]{Ricci2015}
{Ricci}, T.~V., {Steiner}, J.~E., \& {Menezes}, R.~B. 2015, \mnras, 451, 3728

\bibitem[{{Rose} {et~al.}(2018){Rose}, {Tadhunter}, {Ramos Almeida},
  {Rodr{\'\i}guez Zaur{\'\i}n}, {Santoro}, \& {Spence}}]{Rose2018}
{Rose}, M., {Tadhunter}, C., {Ramos Almeida}, C., {et~al.} 2018, \mnras, 474,
  128

\bibitem[{{Santoro} {et~al.}(2020){Santoro}, {Tadhunter}, {Baron}, {Morganti},
  \& {Holt}}]{Santoro2020}
{Santoro}, F., {Tadhunter}, C., {Baron}, D., {Morganti}, R., \& {Holt}, J.
  2020, \aap, 644, A54

\bibitem[{{Sarzi} {et~al.}(2006){Sarzi}, {Falc{\'o}n-Barroso}, {Davies},
  {Bacon}, {Bureau}, {Cappellari}, {de Zeeuw}, {Emsellem}, {Fathi},
  {Krajnovi{\'c}}, {Kuntschner}, {McDermid}, \& {Peletier}}]{Sarzi2006}
{Sarzi}, M., {Falc{\'o}n-Barroso}, J., {Davies}, R.~L., {et~al.} 2006, \mnras,
  366, 1151

\bibitem[{{Sarzi} {et~al.}(2010){Sarzi}, {Shields}, {Schawinski}, {Jeong},
  {Shapiro}, {Bacon}, {Bureau}, {Cappellari}, {Davies}, {de Zeeuw}, {Emsellem},
  {Falc{\'o}n-Barroso}, {Krajnovi{\'c}}, {Kuntschner}, {McDermid}, {Peletier},
  {van den Bosch}, {van de Ven}, \& {Yi}}]{Sarzi2010}
{Sarzi}, M., {Shields}, J.~C., {Schawinski}, K., {et~al.} 2010, \mnras, 402,
  2187

\bibitem[{{Schilizzi} {et~al.}(1983){Schilizzi}, {Fanti}, {Fanti}, \&
  {Parma}}]{Schilizzi1983}
{Schilizzi}, R.~T., {Fanti}, C., {Fanti}, R., \& {Parma}, P. 1983, \aap, 126,
  412

\bibitem[{{Singh} {et~al.}(2013){Singh}, {van de Ven}, {Jahnke}, {Lyubenova},
  {Falc{\'o}n-Barroso}, {Alves}, {Cid Fernandes}, {Galbany},
  {Garc{\'\i}a-Benito}, {Husemann}, {Kennicutt}, {Marino}, {M{\'a}rquez},
  {Masegosa}, {Mast}, {Pasquali}, {S{\'a}nchez}, {Walcher}, {Wild}, {Wisotzki},
  \& {Ziegler}}]{Singh2013}
{Singh}, R., {van de Ven}, G., {Jahnke}, K., {et~al.} 2013, \aap, 558, A43

\bibitem[{{Speranza} {et~al.}(2021){Speranza}, {Balmaverde}, {Capetti},
  {Massaro}, {Tremblay}, {Marconi}, {Venturi}, {Chiaberge}, {Baldi}, {Baum},
  {Grandi}, {Meyer}, {O'Dea}, {Sparks}, {Terrazas}, \&
  {Torresi}}]{Speranza2021}
{Speranza}, G., {Balmaverde}, B., {Capetti}, A., {et~al.} 2021, \aap, 653, A150

\bibitem[{{Veale} {et~al.}(2017){Veale}, {Ma}, {Thomas}, {Greene}, {McConnell},
  {Walsh}, {Ito}, {Blakeslee}, \& {Janish}}]{Veale2017}
{Veale}, M., {Ma}, C.-P., {Thomas}, J., {et~al.} 2017, \mnras, 464, 356

\bibitem[{{Veilleux} {et~al.}(2005){Veilleux}, {Cecil}, \&
  {Bland-Hawthorn}}]{Veilleux2005}
{Veilleux}, S., {Cecil}, G., \& {Bland-Hawthorn}, J. 2005, \araa, 43, 769

\bibitem[{{Veilleux} {et~al.}(2020){Veilleux}, {Maiolino}, {Bolatto}, \&
  {Aalto}}]{Veilleux2020}
{Veilleux}, S., {Maiolino}, R., {Bolatto}, A.~D., \& {Aalto}, S. 2020, \aapr,
  28, 2

\bibitem[{{Veilleux} \& {Osterbrock}(1987)}]{Veilleux1987}
{Veilleux}, S. \& {Osterbrock}, D.~E. 1987, \apjs, 63, 295

\bibitem[{{Venturi} {et~al.}(2021){Venturi}, {Cresci}, {Marconi}, {Mingozzi},
  {Nardini}, {Carniani}, {Mannucci}, {Marasco}, {Maiolino}, {Perna},
  {Treister}, {Bland-Hawthorn}, \& {Gallimore}}]{Venturi2021}
{Venturi}, G., {Cresci}, G., {Marconi}, A., {et~al.} 2021, \aap, 648, A17

\bibitem[{{Younes} {et~al.}(2012){Younes}, {Porquet}, {Sabra}, {Reeves}, \&
  {Grosso}}]{Younes2012}
{Younes}, G., {Porquet}, D., {Sabra}, B., {Reeves}, J.~N., \& {Grosso}, N.
  2012, \aap, 539, A104

\end{thebibliography}

%
%

\begin{appendix}

\section{Additional comments for the galaxies}
\label{Appendix:IndComm}

    \noindent Here we compare the overall kinematical results of the ionised gas of our sources with respect to our previous works \cite{Masegosa2011, Cazzoli2018, HM2020, HM2022}. \\

    \noindent  \textit{NGC\,0266}: The morphology of the H$\alpha$ emission seen from the primary (and unique) component from MEGARA data closely resembles that detected with the NOT narrow band images (see HM22). Given its morphology, in HM22 the ionised gas was classified as a bubble emerging from the nucleus. However the kinematic properties derived with MEGARA for this emission are not characteristic of an outflow (see Fig.~\ref{Fig5:KinMaps_NGC0266}). 
    In the long slit spectroscopic data analysed in \cite{Cazzoli2018}, they derived the presence of two kinematical components, being the secondary blueshifted, with a width consistent with an outflow (v$^{\rm C18}_{\rm gas,S} = -296 \pm 19$\,km\,s$^{-1}$ and $\sigma^{\rm C18}_{\rm gas,S} = 475 \pm 50$\,km\,s$^{-1}$). However, these components were modelled using the [O\,I] line instead of [S\,II], as we use in our analysis of the MEGARA data (low S/N for modelling using [O\,I]), and discrepancies between the kinematics of both lines may exist due to intrinsic differences on their origin \citep[see][]{Cazzoli2018, HM2020}. In fact, with MEGARA a single component is sufficient to obtain a good fit of the lines, with a width in MEGARA consistent with that of the long-slit spectrum ($\sigma^{\rm PSF}_{\rm gas,P} = 182 \pm 14$\,km\,s$^{-1}$ and $\sigma^{\rm C18}_{\rm gas,N} = 165 \pm 17$\,km\,s$^{-1}$, respectively). Overall, the MEGARA maps show a perturbed velocity and a spiral-like H$\alpha$ flux most likely produced by internal effects of the galaxy rather than by outflows. This galaxy is known to have a central bar \citep{Epinat2008} that may be causing a gravitational torque and perturbing the gas. \\
        
    \noindent \textit{NGC\,0315}: The derived velocity dispersion for the primary component in the PSF region of MEGARA data and with the ground-based long-slit spectroscopic data (for the [S\,II] lines) are discrepant, although for the secondary component are consistent within the 1$\sigma$ error ($\sigma^{\rm C18}_{\rm gas,N} = 88 \pm 5$\,km\,s$^{-1}$ and $\sigma^{\rm C18}_{\rm gas,S} = 485 \pm 50$\,km\,s$^{-1}$; for MEGARA $\sigma^{\rm PSF}_{\rm gas,P}=195\pm38$\,km\,s$^{-1}$ and $\sigma^{\rm PSF}_{\rm gas,S}=480\pm120$\,km\,s$^{-1}$; see Table~\ref{Table:resultsKin}). The outflow classification in \cite{Cazzoli2018} was based on the width and the blueshifted velocities of the secondary component (v$^{\rm C18}_{\rm gas,S}=-209\pm 148$ km\,s$^{-1}$), for which we find redshifted velocities in MEGARA ($v_{\rm gas,S}=96\pm 77$ km\,s$^{-1}$). In any case, this secondary component is unresolved in the PSF region, explaining the general \lq Core-halo\rq\,morphology seen in the imaging data, with no other peculiar morphology (see HM22).\\
    
    \noindent \textit{NGC\,3226}: It was classified as \lq Dusty\rq\,with the imaging data in HM22 as there is a dust lane located eastward from the centre (see sharp-divided image in Fig.~2 from \citealt{Masegosa2011}), despite the innermost contours of the H$\alpha$ image show a similar morphology to the secondary component in MEGARA data (see Fig.~2 in \citealt{Masegosa2011}). In \cite{Cazzoli2018}, the secondary component was broad ($\sigma^{\rm C18}_{\rm gas,S} \sim$\,620\,km\,s$^{-1}$) and blueshifted ($v^{\rm C18}_{\rm gas,S} \sim -150$\,km\,s$^{-1}$), with a velocity consistent with the results from MEGARA within 1.5$\sigma$, but differing in the velocity dispersion (consistent only within 4$\sigma$). These differences could be produced due to the use of the [O\,I] line to model the remaining emission lines for the long-slit spectrum instead of the [S\,II] lines as we did with MEGARA data (as for NGC\,0266), given that they fell in the limits of the wavelength range (see Fig.~B7 in \citealt{Cazzoli2018}). Within the MEGARA data, the outflow may be probed by the unresolved, blueshifted secondary component. \\
    
    \noindent  \textit{NGC\,3245}: The H$\alpha$ emission in \cite{Masegosa2011} is extended in the north-south direction, aligned with the slit used with the Palomar spectroscopic data in \cite{HM2020}. The orientation and extension of the ionised gas in the imaging data is consistent with the primary component derived in the MEGARA data. 
    \cite{HM2020} detected a secondary component in the long-slit spectroscopic data, rather narrow ($\sigma^{\rm HM20}_{\rm gas,S}$ of $136\pm 21$\,km\,s$^{-1}$), and redshifted ($94\pm 17$\,km\,s$^{-1}$). However, in the MEGARA data this secondary component is barely resolved out of the PSF region, being generally blueshifted ($-214\pm 142$\,km\,s$^{-1}$) and with a width consistent with that derived for the long-slit spectrum ($198\pm 172$\,km\,s$^{-1}$). In general, the kinematics of the primary and secondary component are very similar to that of NGC\,3226 (see Sect.~\ref{Sect:Discussion}). \\

    \noindent \textit{NGC\,4278}: The ionised gas in the imaging data is compact with no particular signatures. This is in contrast to the H$\alpha$ or [O\,III] flux distributions in MEGARA (see Fig.~\ref{Fig5:KinMaps_NGC4278}), that are slightly tilted towards the gas major axis (see Table~\ref{Table:resultsKin}). 
    The spectrum studied in \cite{Cazzoli2018} was modelled with two components using simultaneously [S\,II] and [O\,I] lines, although the latter were not well constrained. The primary component with [S\,II] was almost at restframe (6$\pm$7\,km\,s$^{-1}$), with $\sigma^{\rm C18}_{\rm gas,N} = 177\pm 35$\,km\,s$^{-1}$ ([O\,I] modelling showed similar results) consistent with the MEGARA data for the same component ($\sim 150$\,km\,s$^{-1}$; see Table~\ref{Table:resultsKin}). The secondary component however was redshifted ($77\pm 34$\,km\,s$^{-1}$), and wider $\sigma^{\rm C18}_{\rm gas,S} = 240 \pm 4$\,km\,s$^{-1}$ ([O\,I] not well constrained with values of $669\pm 67$\,km\,s$^{-1}$). \\

    \noindent \textit{NGC\,4750}: The H$\alpha$ imaging data from HM22 has a similar morphology than the primary component in MEGARA data (see Fig.~\ref{Fig5:Compare}). Both works detect an enhanced region $\sim$4\arcsec\,north-east from the centre, that given the morphologies and kinematics, most likely corresponds to a star forming region that is spatially resolved in the HST image (see Fig.~\ref{Fig5:Compare}). The spectroscopic detection in \cite{Cazzoli2018} is based on the kinematics derived from the [O\,I] line. They found two kinematical components in the emission lines, with the primary being explained with rotational movements, and the secondary, blueshifted (v$^{\rm C18}_{\rm gas,S} = -296\pm$\,21\,km\,s$^{-1}$), broad ($\sigma^{\rm C18}_{\rm gas,S} = 380\pm42$\,km\,s$^{-1}$) component as a possible outflow. For our MEGARA data, the secondary component is barely resolved out of the PSF region in a bubble-like morphology emerging westwards from the centre up to $\sim$\,2.5\arcsec. It is blueshifted ($\langle$v$^{\rm MEGARA}_{\rm gas,S}\rangle = -92\pm$\,50\,km\,s$^{-1}$) and $\rangle \sigma \langle$\,297\,$\pm$\,41\,km\,s$^{-1}$ (see Table~\ref{Table:resultsKin}), consistent with the derived value in \cite{Cazzoli2018}. Considering the morphological and the kinematical signatures, this secondary component is likely tracing the emission from a compact, ionised outflow. The primary component show a low-$\sigma$ region north-east from the nucleus, consistent with star-forming regions considering the line ratios of the emission lines (see Figs.~\ref{Fig5:Compare} and~\ref{Fig:NGC4750_BPTs}). \\
    
    \noindent \textit{NGC\,5055}: The H$\alpha$ ionised gas in the imaging data (see Fig.~2 in \citealt{Masegosa2011}) is concentrated around the nucleus, where we also see a centrally peaked flux with MEGARA. This galaxy is a spiral system with both spiral arms and star forming regions falling within the MEGARA and HST FoV. The apparent asymmetry in the flux and velocity dispersion maps may be caused by the strong dust features produced by the spiral arms, that are more prominent to the south as already seen in \cite{Masegosa2011}, partially obscuring the nuclear region. 
    The star forming regions are proved by the enhanced H$\alpha$-flux regions at $\sim$2.5\arcsec\,north-east and $\sim$\,3\arcsec\,south-west from the centre in the MEGARA flux map, corresponding to low $\sigma$ in the corresponding map ($\sigma <$\,50\,km\,s$^{-1}$; see Fig.~\ref{Fig5:KinMaps_NGC5055}). These were reported as an extended emission at a PA$\sim$30$^{\circ}$ in \cite{Masegosa2011}.
    In general, considering the velocity map of the primary component, the detected ionised gas emission is more likely associated to a regular gas disc.

\section{Line ratio maps for the individual ionised gas components}
\label{Appendix:BPTs}
  
  \noindent We include here the line ratio maps for all the individual components of the ionised gas for all the targets in the sample, except NGC\,1052 (see C22). For NGC\,3245 and NGC\,5055, the [O\,I] lines are undetected, thus we only show the log([S\,II]/H$\alpha$) and log([N\,II]/H$\alpha$) maps.
  
\begin{figure*}
    \includegraphics[width=\textwidth]{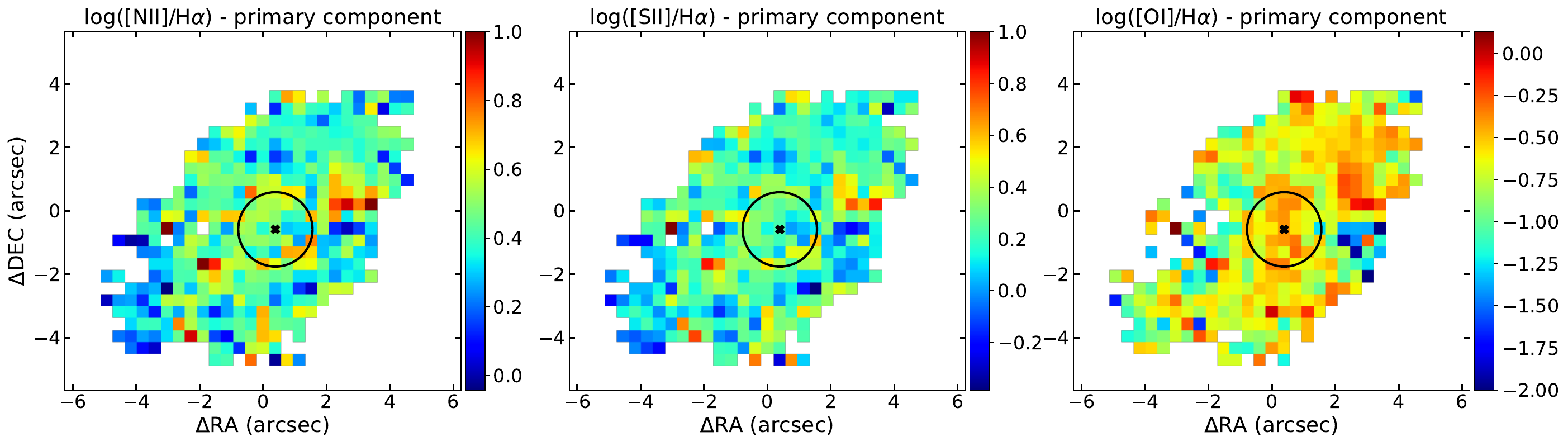}
    \caption{Line ratios for the primary component of the ionised gas in NGC\,0266. From left to right, log([N\,II]/H$\alpha$), log([S\,II]/H$\alpha$) and log([O\,I]/H$\alpha$).}
    \label{Fig:NGC0266_BPTs}
\end{figure*}

\begin{figure*}
    \includegraphics[width=\textwidth]{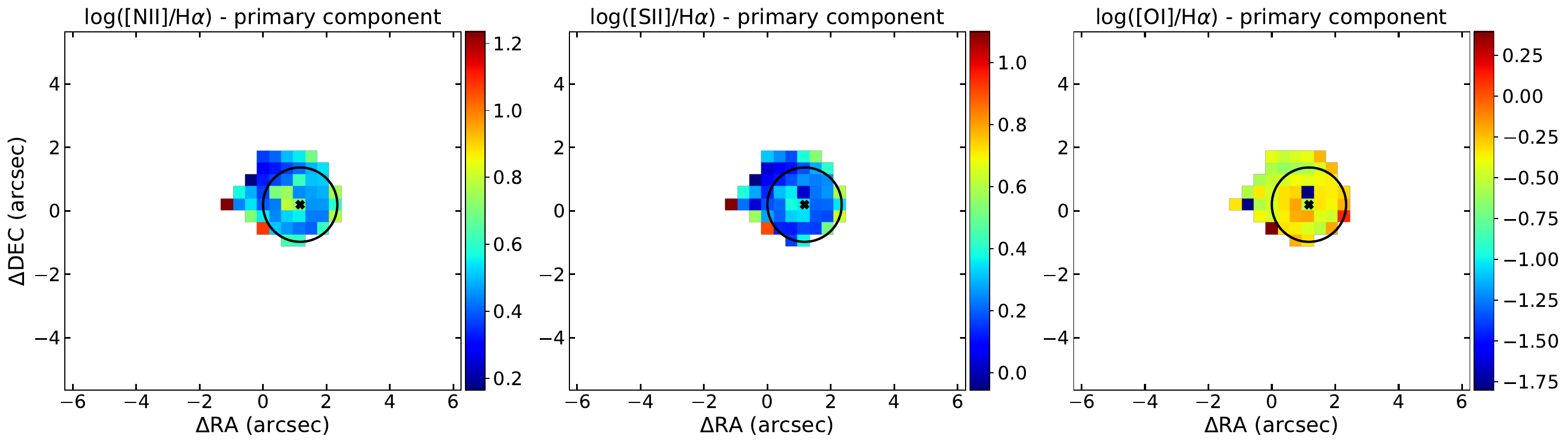}
    \includegraphics[width=\textwidth]{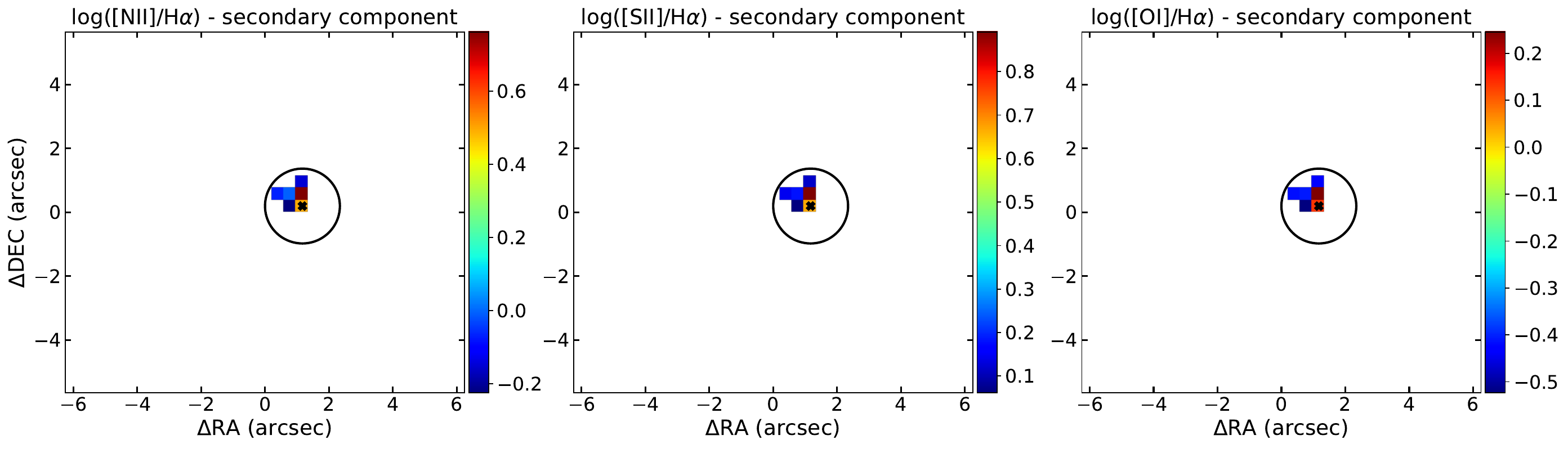}
    \caption{Line ratios for the primary (upper panels) and secondary (bottom panels) components of the ionised gas in NGC\,0315.}
    \label{Fig:NGC0315_BPTs}
\end{figure*}

\begin{figure*}
    \includegraphics[width=\textwidth]{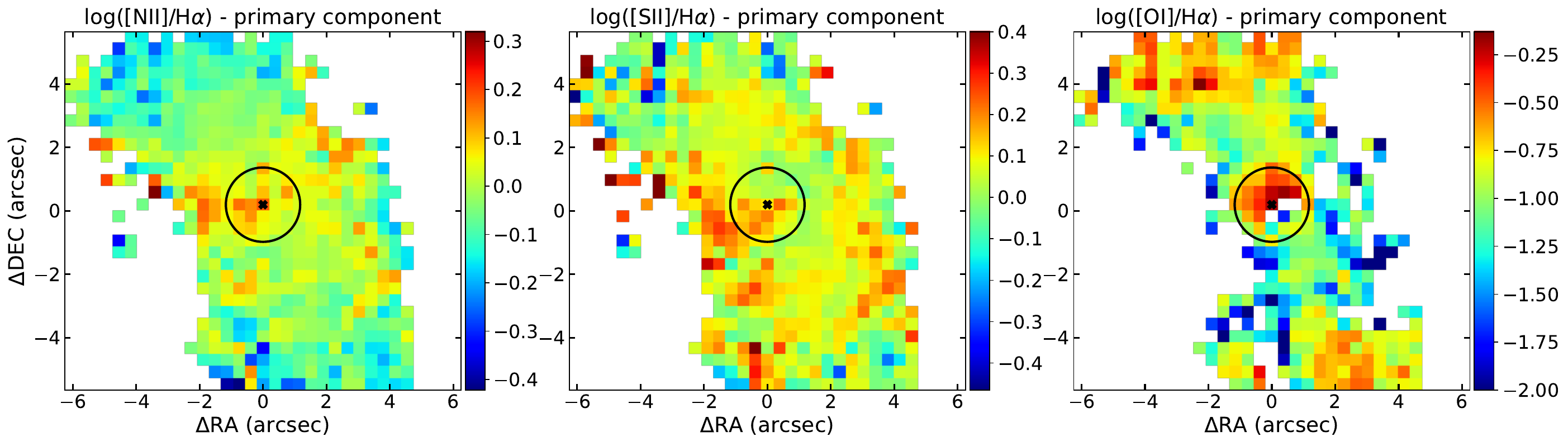}
    \includegraphics[width=\textwidth]{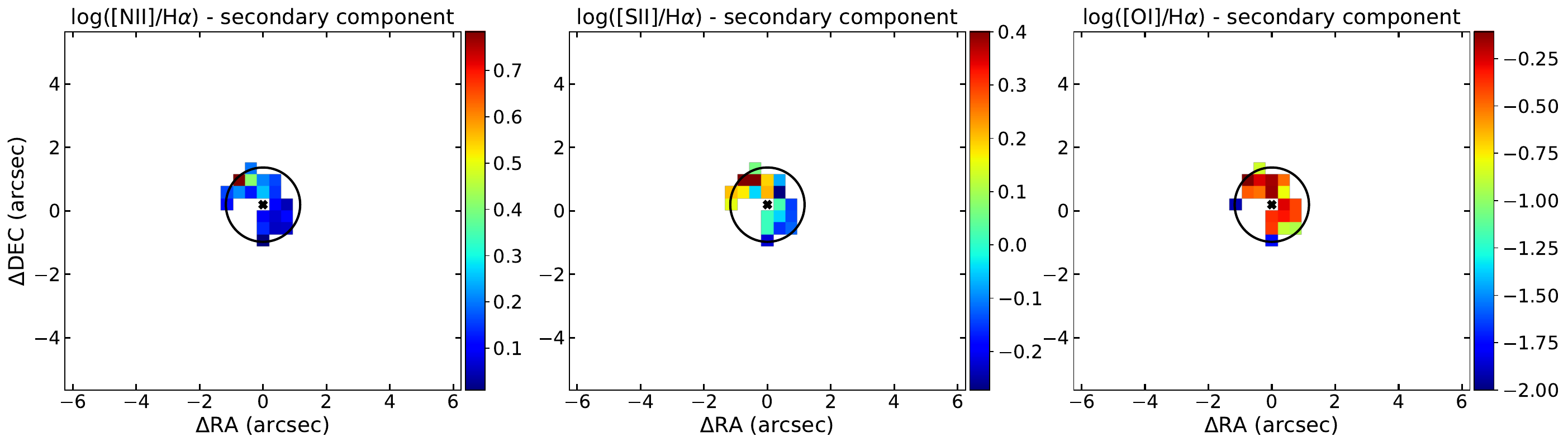}
    \caption{Line ratios for the primary (upper panels) and secondary (bottom panels) components of the ionised gas in NGC\,3226.}
    \label{Fig:NGC3226_BPTs}
\end{figure*}

\begin{figure*}
    \centering
    \includegraphics[width=\textwidth]{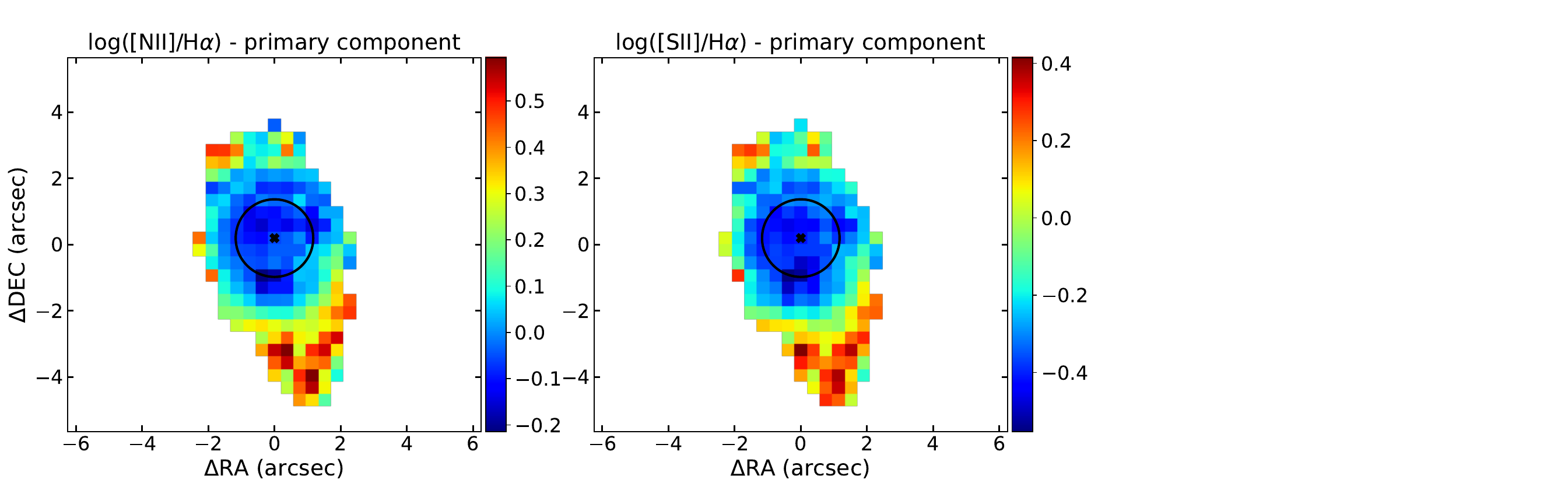}
    \includegraphics[width=\textwidth]{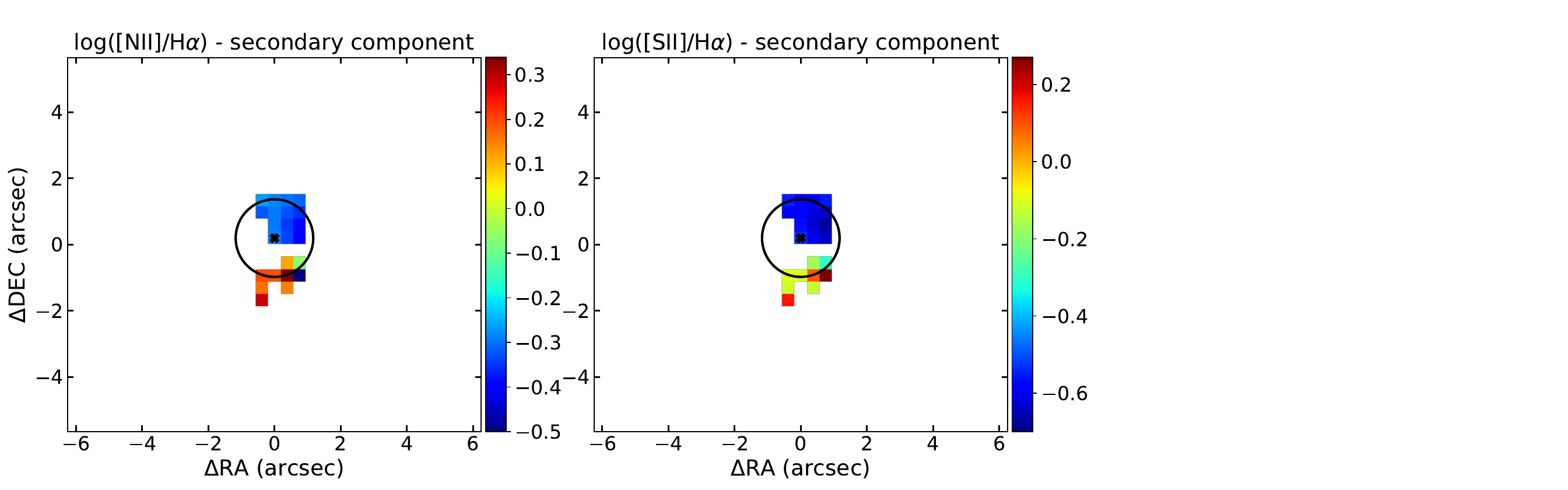}
    \caption{Line ratios for the primary (upper panels) and secondary (bottom panels) components of the ionised gas in NGC\,3245.}
    \label{Fig:NGC3245_BPTs}
\end{figure*}

\begin{figure*}
    \centering
    \includegraphics[width=\textwidth]{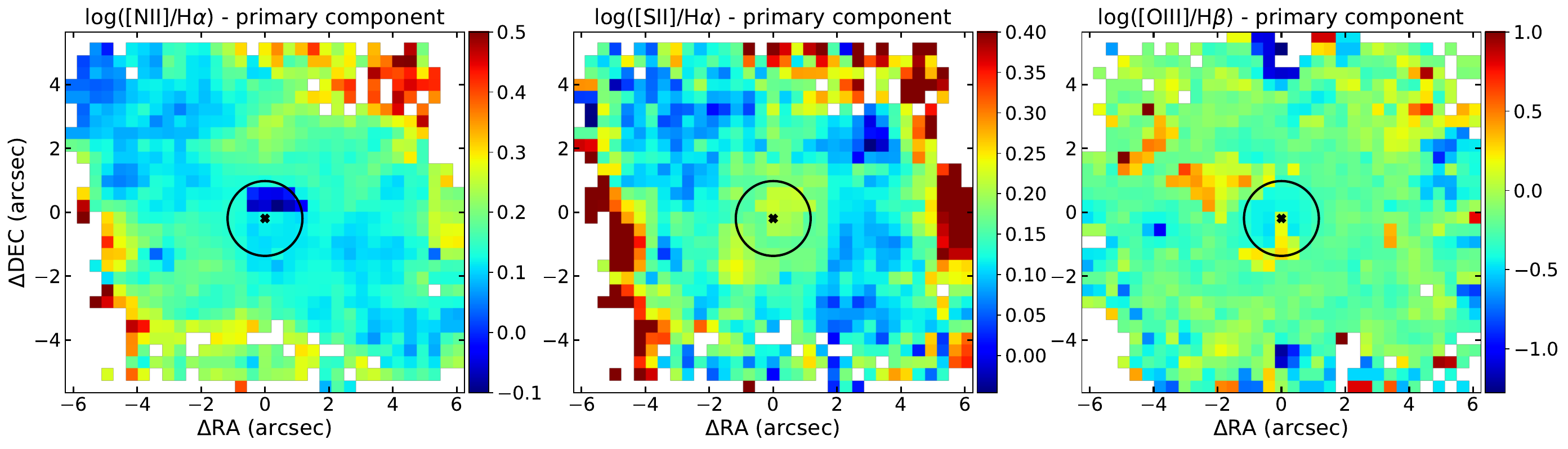}
    \includegraphics[width=\textwidth]{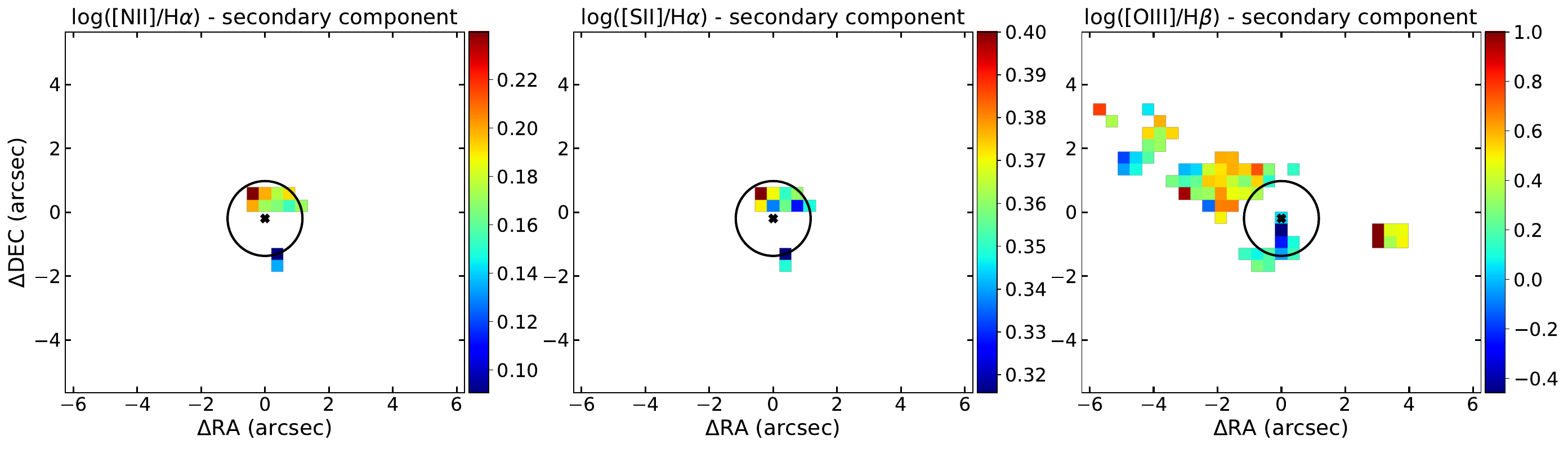}
    \caption{Line ratios for the primary and secondary components of the ionised gas in NGC\,4278.}
    \label{Fig:NGC4278_BPTs}
\end{figure*}

\begin{figure*}
    \includegraphics[width=\textwidth]{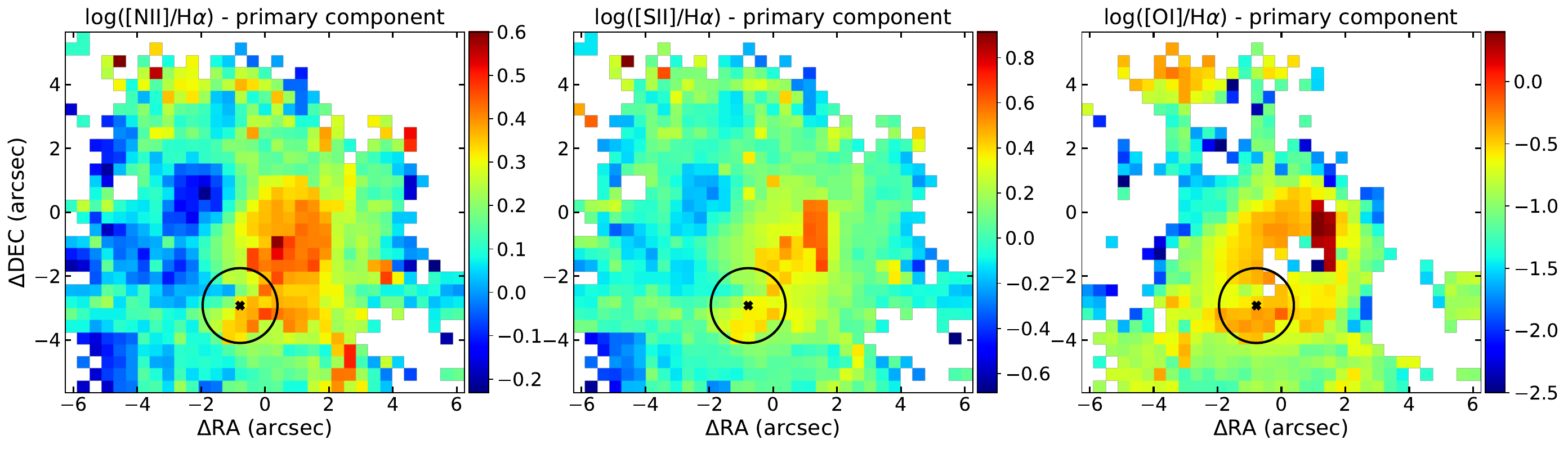}
    \includegraphics[width=\textwidth]{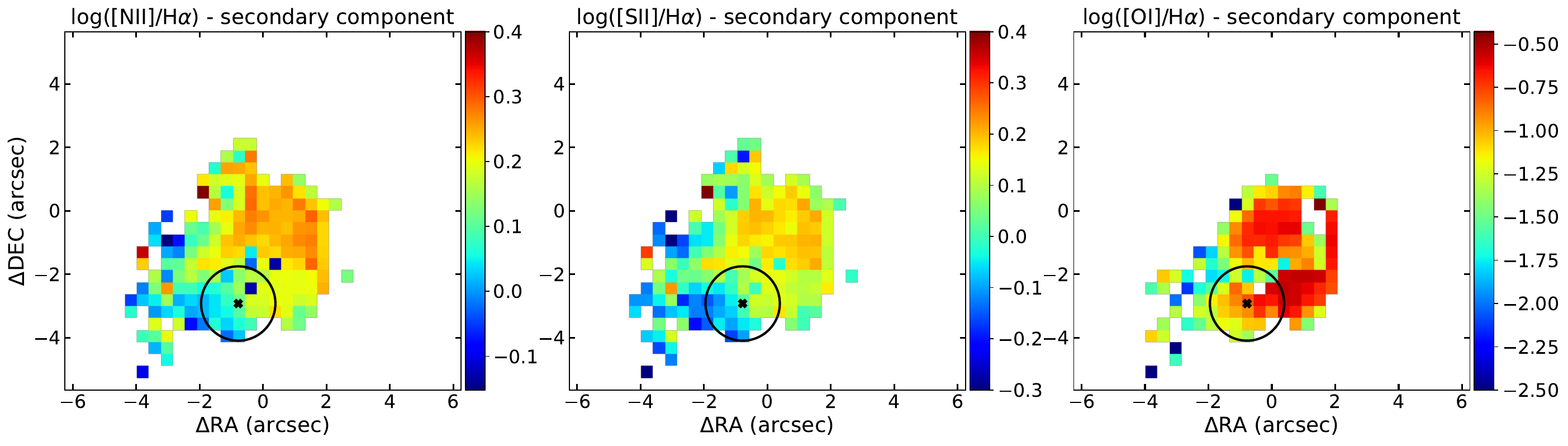}
    \includegraphics[width=\textwidth]{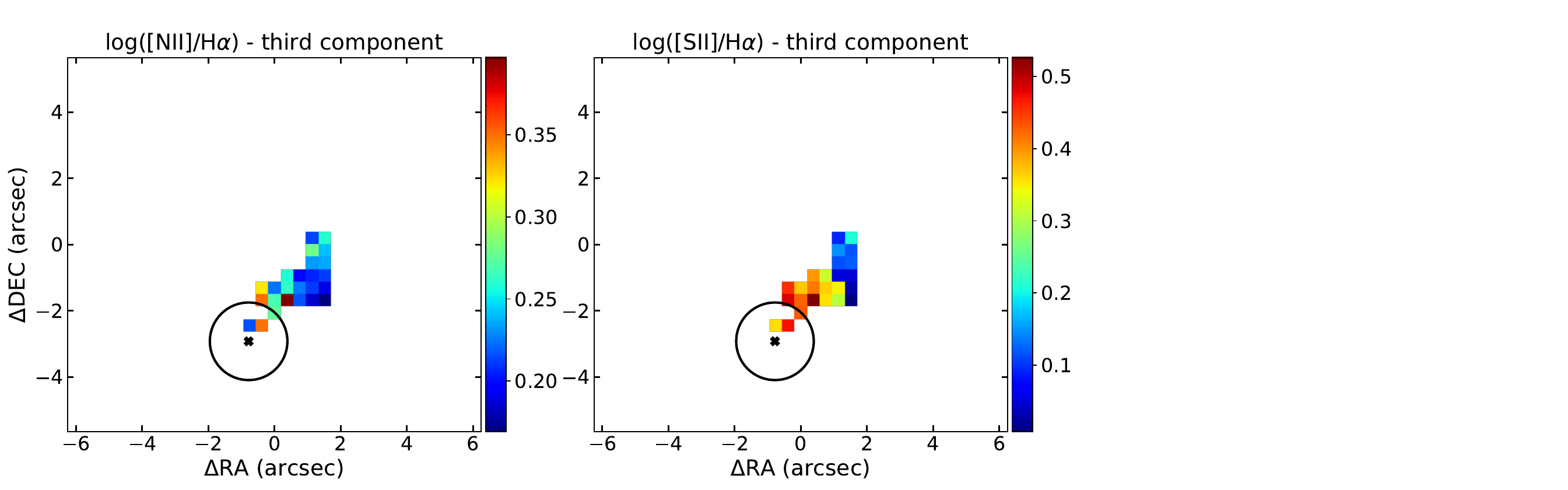}
    \caption{Line ratios for the primary, secondary and tertiary components of the ionised gas in NGC\,4438.}
    \label{Fig:NGC4438_BPTs}
\end{figure*}

\begin{figure*}
    \centering
    \includegraphics[width=\textwidth]{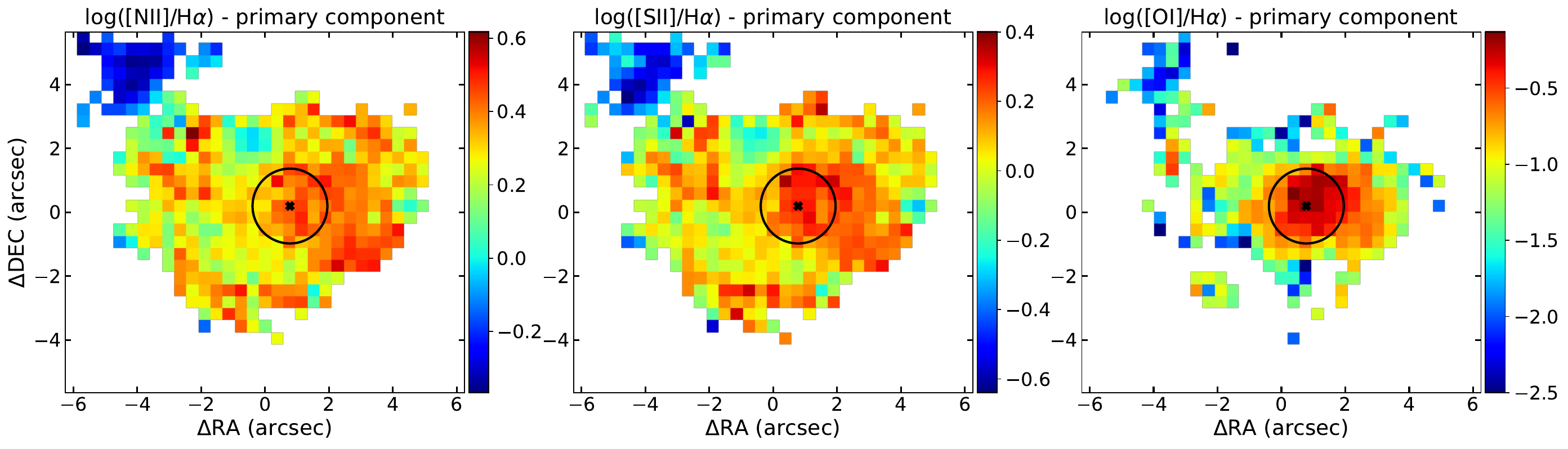}
    \includegraphics[width=.7\columnwidth]{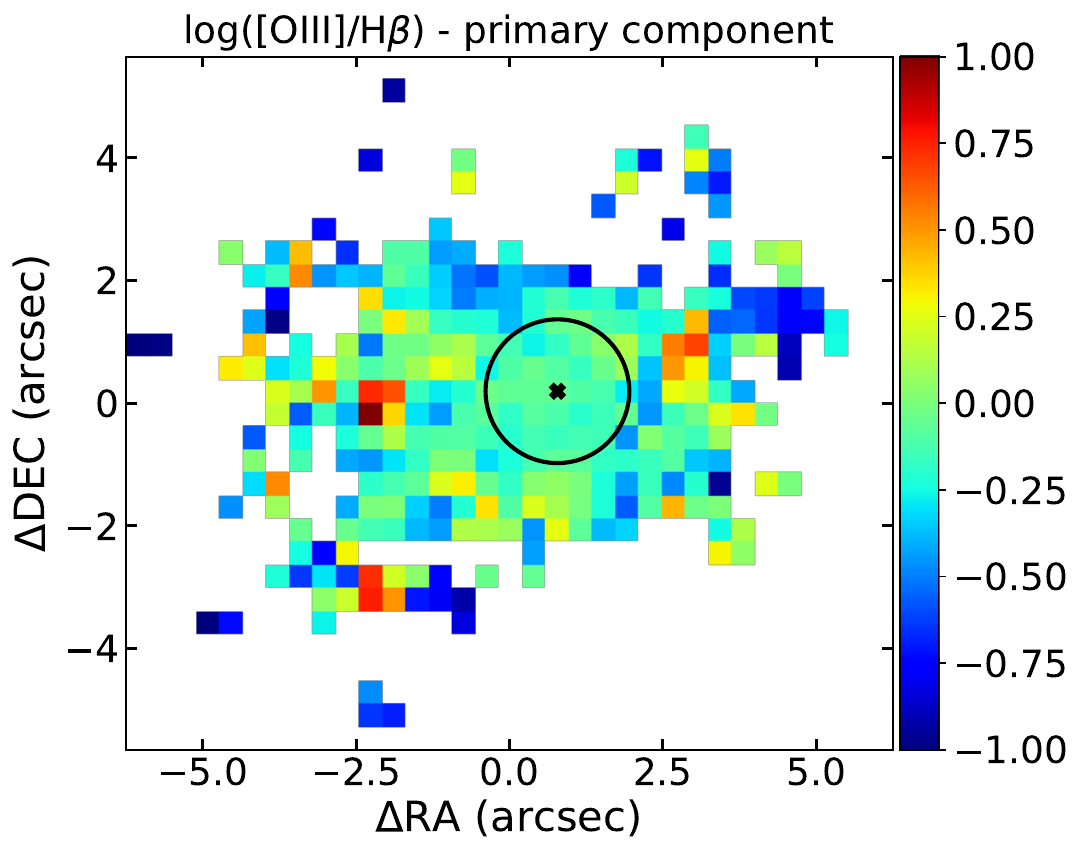}\\
    \includegraphics[width=\textwidth]{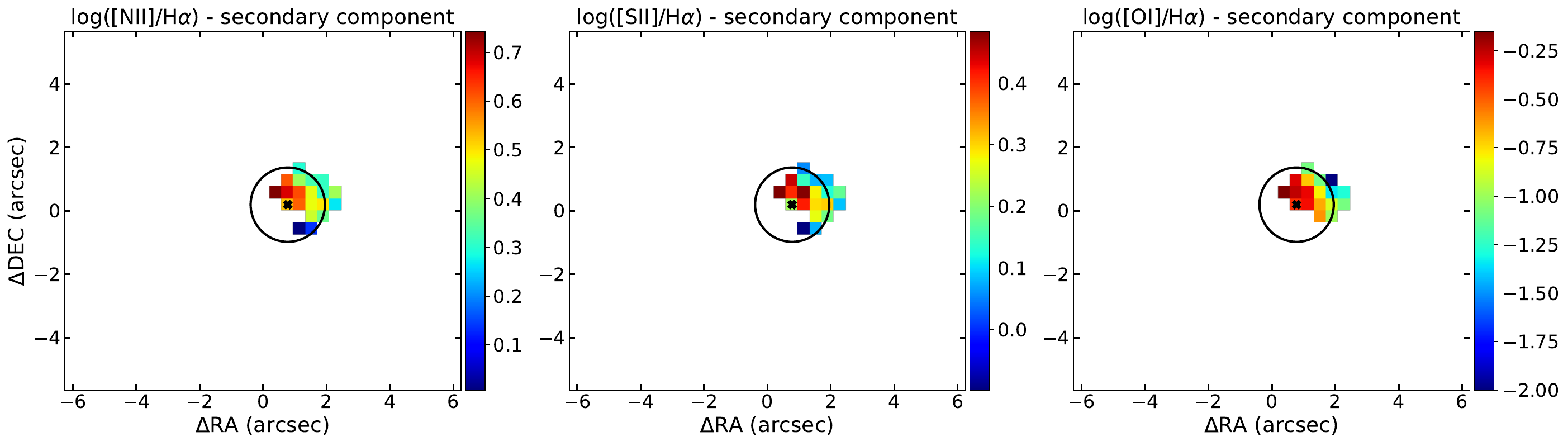}
    \caption{Line ratios for the primary (first and second panels) and secondary (bottom panel) components of the ionised gas in NGC\,4750.}
    \label{Fig:NGC4750_BPTs}
\end{figure*}

\begin{figure*}
    \centering
    \includegraphics[width=\textwidth]{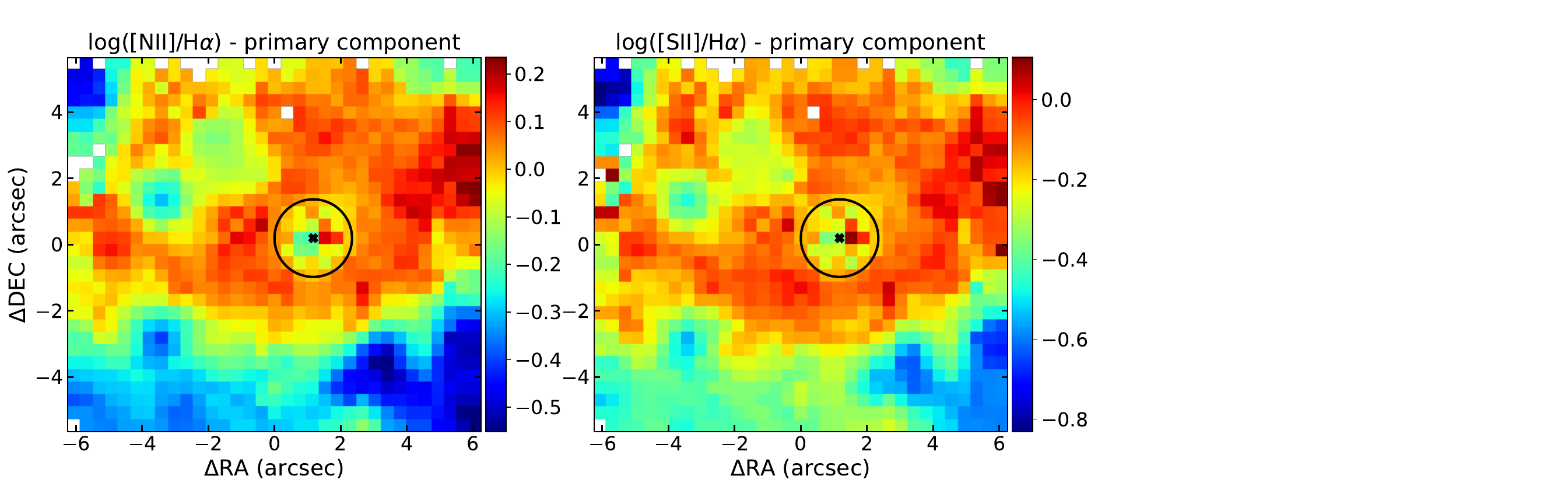}
    \caption{Line ratios for the primary component of the ionised gas in NGC\,5055.}
    \label{Fig:NGC5055_BPTs}
\end{figure*}

\section{P-V and P-S plots for the stellar component of the galaxies}
\label{AppendixA:PV-PS}

 \noindent We include here the position-velocity and position-velocity dispersion of all the galaxies showing a rotational pattern in the stellar component. The values are obtained from a simulated 1\arcsec-width long slit oriented along the stellar major axis, as done for NGC\,1052 in C22. We estimated the kinematic centre assuming a symmetry between the approaching and receding sides of the velocity curves. The kinematic centre and the photometric centre are the same for NGC\,0266, NGC\,3245 and NGC\,5055, and is shifted by $\sim$\,0.5\arcsec for NGC\,3226, NGC\,4278, NGC\,4438 and NGC\,4750. This means that both centres (i.e. photometric and kinematic) are located within the PSF region for all the targets.
 
 \begin{figure}
    \centering
    \includegraphics[width=\columnwidth]{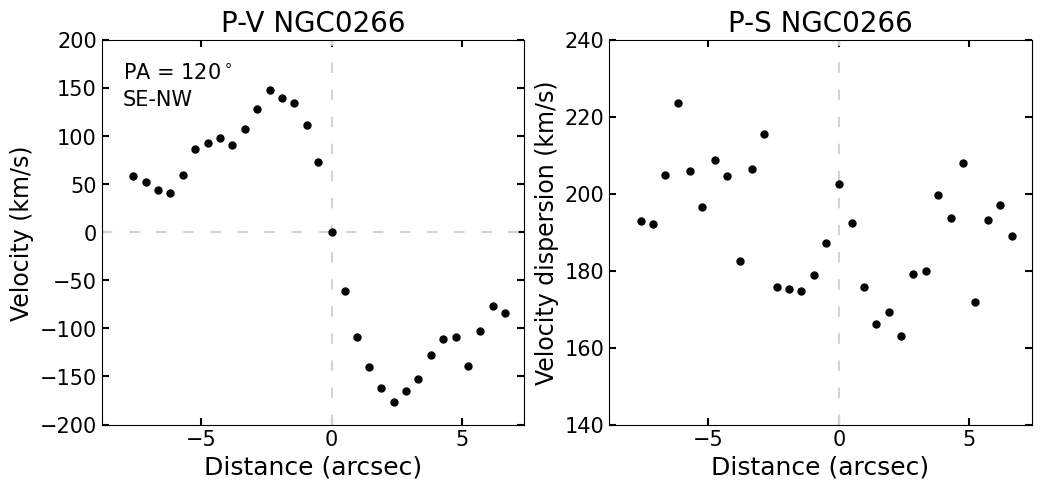}
    \includegraphics[width=\columnwidth]{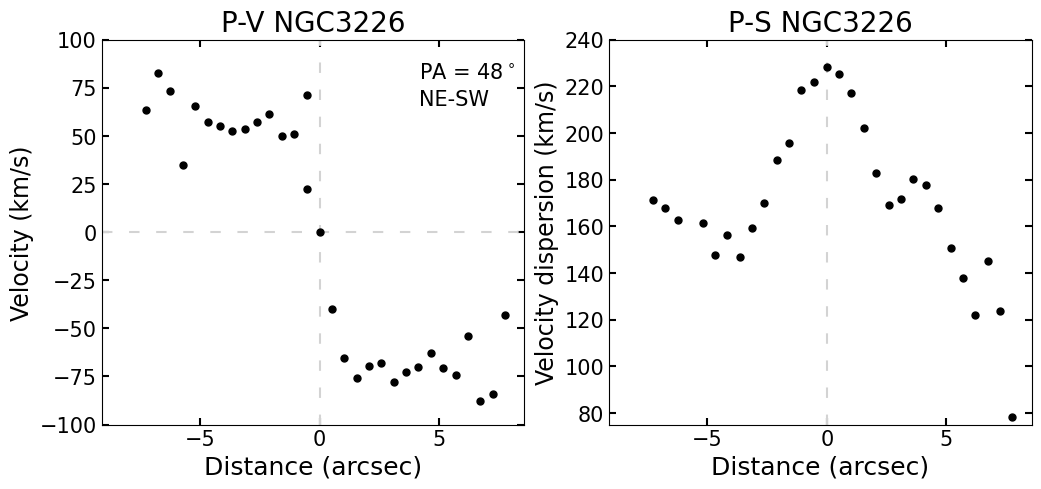}
    \includegraphics[width=\columnwidth]{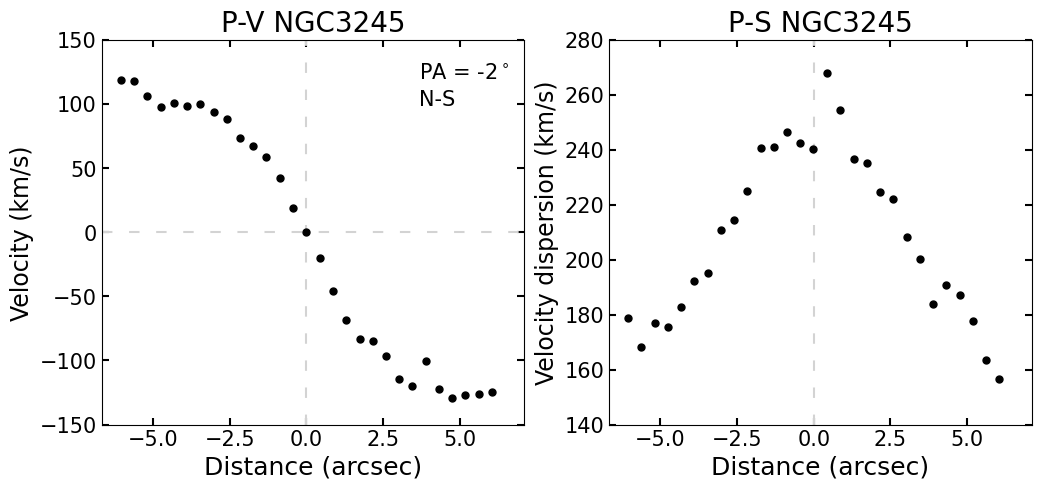}
    \includegraphics[width=\columnwidth]{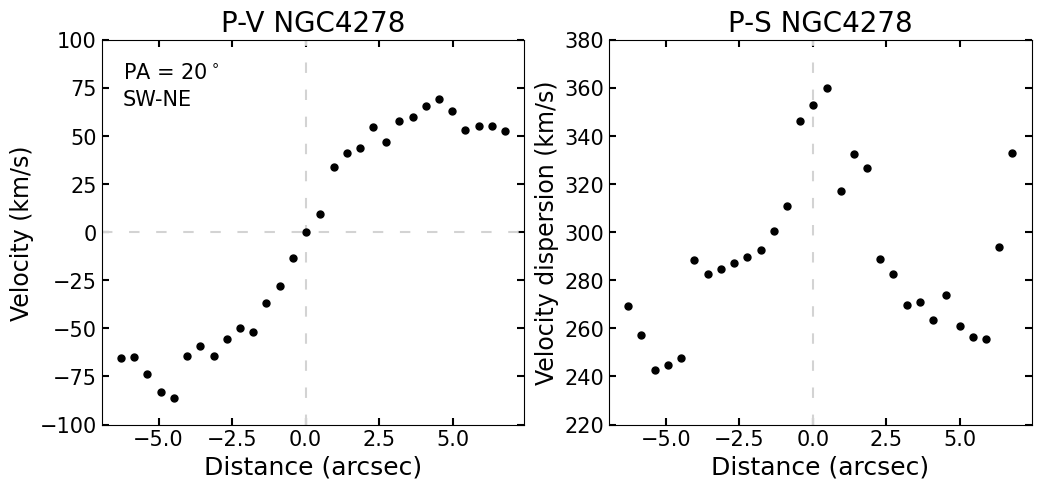}
    \includegraphics[width=\columnwidth]{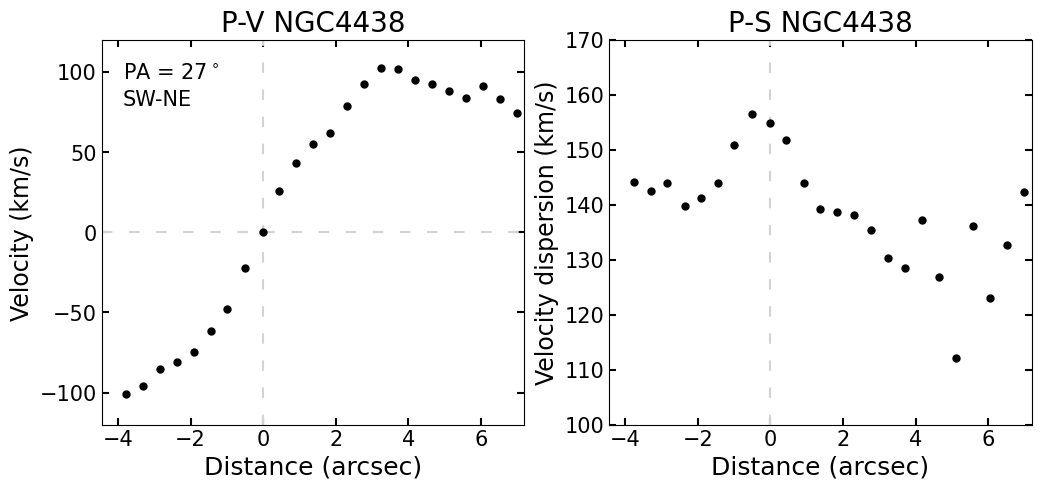}
    \caption{Position-velocity and position-velocity dispersion of the stellar component of NGC\,0266, NGC\,3226, NGC\,3245, NGC\,4278 and NGC\,4438. The x-axis represents the distance with respect to the kinematic centre. The velocities are corrected from the systemic velocity of the galaxy.}
    \label{Fig:PV-PS_1}
\end{figure}

 \begin{figure}
    \centering
    \includegraphics[width=\columnwidth]{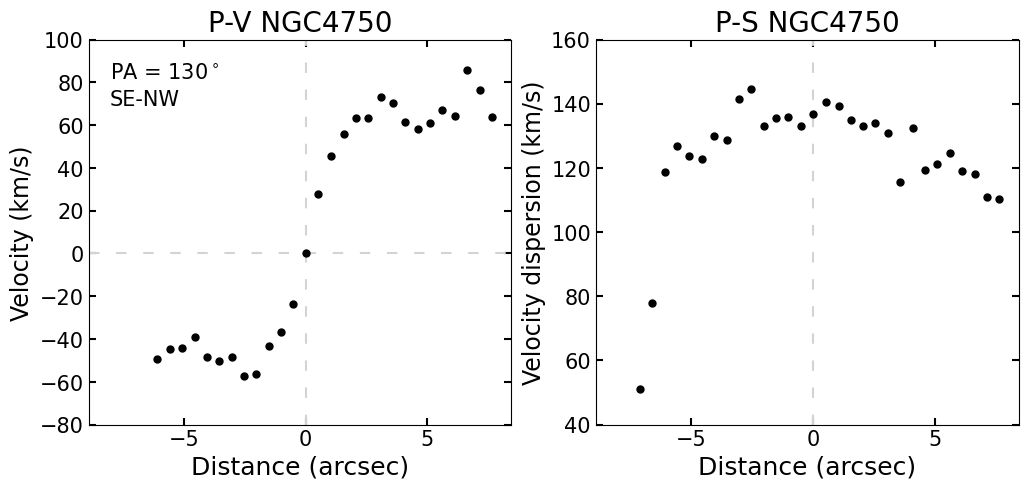}
    \includegraphics[width=\columnwidth]{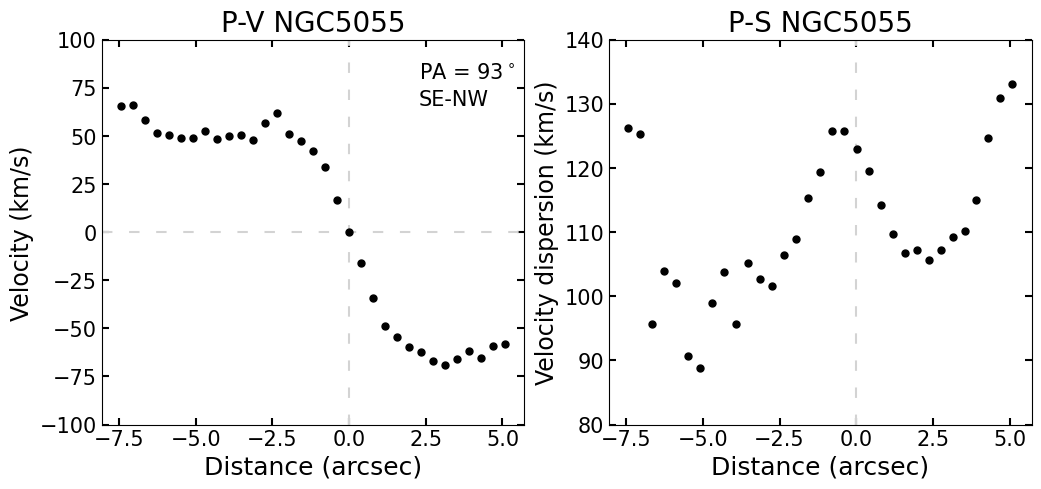}
    \caption{Same as Fig.~\ref{Fig:PV-PS_1} for NGC\,4750 and NGC\,5055.}
    \label{Fig:PV-PS_2}
\end{figure}

\end{appendix}


\end{document}